\documentclass[usenatbib]{mn2e}

\pdfoutput=1
\usepackage{graphicx}
\usepackage{amsmath}
\usepackage{amssymb}
\usepackage{float}
\usepackage{pdflscape}
\usepackage{placeins}
\usepackage{caption3}
\usepackage{capt-of}
\usepackage{longtable}
\usepackage[T1]{fontenc}
\usepackage{aecompl}
\newcommand{\Mpc}{\rm\; Mpc}

\newcommand{\km}{\rm\; km}

\newcommand{\s}{\rm\; s}

\newcommand{\Msun}{\hbox{$\rm\thinspace M_{\odot}$}}

\newcommand{\erg}{\rm\; erg}

\newcommand{\ergps}{\hbox{$\erg\s^{-1}\,$}}

\newcommand{\kmps}{\hbox{$\km\s^{-1}\,$}}

\newcommand{\kmpspMpc}{\hbox{$\kmps\Mpc^{-1}\,$}}

\providecommand{\e}[1]{\ensuremath{\times 10^{#1}}}
\providecommand{\err}[2]{\ensuremath{^{+#1}_{-#2}}}
\newcommand{\appropto}{\mathrel{\vcenter{
  \offinterlineskip\halign{\hfil$##$\cr
    \propto\cr\noalign{\kern2pt}\sim\cr\noalign{\kern-2pt}}}}}

\title[A Relationship Between Halo Mass, Cooling and AGN Heating]{A Relationship Between Halo Mass, Cooling, AGN Heating, and the Coevolution of Massive Black Holes}
\author[R. A. Main et al.]
    {\parbox[]{7.in}{R.~A.~Main$^{1,2,3}$, B.~R.~McNamara$^{1,4,5}$, P.~E.~J.~Nulsen$^{5,7}$, H.~R.~Russell$^{1,6}$, A.~N.~Vantyghem$^{1}$ \\
\footnotesize
$^1$Department of Physics and Astronomy, University of Waterloo, 200 University Avenue West, Waterloo, ON N2L 3G1, Canada \\
$^2$Department of Astronomy and Astrophysics, University of Toronto, 50 St. George Street, Toronto, ON M5S 3H4, Canada \\
$^3$Canadian Institute for Theoretical Astrophysics, University of Toronto, ON M5S 3H4, Canada \\
$^4$Perimeter Institute for Theoretical Physics, 31 Caroline Street North, Waterloo, ON N2L 2Y5, Canada \\
$^5$Harvard-Smithsonian Center for Astrophysics, 60 Garden Street, Cambridge, MA 02138, USA \\
$^6$Institute of Astronomy, Madingley Road, Cambridge, CB3 0HA\\
$^7$ICRAR, University of Western Australia, 35 Stirling Hwy, Crawley, WA 6009, Australia\\
  }
}

\begin{document}
\maketitle

\begin{abstract}

We derive X-ray mass, luminosity, and temperature profiles for 45 galaxy clusters to explore relationships between halo mass,  AGN feedback, and central cooling time.  
We find that radio--mechanical feedback power (referred to here as ``AGN power") in central cluster galaxies correlates with halo mass as P$_{\rm mech}$ $\propto$ M$^{1.55\pm0.26}$, but only in halos with central atmospheric cooling times shorter than 1 Gyr.
The trend of AGN power with halo mass is consistent with the scaling expected from a self-regulating AGN feedback loop, as well as with galaxy and central black hole co-evolution along the $M_{\rm BH} - \sigma$ relation.  AGN power in clusters with central atmospheric cooling times longer than $\sim 1$ Gyr typically lies two orders of magnitude below those with shorter central cooling times.   Galaxies centred in clusters with long central cooling times nevertheless experience ongoing and occasionally powerful AGN outbursts.
We further investigate the impact of  feedback on cluster scaling relations.  
We find $L-T$, and $M-T$ relations in clusters with direct evidence of feedback which are steeper than self-similar, but not atypical compared to previous studies of the full cluster population. 
While the gas mass rises, the stellar mass remains nearly constant with rising total mass, consistent with earlier studies.
This trend is found regardless of central cooling time, implying tight regulation of star formation in central galaxies as their halos grew, and long-term balance between AGN heating and atmospheric cooling.  Our scaling relations are presented in forms that can be incorporated easily into galaxy evolution models.

\end{abstract}


\section{Introduction}

Energetic feedback from active galactic nuclei (AGN) has governed the growth of bulge galaxies from the quasar era (Silk \& Rees 1997) through late times (\citealt{Croton_2006}, reviewed by \citealt{Mcnamara_2007, mcnamara_2012, Fabian_2012}).  In the early Universe, quasar winds regulated the growth of galaxies by staving off accretion of molecular clouds, quenching both star formation and accretion onto the quasar itself (\citealt{Silk_1998}, reviewed by \citealt{Alexander_2012}).  
As the Universe aged through redshift two to the present, the descendants of quasars -- elliptical and brightest cluster galaxies (BCGs) -- developed hot atmospheres of gas now observed in X-rays.  Hot atmospheres serve as repositories of fuel from which elliptical galaxies and their supermassive black holes form.  When an atmosphere's cooling timescale becomes shorter than the age of the system,  the gas is expected to cool and accrete onto the galaxy fuelling both star formation
and the AGN \citep{Fabian_1994}.  A significant fraction of giant elliptical galaxies harbour radio AGN signalling ongoing accretion onto  
massive black holes \citep{Heckman_2014}.  Gravitational binding energy is released in the form of radio jets and winds, which drive shock waves and buoyantly-rising bubbles into the surrounding atmospheres (e.g., \citealt{Churazov_2000, McNamara_2000, birzan_2004, Dunn_2005}). The energy released in this is captured by the intracluater medium, suppressing cooling flows and regulating star formation (e.g., \citealt{Nulsen_2005, Forman_2007, Blanton_2011, Randall_2011}).   
In addition to heating, radio jets drive hot plasma \citep{Simionescu_2008, Kirkpatrick_2011} and cold molecular 
outflows at rates of tens to hundreds of solar masses per year \citep{Edge_2001, Salome_2003, Russell_2014, McNamara_2014, Morganti_2015}.
Flow rates of this magnitude rival or exceed the star formation rates in central galaxies \citep{Kirkpatrick_2015}, thus they must be a significant aspect of the co-evolution of galaxies and massive black holes \citep{Heckman_2014}.  Much is not understood.  For example, whether the outflowing gas leaves the galaxy entirely
or returns to fuel future star formation and AGN activity is unclear \citep{McNamara_2014, Morganti_2015}.   Nevertheless, radio-mechanical feedback is clearly an important \citep{Croton_2006, bower_2006} and long-term phenomenon that has persisted for at least the past 7 Gyr  \citep{Ma_2011, Hlavacek_2013, McDonald_2013}, or a substantial fraction of cluster ages.

The degree to which feedback affects cluster atmospheres as a whole is unclear.  The most compelling, albeit indirect, evidence of AGN heating are departures of cluster scaling relations from simple power-law forms expected were clusters evolving under the influence of gravity alone \citep{Kaiser_1986, Evrard_1991}.
For example, the observed $L-T$ and $L-M$ relations in clusters are steeper than the $L \propto T^{2} $, $L \propto M^{4/3}$ relations expected from self-similar evolution (see \citealt{giodini_2013} for a review). Modeling has shown that the observed scaling relations can be reproduced when the energy released by AGN
feedback and radiative cooling are included \citep{Bialek_2001, Babul_2002, Tozzi_2001, Voit_2003, Mccarthy_2010, Short_2010, Planelles_2013}.  While the early ``preheating"  models of Kaiser, Evrard, and Henry conflict with observations of the Ly$\alpha$ forrest and the entropy profiles of cluster cores  (reviewed by \citealt{Kravtsov_2012}), recent observations indicate that continual heating by radio AGN,  preferentially in lower mass clusters,  may be able to supply the $\sim 1$ keV per particle of heat needed
to account for the observed scaling \citep{Ma_2013}.  

The declining fraction of baryons residing in stars above halo masses of $\sim 10^{12} \Msun$ (eg. \citealt{David_1995, Behroozi_2013, Gonzalez_2013}) may be another indication that AGN heating is significant on large scales.  As halos become more massive, a progressively smaller fraction of the total mass is contained in stars despite the increasing fuel supply.  At the same time, the total baryonic mass fraction (gas + stars) approaches the cosmic value with increasing halo mass.  The stellar mass fraction is a measure of the integrated star formation history. As such, observations that the stellar mass fraction declines with halo mass implies the quenching of star formation is dependent on halo mass.  It also indicates that the most massive halos are able to retain their baryons while less massive halos lose them (\citealt{Balogh_2000}, \citealt{kravtsov_2014}). The prevalence of radio-mechanical feedback  in giant elliptical galaxies \citep{Best_2007}, and high power output in galaxy clusters \citep{Vantyghem_2014, vikh_2006, Mcdonald_2015} makes
radio-mechanical  AGN heating an appealing mechanism to prevent cooling and star formation, and perhaps an agent ejecting gas from lower mass halos.

In order to determine the degree to which radio-AGN are affecting the baryonic mass fraction and scaling relations, the relationship between halo mass and jet mechanical power
must be explored.  The measurement of cluster masses and scaling relations using a variety of techniques has been a burgeoning topic
for decades, but primarily in the context of cosmology \citep{vikh_2006, Allen_2006, Mahdavi_2013, Mantz_2010} and large scale structure formation \citep{Clowe_2006, Hoekstra_2004}.  We address here the relationship between
hydrostatic cluster masses and radio-mechanical feedback from supermassive black holes using a complete, flux-limited sample of clusters.  
We focus here for the first time on relationships between jet power and halo mass. We adopt standard methodologies using X-ray cavities to estimate mechanical AGN jet power where possible, and use the relation between synchrotron luminosities and mechanical AGN power from \citet{Cavagnolo_2010} to cover our full sample.
Using archival Chandra X-ray data and the Two Micron All-Sky Survey\footnote{http://www.ipac.caltech.edu/2mass/} (2MASS, \citealt{Skrutskie_2006}), we investigate cluster scaling relations within our sample and compare them to previous measurements of the cluster population as a whole. We measure the scaling of gas mass with halo mass at $R_{2500}$ and the relationship between stellar masses of BCGs and halo mass. We investigate how AGN power scales with several cluster properties, focussing on the critically important atmospheric cooling. We focus on feedback above and below the central cooling time threshold at 1Gyr, beyond which thermal conduction is insufficient at compensating for radiative losses (eg: \citealt{Voit_2015}). We investigate the effect AGN feedback may have on on-going cluster scaling relations. We assume a cosmology with $\Omega_{M}=0.3$, $\Omega_{\Lambda}=0.7$, H$_{0} = 70\kmpspMpc$ (an approximation of the concordance cosmology, i.e. \citealt{Bennett_2014}). All errors are quoted at the 1$\sigma$ level.

\section{Sample and Data Reduction}

\subsection{Sample}

We investigate the effects of radio-mode feedback in the HIghest FLUx Galaxy Cluster Sample (HIFLUGCS), a complete flux limited X-ray sample of 64 galaxy clusters at galactic latitude $|b| > 20^{o}$ with X-ray flux $f_{x} > 2.0\e{-11} \ergps \text{cm}^{-2}$ in the 0.1-2.4 keV band \citep{Reiprich_2002}. All of the clusters in HIFLUGCS have been observed with Chandra, and 26 of these systems have observed X-ray cavities \citep{birzan_2012}. All 64 clusters in HIFLUGCS have detections or upper limits of 1.4 GHz radio emission of central radio sources. We use the 23 X-ray cavity systems with central cooling times below one Gyr to investigate the scaling relations of AGN mechanical power with cluster properties, since the AGN power in these systems is correlated with the cluster-scale properties of the ICM (\citealt{Mittal_2009}, section \ref{sec:ncc}). We refer to this subset of 23 X-ray cavity systems as our primary sample. We use the 1.4 GHz emission to investigate AGN heating using mechanical power in the full HIFLUGCS sample in Section \ref{sec:ncc}. Values used for the HIFLUGCS systems not in our primary sample are taken from the literature, and are referenced throughout.

We include an additional 27 cool-core clusters with radio cavities embedded in their X-ray atmospheres systems to extend the dynamic range of our sample. We refer to this as the extended sample throughout. This extended sample is comprised of the systems in \citet{rafferty_2006} which are not in HIFLUGCS, and the clusters from the Massive Cluster Survey (MACS, \citealt{Ebeling_2001}) with clearly-defined cavities from \citet{Hlavacek_2012}. We additionally include the giant elliptical galaxies NGC5813 and NGC5846. Table \ref{table:obslist} lists our sample and cleaned Chandra exposure times of all observations used.

\subsection{Chandra Data Reduction}
All Chandra observations were reprocessed with \textsc{ciao} 4.6 using \textsc{caldb} 4.5.9. Events were corrected for the time-dependent gain and charge transfer inefficiency and then filtered to remove those with bad grades. 
The improved background screening available in \textsc{vfaint} mode was used whenever possible.
Background light curves were extracted from the level 2 event files, and were filtered for flares using the \textsc{lc\_clean}\footnote{http://cxc.cfa.harvard.edu/contrib/maxim/acisbg/} routine of M. Markevitch. Blank-sky backgrounds were extracted for each observation, processed identically to the event files, and reprojected to the sky position of the corresponding event files. The blank-sky backgrounds were normalized to match the 9.5-12.0 keV flux in the data set. All observations used, and final cleaned exposure times are detailed in Table \ref{table:obslist}.

\begin{table*}

\caption{Target list and properties of our sample.}
\begin{tabular}{l c c c c c c c c}
\hline
System & z & Exposure$^{a}$ & N$_{\rm H}$$^{b}$ & m$_{K}$ & M$_{K}$ & r$_{e}$$^{c}$ & BCG Stellar Mass & M$_{BH}$\\
& & (ks) & ($10^{22}$cm$^{-2}$) & & & (kpc) & (10$^{11}$ M$_{\odot}$) & (10$^{9}$ M$_{\odot}$) \\
\hline
2A0335+096 & 0.035 & 82.3 & 0.224 & 9.81$\pm0.05$ & -26.18$\pm0.05$ & 44.3 & $5.33\pm0.26$ & 1.2$\err{2.7}{0.7}$ \\
A85 & 0.055 & 37.4 & 0.039$^{*}$ & 10.09$\pm0.04$ & -26.72$\pm0.04$ & 50.8 & $8.83\pm0.36$ & 2.0$\err{4.2}{1.1}$ \\
A133 & 0.057 & 109.4 & 0.0153 & 10.51$\pm0.06$ & -26.35$\pm0.06$ & 28.9 & $6.28\pm0.33$ & 1.4$\err{3.1}{0.8}$ \\
A262 & 0.016 & 108.6 & 0.089$^{*}$ & 8.75$\pm0.03$ & -25.66$\pm0.03$ & 28.9 & $3.31\pm0.08$ & 0.8$\err{1.7}{0.4}$ \\
A478 & 0.088 & 135.2 & 0.281$^{*}$ & 11.31$\pm0.07$ & -26.68$\pm0.07$ & 67.8 & $8.49\pm0.51$ & 1.9$\err{4.1}{1.0}$ \\
A496 & 0.033 & 52.4 & 0.04 & 9.81$\pm0.02$ & -26.26$\pm0.04$ & 41.9 & $5.79\pm0.22$ & 1.3$\err{2.9}{0.7}$ \\
A1795 & 0.063 & 391.5 & 0.041$^{*}$ & 10.60$\pm0.08$ & -26.47$\pm0.08$ & 69.7 & $6.99\pm0.52$ & 1.6$\err{3.4}{0.9}$ \\ 
A2029 & 0.077 & 74.7 & 0.033 & 10.11$\pm0.05$ & -27.42$\pm0.05$ & 72.2 & $16.7\pm0.8$ & 3.6$\err{7.7}{1.9}$ \\
A2052 & 0.036 & 486.8 & 0.027 & 9.55$\pm0.06$ & -26.33$\pm0.06$ & 51.2 & $6.17\pm0.31$ & 1.4$\err{3.0}{0.8}$ \\
A2199 & 0.030 & 119.6 & 0.039$^{*}$ & 9.17$\pm0.03$ & -26.36$\pm0.03$ & 38.9 & $6.31\pm0.19$ & 1.5$\err{3.1}{0.8}$ \\
A2204 & 0.152 & 69.7 & 0.061 & 12.24$\pm0.19$ & -26.73$\pm0.19$ & 81.8 & $8.9\pm1.5$ & 2.0$\err{4.3}{1.1}$ \\
A2597 & 0.085 & 108.5 & 0.0246 & 9.66$\pm0.04$ & -25.63$\pm0.04$ & 40.6 & $3.24\pm0.31$ & 0.8$\err{1.7}{0.4}$ \\
A4059 & 0.048 & 84.9 & 0.012 & 9.82$\pm0.05$ & -26.67$\pm0.05$ & 58.4 & $8.44\pm0.40$ & 1.9$\err{4.1}{1.0}$ \\
Centaurus & 0.011 & 179.2 & 0.0854 & 7.14$\pm0.02$ & -26.33$\pm0.02$ & 36.0 & $6.15\pm0.14$ & 1.4$\err{3.0}{0.8}$ \\
Hydra A & 0.055 & 152.6 & 0.0425 & 10.90$\pm0.06$ & -25.91$\pm0.06$ & 33.8 & $4.18\pm0.24$ & 1.0$\err{2.1}{0.5}$ \\
MKW3S & 0.045 & 48.3 & 0.0286 & 10.85$\pm0.06$ & -25.54$\pm0.06$ & 32.8 & $2.96\pm0.17$ & 0.7$\err{1.5}{0.4}$ \\
NGC507 & 0.017 & 37.8 & 0.06 & 8.30$\pm0.02$ & -25.94$\pm0.02$ & 28.4 & $4.31\pm0.09$ & 1.0$\err{2.2}{0.5}$ \\
NGC1399 & 0.0048 & 145.2 & 0.0138 & 6.31$\pm0.03$ & -25.26$\pm0.03$ & 22.0 & $2.29\pm0.06$ & 0.6$\err{1.2}{0.3}$ \\
NGC1550 & 0.012 & 89.0 & 0.12$^{*}$ & 8.77$\pm0.03$ & -24.89$\pm0.03$ & 15.6 & $1.62\pm0.04$ & 0.4$\err{0.9}{0.2}$ \\
NGC4636 & 0.0031 & 133.4 & 0.0185 & 6.42$\pm0.04$ & -24.20$\pm0.04$ & 20.8 & $0.87\pm0.03$ & 0.4$\err{0.8}{0.2}$ \\
NGC5044 & 0.0093 & 82.5 & 0.051 & 7.71$\pm0.02$ & -25.31$\pm0.02$ & 16.3 & $2.40\pm0.04$ & 0.6$\err{1.2}{0.3}$ \\
PKS1404-267 & 0.022 & 83.5 & 0.0431 & 9.57$\pm0.03$ & -25.28$\pm0.03$ & 22.8 & $2.34\pm0.07$ & 0.6$\err{1.2}{0.3}$ \\
Sersic159/03 & 0.058 & 90.3 & 0.039$^{*}$ & 10.59$\pm0.10$ & -26.33$\pm0.10$ & 73.4 & $6.13\pm0.55$ & 1.4$\err{3.0}{0.8}$ \\
\hline
3C388 & 0.0917 & 32.2 & 0.0562 & 11.67$\pm0.06$ & -26.24$\pm0.06$ & 41.0 & - & 1.3$\err{2.8}{0.7}$ \\
3C401 & 0.201 & 30.8 & 0.0535 & - & - & - & - & - \\
4C55.16 & 0.241 & 64.5 & 0.0449 & 13.84$\pm0.13$ & -26.09$\pm0.13$ & 45.7 & - & 1.2$\err{2.5}{0.6}$ \\
A1835 & 0.253 & 177.0 & 0.02 & 12.67$\pm0.14$ & -27.37$\pm0.14$ & 70.8 & - & 3.4$\err{7.4}{1.8}$ \\
Cygnus A & 0.056 & 182.1 & 0.28$^{*}$ & 10.28$\pm0.06$ & -26.70$\pm0.06$ & 52.4 & - & 1.9$\err{4.1}{1.0}$ \\
HCG62 & 0.0137 & 115.5 & 0.0355 & 8.63$\pm0.03$ & -25.36$\pm0.03$ & 24.5 & - & 0.6$\err{1.3}{0.3}$ \\
Hercules A & 0.154 & 95.0 & 0.06 & 12.55$\pm0.11$ & -24.46$\pm0.09$ & 56.1 & - & 1.6$\err{3.4}{0.8}$ \\
MACSJ0159.8-0849 & 0.405 & 64.4 & 0.020 & - & - & - & - & - \\
MACSJ0242.5-0253 & 0.314 & 8.0 & 0.029 & - & - & - & - & - \\
MACSJ0429.6-0253 & 0.399 & 19.3 & 0.043 & - & - & - & - & - \\
MACSJ0547.0-3904 & 0.319 & 19.2 & 0.037 & - & - & - & - & - \\
MACSJ0913.7+4056 & 0.442 & 69.3 & 0.016 & - & - & - & - & - \\
MACSJ1411.3+5212 & 0.464 & 74.1 & 0.0138 & - & - & - & - & - \\
MACSJ1423.8+2404 & 0.543 & 105.5 & 0.022 & - & - & - & - & - \\
MACSJ1532.8+3021 & 0.345 & 84.5 & 0.0248 & - & - & - & - & - \\
MACSJ1720.2+3536 & 0.391 & 53.7 & 0.036 & - & - & - & - & - \\
MACSJ1931.8-2634 & 0.352 & 91.0 & 0.08 & - & - & - & - & - \\
MACSJ2046.0-3430 & 0.423 & 44.1 & 0.047 & - & - & - & - & - \\
MACSJ2140.2-2339 & 0.313 & 33.2 & 0.036 & - & - & - & - & - \\
MS0735.6+7421 & 0.216 & 447.2 & 0.031 & 13.32$\pm0.17$ & -26.37$\pm0.17$ & 59.1 & - & 1.5$\err{3.1}{0.8}$ \\
NGC5813 & 0.0065 & - & - & 7.41$\pm0.03$ & -24.83$\pm0.03$ & 21.8 & - & 0.4$\err{0.8}{0.2}$ \\
NGC5846 & 0.0057 & - & - & 6.94$\pm0.02$ & -25.02$\pm0.02$ & 14.5 & - & 0.5$\err{1.0}{0.2}$ \\
Perseus & 0.018 & 449.5 & 0.133 & 8.13$\pm0.04$ & -26.33$\pm0.04$ & 49.4 & - & 1.4$\err{3.0}{0.8}$ \\
PKS0745-191 & 0.103 & 116.0 & 0.415 & 11.49$\pm0.09$ & -26.82$\pm0.09$ & 52.0 & - & 2.2$\err{4.6}{1.1}$ \\
RBS797 & 0.354 & 38.3 & 0.0256 & - & - & - & - & - \\
Zw2701 & 0.215 & 95.8 & 0.007 & 13.42$\pm0.17$ & -26.25$\pm0.17$ & 36.0 & - & 1.3$\err{2.8}{0.7}$ \\
Zw3146 & 0.291 & 34.0 & 0.0224 & 13.88$\pm0.28$ & -26.46$\pm0.28$ & 65.5 & - & 1.6$\err{3.4}{0.8}$ \\
\hline
\end{tabular}
\\
Above the first horizontal line comprises our Primary Sample.  The full table comprises the Extended sample. Notes:  $^{a}$Total cleaned exposure time. $^{b}$N$_{\rm H}$ values marked with an asterisk are significantly different from the galactic value of \citet{kalberla}. $^{c}$The extrapolation radius of the 2MASS K-band light profile.

\label{table:obslist}

\end{table*}

\section{Analysis}

\subsection{X-ray Analysis}

\subsubsection{Spectral Extraction}
Spectra were extracted from concentric circular annuli centred on the cluster centre. For systems with central surface brightness cusps, the centre was taken to be the position of the brightest pixel. For systems with no obvious central surface brightness peak, the cluster centre was taken to be the centroid of the X-ray emission, computed iteratively up to three times. 
Point sources were detected using the \textsc{wavdetect} \citep{Freeman_2002} wavelet algorithm in \textsc{ciao}, along with an image of the point spread function (PSF) to account for the degradation of the off-axis PSF. These point sources were confirmed by eye, and masked out in further analysis. Any known extended sources, or cluster substructure associated with mergers was also masked out of further analysis. The masked regions typically accounted for only a few percent of the total area.

Annuli enclose a minimum of $\sim$3000 counts permitting temperatures to be measured accurately in deprojection. Fewer source counts were required for low-temperature systems as emission lines make their temperature easier to determine. For systems with a very high number of source counts, we are limited by computational time of spectral fitting and error propagation rather than source counts. For these systems, the number of counts per annulus is chosen such that there are no more than $\sim$10-12 annuli per system. Spectra were extracted from these annuli in \textsc{ciao}, and were grouped to have at least 30 counts per energy channel. Weighted redistribution matrix files (RMFs) were extracted using \emph{mkacisrmf}, and weighted auxiliary response files (ARFs) were created using \emph{mkwarf}. Spectra for observations on the same chip were summed together for observations of similar time period, and were kept separate otherwise. The ARFs and RMFs for the summed spectra were weighted according to the relative number of counts in each spectrum.  The area lost to masked point sources, chip gaps, and so forth were accounted for.

To check for residual soft background emission, spectra were extracted from regions without any cluster emission. These spectra were then compared with the blank-sky backgrounds for consistency. In the case that the soft background was inconsistent with the blank-sky background, the residual emission was modelled by one or two soft thermal models with $Z/Z_{\odot}=1$, $z=0$ as described in \citet{vikh_2005}. The normalization was allowed to be negative to account for an over-subtraction of the soft backgrounds. In systems where the cluster emission is non-negligible over the entire detector, the additional soft background component is fit simultaneously with the cluster emission in the outermost annulus. The model for the soft background emission was scaled by the area of each annulus and added as a \textsc{corrfile} to each spectrum.

\subsubsection{Mass Profiles: Hydrostatic Method}

To calculate hydrostatic masses spectra were analysed in \textsc{xspec} using the \textsc{nfwmass} mixing model (\citealt{nulsen_clmass}). This method assumes that the cluster is spherically symmetric, and that the X-ray emitting cluster gas is in hydrostatic equilibrium. The effects of non-thermal pressure at $R_{2500}$ are expected to bias mass measurements low on the order of $10\%-20\%$ \citep{Nagai_2007}, but have also been found to be consistent with zero \citep{vikhlinin_2009,Mahdavi_2013}. We use the additional assumption that the underlying gravitational potential (i.e. including both dark matter and baryonic mass) follows the NFW profile \citep{navarro_1997}
\begin{equation} \label{eq:NFW}
\rho(r) = \frac{\rho_{0}}{r/r_{s} (1 + r/r_{s})^{2} },
\end{equation}
where $\rho_{0}$ is a characteristic density, and $r_{s}$ is the scale radius. The NFW profile has been found to be an accurate description of cluster mass profiles, including systems with significant feedback \citep{Point_2005,vikh_2006,Schmidt_2007,Gitti_2007}. Under the assumption of spherical symmetry, the enclosed mass as a function of radius is described in terms of observables as
\begin{equation}
M(r) = \frac{-kTr}{G\mu m_{H}} \left( \frac{d \text{ log } n_{e} }{d \text{ log } r} + \frac{d \text{ log } T}{d \text{ log } r} \right).
\end{equation} 
The full radial information of the temperature, gas density, and gravitating density are then given by the free parameters of the spectral fits: $r_{s}$, $A = 4\pi G \rho_{0} r_{s}^{2} \mu m_{H}$, and a temperature for each annulus. As $r_{s}$ and $A = 4\pi G \rho_{0} r_{s}^{2} \mu m_{H}$ are not independent variables, error bars on the total mass are obtained from their $1-\sigma$ confidence ellipses.

In conjunction with the \textsc{nfwmass} model, each region was fit in the energy range $0.6-7.0$ keV by a \textsc{phabs(apec)} model in \textsc{xspec} version 12.8.0 \citep{arnaud_xspec}, using \textsc{apec} version 2.0.2 \citep{apec}. The inner regions of each cluster were excluded within the volume of the X-ray cavities, to avoid biasing the spectral fits due to multiphase gas and substructure. The spectra were fit using $\chi^{2}$ minimization. The values of \citet{angr} were used for the solar abundances to calculate metallicity in our spectral fits. The hydrogen column density ($N_{H}$) values were fixed to the galactic values of \citet{kalberla}, except in systems where the best-fit value was found to be significantly different. In these cases, we used the best-fit $N_{H}$ to fit the spectra beyond the cool-core.  The $N_{H}$ values used are listed in Table \ref{table:obslist}. Abundances were allowed to vary in the spectral fits, but were tied, as necessary, between adjacent annuli in the outer regions where the metallicity is nearly constant \citep{deGrandi_2001}.

Galaxy cluster masses are typically measured to R$_{\Delta}$,  the radius within which the enclosed mass is $\Delta$ times the critical mass density at the cluster redshift. Then, 
\begin{equation} \label{eq:mcrit}
M_{\Delta} = \frac{4 \pi R_{\Delta}^{3}}{3} \Delta \rho_{c},
\end{equation}
where $\rho_{c} = 3H^{2} / 8\pi G $, $H = H_{0}E(z)$, and $E(z) = [\Omega_{M}(1+z)^{3} + \Omega_{\Lambda}]^{1/2}$ in a flat $\Lambda$-CDM cosmology. We determine masses out to $R_{2500}$, as the Chandra data extend beyond it for the majority of our sample, and $M_{2500}$ is then used as a proxy for cluster mass. The values of $M_{2500}$ and $R_{2500}$ are determined by numerically solving equation \ref{eq:mcrit}, given the NFW profile defined by our best fit $A$ and $r_{s}$ values. In some nearby systems, extrapolation of the NFW profile was needed to reach $R_{2500}$. In these systems, the errors on $M_{2500}$ are likely underestimated. All calculated values of $M_{2500}$ are given in Table \ref{table:properties}, and the systems where extrapolation to $R_{2500}$ was needed are noted. The reduced $\chi^{2}$ values of our spectral fits range between 0.89 and 1.32 with a mean value of 1.09. The Perseus cluster is an outlier with a reduced $\chi^{2}$ value of 3.11. Perseus has exceptionally high quality data due both to being the brightest X-ray cluster in the sky and having deep exposures, and requires a multi-temperature model with multiple element abundances allowed to vary to accurately fit its spectra (e.g., \citealt{Sanders_2004}). Our total mass profiles and gas mass profiles are
presented in Appendix A.

\begin{figure}
\begin{minipage}{1.0\columnwidth}
\centering
\includegraphics[width=1.0\columnwidth]{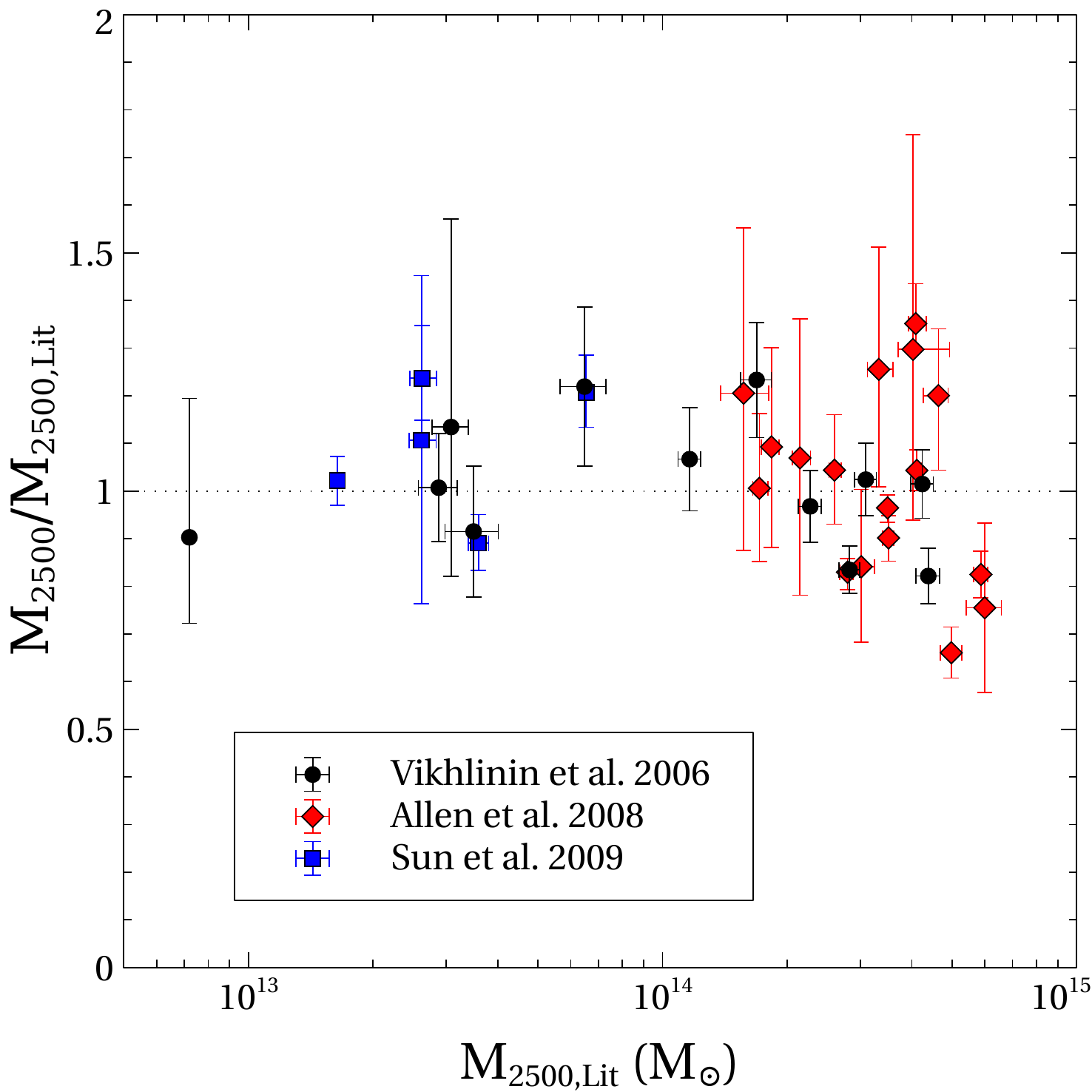}

\caption{Comparison of a subsample of our masses to values in the literature.}

\label{figure:vikh_comp}

\end{minipage}
\end{figure}

As a calibration, we compare our mass measurements with other Chandra based hydrostatic measurements from the literature. We determine $M_{2500}$ for the sample of clusters analyzed in \citet{vikh_2006}, except for Abell 2390 which was unconstrained. We also compare our $M_{2500}$ values to those of \citet{Allen_2008} and \citet{Sun_2009} in the systems that overlap our sample. We convert the mass measurements from the literature to a consistent cosmology, and compare our measurements in Figure \ref{figure:vikh_comp}. We find that our mass results are consistent within $R_{2500}$ with no mass-dependent bias. 

\subsubsection{Mass Profiles: Remaining HIFLUGCS Systems}

We derive $M_{2500}$ for the remaining HIFLUGCS systems (i.e., HIFLUGCS systems not in our primary sample) using the $\beta$, $r_{c}$, T$_{x}$ and M$_{500}$ parameters in table 4 of \citet{Reiprich_2002}, which fully describe the mass profile. $ROSAT$, $ASCA$, and $Einstein$ observations  were used to make these mass measurements, and likely have different systematics than our Chandra mass measurements. To correct for this, we compare how our mass measurements differ in the 24 systems where we have already determined $M_{2500}$ using Chandra data. We find that the average ratio of our measurements to be $M_{2500}/M_{2500,\rm Reiprich} = 1.18$ with a standard deviation of $0.33$. We apply this ratio to the $M_{2500}$ values derived from \citet{Reiprich_2002} to estimate $M_{2500}$ for the remainder of the HIFLUGCS systems.  Measurements with this level of accuracy are sufficient, as these additional mass values are being used to investigate broad, qualitative trends.

\subsubsection{Gas Mass}

The gas density profiles were derived from the same set of spectra as our total masses, but including spectra in the central region of the cluster. The spectra were fit in the $0.6-7.0$ keV range with using a \textsc{projct}(\textsc{phabs}*\textsc{apec}) model in \textsc{xspec}, to obtain deprojected gas density profiles. The density profiles were then integrated in a piecewise manner from the centre of the cluster to obtain the radial gas mass distribution. From the gas mass profile, we derived $M_{g}$ within R$_{2500}$ (shown in Table \ref{table:properties}). In systems where the data do not reach $R_{2500}$, $M_{g}$ at $R_{2500}$ was estimated by extrapolating an NFW profile fit to the inner profiles. The emissivity measurements of the outer bins may be artificially high as we assume no cluster emission beyond (see the gas mass profiles in Appendix A). Due to this effect, we have extrapolated our gas mass profiles for which only the final bin is beyond $R_{2500}$.

\subsubsection{Luminosity and Temperature}

Other quantities of interest include the bolometric luminosity and temperature of our systems. To this end, spectra were extracted from apertures out to $R_{2500}$ in each system, with the inner 0.3 $R_{2500}$ excluded. These are used as proxies for the virial temperature and luminosity, since many of our systems do not have data far beyond $R_{2500}$. These spectra were fit with a \textsc{phabs(apec)} model in \textsc{xspec}, along with an additional \textsc{apec} model for the excess soft background when needed. The \textsc{cflux} model in \textsc{xspec} was used to calculate the unabsorbed flux in the energy range of 0.05 keV - 50 keV used in the luminosity measurement.

\subsubsection{Central Cooling Time}

\citet{Hudson_2010} showed the central cooling time to be the best means of distinguishing a cool-core in systems with high quality X-ray data. When investigating the full HIFLUGCS sample, we separated the systems into two populations with central cooling times lying above and below $ 1\text{ Gyr}$.  Central cooling time below $1\text{ Gyr}$ corresponds 
observationally to the cooling time threshold for the onset of star formation in central galaxies and is a strong indication of a cool-core \citep{Rafferty_2008, Cavagnolo_2008, Mittal_2009,Hudson_2010}. Theoretically, this timescale is a rough threshold beyond which cooling in the centres of clusters cannot be locally suppressed \citep{Sharma_2012, Voit_2015}. Cooling time profiles at the centres of clusters continue to decrease with radius \citep{Panagoulia_2014}, so it is important to define the central cooling time in a way that is unbiased by resolution. We use the $t_{c}$ values of \citet{Hudson_2010}, where the central cooling times are all determined at a consistent radius of 0.004$R_{500}$. This radius is resolved in all systems. 

\subsection{Cavity Energetics}

The mechanical energy output from the central AGN can be directly measured from cavities observed in X-ray data \citep{McNamara_2000, Churazov_2000, birzan_2004}. Assuming the cavities are in pressure balance with the surrounding ICM, the minimum mechanical energy required to inflate the cavity is
\begin{equation}
E_{\rm cav} = \frac{\gamma}{\gamma - 1} pV,
\end{equation}
where $p$ is the pressure surrounding the cavity, $V$ is the volume of the cavity, and $\gamma$ is the adiabatic index of the medium filling the cavity. It is commonly assumed that the cavities are filled with relativistic plasma, as they are created by radio synchrotron emitting jets. In this case, $\gamma = 4/3$, and the energy output is $E_{\rm cav} = 4pV$. The mean power of the AGN is then estimated by dividing by the outburst age. AGN powers derived in this manner generally underestimate the total power \citep{Mcnamara_2007}.
In this study, the buoyancy timescale of the cavities is used in all power calculations \citep{McNamara_2000, Churazov_2000, Churazov_2001}, calculated as
\begin{equation}
t_{\rm buoy} \simeq R\sqrt{SC/(2gV)},
\end{equation}
where $S$ is the cross section of the bubble, and $C$ is the drag coefficient taken as $C=0.75$. The gravitational acceleration is calculated as $g \approx 2\sigma^{2}/R$, where $\sigma$ is the stellar velocity dispersion. We use the buoyant timescale for consistency, since all cavity systems in our sample have existing power estimates in the literature using this timescale. It is likely an overestimate of cavity ages, as bubbles are expected to expand supersonically in the early stages of being driven by the jet. Cavity energies and power measurements are listed with references in Table \ref{table:properties}. 

\subsubsection{Radio Luminosity}

X-ray cavities are inflated by radio jets so we use the luminosity of central radio sources as a proxy for AGN activity. We have obtained monochromatic, 1.4 GHz radio luminosities of central radio sources for all of the HIFLUGCS objects from \citet{Mittal_2009} and \citet{birzan_2012}. In these studies, a radio source was deemed central if it is within 50$h^{-1}_{71}$kpc of the X-ray peak. In systems without detected central radio sources, upper limits were estimated using the NVSS images by \citet{birzan_2012}.

\subsection{Stellar and Black Hole Mass}
\label{sec:bhmass}

Apparent K-band luminosities taken from the 2MASS Extended Source Catalog are used to estimate stellar masses of the BCGs.   
Stellar magnitudes were corrected for galactic extinction \citep{schlegel_1998}, evolution, and K-corrections  \citep{poggianti_1997}.  Absolute magnitudes were calculated assuming our adopted cosmology, and redshifts in Table \ref{table:obslist}. 
We convert K-band luminosities to stellar mass using the relation $\text{log}(M/L_{K}) = -0.206 + 0.135(B-V)$ from \citet{Bell_2003}, a model fit to a large, varied sample of galaxies.  Since we do not have $B-V$ measurements for all of our systems, we adopt a $B-V$ value of 1.0, a good estimate in massive BCGs \citep{Baldry_2008}. In addition, our results are insensitive to changes of order $0.1$ in the $B-V$ colours used. We adopted the mass-to-light ratio and color found by Hogan et al. (2016, in preparation) that simultaneously reproduces hydrostatic mass profiles, similar to those measured here, and those derived from stellar velocity dispersion profiles and weak lensing profiles. In addition to total stellar masses of the BCGs, we estimate stellar mass within 20 kpc from the 2MASS radial K-band light profiles to compare star formation histories of our systems at a consistent radius.  We convert the standard 2MASS circular apertures between 5 and 70 arcseconds to kpc using our adopted cosmology and the redshifts listed in Table 1, and interpolate the magnitudes between apertures to reach an estimate of the K-band magnitude within 20kpc.

K-band luminosities \citep{Graham_2007} were used to estimate the black hole masses as, 
\begin{equation}
\text{log}\left(\frac{M_{\rm BH}}{M_{\odot}}\right) = -0.38(\pm0.06) (M_{K} + 24) + 8.26(\pm0.11).
\end{equation}
The scatter of 0.33 dex in this relation is included in the errors on our black hole masses. 

\citet{Lauer_2007} showed that the K-band magnitudes for BCGs from 2MASS do not capture the full extent of the stellar envelope and thus are likely to be
underestimated. However, \citet{Batcheldor_2007} find that black hole masses based on NIR magnitudes agree well with estimates based on the stellar velocity dispersion in BCGs. As such, the accuracy of our black hole estimates are unclear, and we take this into account in our interpretation of our results in section \ref{sec:accretion}.

\begin{table*}

\caption{System properties derived from Chandra X-ray Data.}

\begin{tabular}{l c c c c c c c}
\hline
System & P$_{\rm cav}$$^{a}$ & pV & M$_{2500}$$^{b}$ & Mg$_{2500}$ & kT & L$_{\rm bol}$ & References  \\
& ($10^{42}\ergps$) & ($10^{58}$erg) & ($10^{13}$M$_{\odot}$) & ($10^{13}$M$_{\odot}$) & (keV) & (10$^{42}$ erg s$^{-1}$) \\
\hline
2A0335+096 & $24\err{23}{6}$ & $1.1\err{1.0}{0.3}$ & $10.4\pm0.6$ & $1.04\err{0.03}{0.03}$ & $3.46\err{0.09}{0.09}$ & $130\err{1}{1}$ & [1] \\ 
A85 & $37\err{37}{11}$ & $1.2\err{1.2}{0.4}$ & $20.6\pm1.0$ & $2.40\err{0.06}{0.06}$ & $6.6\err{0.1}{0.1}$ & $471\err{2}{2}$ & [1] \\ 
A133 & $620\err{260}{20}$ & $24\err{11}{1}$ & $12.4\pm0.9$ & $1.04\err{0.06}{0.06}$ & $4.4\err{0.1}{0.1}$ & $96.1\err{0.8}{0.8}$ & [1] \\ 
A262 & $9.7\err{7.5}{2.6}$ & $0.13\err{0.10}{0.03}$ & $3.2\pm0.1$ & $0.283\err{0.009}{0.008}$ & $2.23\err{0.04}{0.04}$ & $27.8\err{0.2}{0.2}$ & [1] \\ 
A478 & $100\err{80}{20}$ & $1.5\err{1.1}{0.4}$ & $43\pm1.4$ & $5.4\err{0.3}{0.3}$ & $7.5\err{0.3}{0.3}$ & $729\err{6}{6}$ & [1] \\ 
A496 & $172$ & $2.34$ & $13.3\pm0.4$ & $1.17\err{0.08}{0.05}$ & $4.5\err{0.1}{0.1}$ & $123\err{1}{1}$ & [2] \\ 
A1795 & $160\err{230}{50}$ & $4.7\err{6.6}{1.6}$ & $23.2\pm0.3$ & $2.8\err{0.2}{0.2}$ & $6.0\err{0.1}{0.1}$ & $366\err{3}{3}$ & [1] \\ 
A2029 & $87\err{49}{4}$ & $4.8\err{2.7}{0.1}$ & $33.8\pm0.8$ & $3.6\err{0.1}{0.1}$ & $8.0\err{0.1}{0.1}$ & $869\err{4}{4}$ & [1] \\ 
A2052 & $150\err{200}{70}$ & $1.7\err{2.3}{0.7}$ & $8.6\pm0.1$ & $0.79\err{0.09}{0.09}$ & $3.1\err{0.05}{0.05}$ & $90\err{1}{1}$ & [1] \\ 
A2199 & $270\err{270}{60}$ & $7.5\err{6.6}{1.5}$ & $17.4\pm0.5$ & $1.65\err{0.01}{0.01}$ & $3.9\err{0.2}{0.2}$ & $118\err{3}{3}$ & [1] \\ 
A2204 & $280\err{50000}{0}$ & $5.6$ & $55.2\pm0.7$ & $6.0\err{0.2}{0.2}$ & $9.2\err{0.3}{0.3}$ & $987\err{7}{7}$ & [3] \\ 
A2597 & $67\err{87}{29}$ & $3.6\err{4.6}{1.5}$ & $14.4\pm0.2$ & $1.33\err{0.04}{0.04}$ & $4.00\err{0.05}{0.05}$ & $175\err{1}{1}$ & [1] \\ 
A4059 & $96\err{89}{35}$ & $3.0\err{2.5}{0.9}$ & $15.3\pm1.0$ & $1.02\err{0.04}{0.05}$ & $4.4\err{0.2}{0.2}$ & $112\err{2}{2}$ & [1] \\ 
Centaurus & $7.4\err{5.8}{1.8}$ & $0.060\err{0.051}{0.015}$ & $^{*}$$8.2\err{0.2}{0.2}$ & $0.056\err{0.007}{0.007}$ & - & - & [1] \\ 
Hydra A & $430\err{200}{50}$ & $64\err{48}{11}$ & $16.0\err{1.2}{3.3}$ & $1.65\err{0.03}{0.03}$ & $3.79\err{0.07}{0.07}$ & $187\err{1}{1}$ & [1] \\ 
MKW3S & $410\err{420}{44}$ & $38\err{39}{4}$ & $12.4\pm1.3$ & $0.96\err{0.02}{0.02}$ & $4.3\err{0.2}{0.1}$ & $78.1\err{0.9}{0.9}$ & [1] \\ 
NGC507 & $19\err{14}{7}$ & $0.35\err{0.06}{0.04}$ & $2.1\pm0.3$ & $0.101\err{0.004}{0.004}$ & $1.41\err{0.04}{0.04}$ & $4.7\err{0.1}{0.1}$ & [4],[5] \\ 
NGC1399 & 1.2$\err{0.5}{0.5}$ & $0.11\err{0.05}{0.05}$ & $^{\dagger}$$2.38\err{0.94}{0.90}$ & - &  - & - & [6] \\
NGC1550 & $15.4$ & $0.17$ & $1.67\pm0.07$ & $0.11\err{0.01}{0.01}$ & $1.27\err{0.03}{0.03}$ & $5.5\err{0.2}{0.2}$ & [2] \\ 
NGC4636 & $2.76\err{0.56}{0.91}$ & $0.012$ & $^{\dagger}$$0.82\err{0.28}{0.28}$ & - & - & - & [4],[5] \\
NGC5044 & $4.2\err{1.2}{2.0}$ & $0.049$ & $1.00\pm0.05$ & $0.063\err{0.006}{0.005}$ & $1.24\err{0.02}{0.02}$ & $4.3\err{0.1}{0.1}$ & [4],[5] \\
PKS1404-267 & $20\err{26}{9}$ & $0.12\err{0.15}{0.05}$ & $3.2\pm0.1$ & $0.225\err{0.005}{0.005}$ & $1.66\err{0.04}{0.05}$ & $11.4\err{0.2}{0.2}$ & [1] \\ 
Sersic159/03 & $780\err{820}{260}$ & $25\err{26}{8}$ & $9.5\pm0.5$ & $0.84\err{0.03}{0.03}$ & $2.53\err{0.04}{0.04}$ & $71.9\err{0.4}{0.4}$ & [1] \\
\hline

3C388 & $200\err{280}{80}$ & $5.2\err{7.5}{2.1}$ & $6.7\pm0.9$ & $0.45\err{0.09}{0.09}$ & $3.39\err{0.24}{0.16}$ & $44\err{1}{1}$ & [1] \\ 
3C401 & $650\err{1200}{420}$ & $11\err{20}{7}$ & $9.2\pm2.7$ & $0.585\err{0.084}{0.066}$ & $2.98\err{0.34}{0.31}$ & $49\err{2}{2}$ & [1] \\ 
4C55.16 & $420\err{440}{160}$ & $12\err{12}{4}$ & $21\pm3$ & $1.93\err{0.08}{0.11}$ & $5.0\err{0.2}{0.2}$ & $304\err{5}{5}$ & [1] \\ 
A1835 & $1800\err{1900}{600}$ & $47\err{50}{16}$ & $48\pm2$ & $5.6\err{0.1}{0.1}$ & $10.2\err{0.2}{0.2}$ & $1540\err{8}{8}$ & [1] \\ 
Cygnus A & $1300\err{1100}{200}$ & $84\err{70}{14}$ & $17.3\pm0.08$ & $1.73\err{0.06}{0.06}$ & $7.61\err{0.21}{0.20}$ & $287\err{2}{2}$ & [1] \\ 
HCG62 & $3.9\err{6.1}{2.3}$ & $0.046\err{0.073}{0.028}$ & $1.70\pm0.04$ & $1.0\err{2.7}{0.8}$ & $1.06\err{0.02}{0.02}$ & $1.97\err{0.05}{0.05}$ & [1] \\ 
Hercules A & $310\err{400}{90}$ & $31\err{40}{9}$ & $20.8\pm1.4$ & $1.64\err{0.22}{0.18}$ & $5.06\err{0.17}{0.17}$ & $179\err{2}{2}$ & [1] \\ 
MACSJ0159.8-0849 & 377 & $13.3$ & $56\pm6$ & $4.5\err{0.3}{0.3}$ & $10.9\err{0.5}{0.6}$ & $1260\err{20}{20}$ & [7] \\ 
MACSJ0242.5-2132 & 353 & $8.96$ & $23\pm6$ & $2.4\err{0.2}{0.4}$ & $6.5\err{1.0}{0.8}$ & $576\err{20}{20}$ & [7] \\ 
MACSJ0429.6-0253 & $83$ & $2.17$ & $20\pm3$ & $2.22\err{0.38}{0.38}$ & $8.1\err{1.5}{1.1}$ & $723\err{30}{30}$ & [7] \\ 
MACSJ0913.7+4056 & $7560$ & $150$ & $29\pm4$ & $2.3\err{0.3}{0.3}$ & $7.3\err{0.7}{0.6}$ & $523\err{20}{20}$ & [7] \\ 
MACSJ1411.3+5212 & $3560$ & $49.1$ & $17\pm3$ & $1.83\err{0.17}{0.17}$ & $6.0\err{0.5}{0.5}$ & $456\err{10}{10}$ & [7] \\ 
MACSJ1423.8+2404 & $1400\err{2500}{900}$ & $61$ & $27\pm3$ & $2.6\err{0.2}{0.2}$ & $8.2\err{0.8}{0.6}$ & $801\err{20}{20}$ & [7] \\ 
MACSJ1532.8+3021 & $2200\err{900}{900}$ & $270$ & $42\pm8$ & $3.6\err{0.3}{0.3}$ & $7.0\err{0.3}{0.3}$ & $867\err{10}{10}$ & [8] \\
MACSJ1720.2+3536 & $380$ & $17.64$ & $25\pm5$ & $3.2\err{0.2}{0.2}$ & $8.13\err{0.61}{0.56}$ & $899\err{20}{20}$ & [7] \\ 
MACSJ1931.8-2634 & $8760$ & $83.4$ & $52\pm14$ & $4.4\err{0.3}{0.3}$ & $7.9\err{0.4}{0.4}$ & $1060\err{10}{10}$ & [7] \\ 
MACSJ2046.0-3430 & $605$ & $22.86$ & $19\pm5$ & $2.2\err{0.1}{0.2}$ & $5.2\err{0.5}{0.4}$ & $451\err{10}{10}$ & [7] \\ 
MACSJ2140.2-2339 & $179$ & $3.43$ & $19\pm2$ & $2.2\err{0.2}{0.2}$ & $6.0\err{0.4}{0.4}$ & $479\err{10}{10}$ & [7] \\ 
MS0735.6+7421 & $6900\err{7600}{2600}$ & $1600\err{1700}{600}$ & $28.5\pm0.9$ & $2.90\err{0.08}{0.08}$ & $6.72\err{0.08}{0.08}$ & $523\err{2}{2}$ & [1] \\
NGC5813 & $3.97\err{1.02}{2.36}$ & 0.064 & $^{\dagger}$$0.088\err{0.004}{0.004}$ & - & - & - & [4],[5] \\
NGC5846 & $0.88\err{0.30}{0.59}$ & 0.0022 & $^{\dagger}$$0.12\err{0.04}{0.04}$ & - & - & - & [4],[5] \\
Perseus & $150\err{100}{30}$ & $19\err{20}{1}$ & $^{*}$$15.0\err{0.3}{0.3}$ & $1.74\err{0.01}{0.01}$ & - & - & [1] \\ 
PKS0745-191 & $1700\err{1400}{300}$ & $69\err{56}{10}$ & $33\pm2$ & $4.0\err{0.2}{0.2}$ & $8.0\err{0.2}{0.2}$ & $1200\err{10}{10}$ & [1] \\  
RBS797 & $3340\err{1410}{1410}$ & $38\err{50}{15}$ & $50\pm5$ & $3.8\err{0.4}{0.4}$ & $9.1\err{0.7}{0.6}$ & $926\err{20}{20}$ & [1] \\  
Zw2701 & $920\err{180}{180}$ & $42.0\err{0.8}{0.8}$ & $19\pm1$ & $1.7\err{0.2}{0.2}$ & $5.5\err{0.2}{0.2}$ & $267\err{4}{4}$ & [9] \\ 
Zw3146 & $5800\err{6800}{1500}$ & $380\err{460}{110}$ & $45\pm10$ & $5.1\err{0.9}{1.2}$ & $8.5\err{0.6}{0.4}$ & $1300\err{20}{20}$ & [1] \\

\hline

\end{tabular}
\label{table:properties}
\\
Above the first horizontal line comprises our Primary Sample.  The full table comprises the Extended sample. Notes: $^{a}$$pV$ and $P_{\rm cav}$ values taken from the references listed, using $E_{\rm cav} = 4pV$. We assume factor of 2 errors in systems lacking quoted error bars. $^{b}$Masses marked with a dagger are derived from \citet{Reiprich_2002}, as described in section \ref{sec:ncc}. References: [1] \citet{rafferty_2006}, [2] \citet{birzan_2012}, [3] \citet{Sanders_2009}, [4] \citet{Cavagnolo_2010}, [5] \citet{Russell_2013}, [6] \citet{Shurkin_2008}, [7] \citet{Hlavacek_2012}, [8] \citet{Hlavacek_2013}, [9] Ma et al. in prep.

\end{table*}

\section{Results and Discussion}

\subsection{Mass Partitioning}
\label{sec:masspart}

\subsubsection{Gas Mass Fraction}
\label{sec:gasmass}
\begin{figure}
\begin{minipage}{1.0\columnwidth}
\centering

\includegraphics[width=1.0\textwidth]{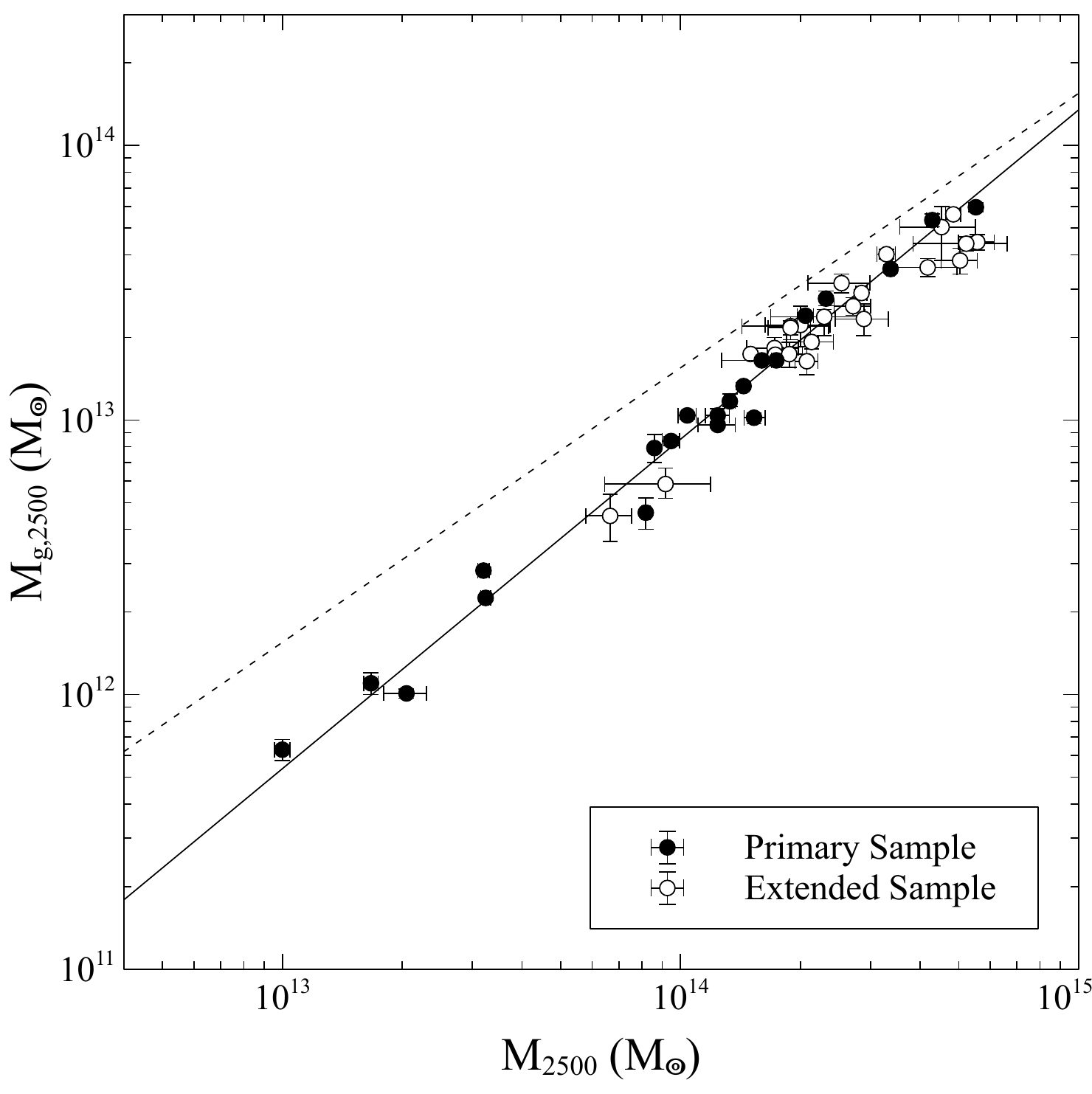} 

\caption{Gas mass within R$_{2500}$ vs. M$_{2500}$. The solid line shows the best-fit powerlaw to the primary sample, given by $M_{g,2500}/10^{13}M_{\odot} = 10^{-1.27}(M_{2500}/10^{13}M_{\odot})^{1.2}$. The dotted line shows the Planck value of $\Omega_{b} / \Omega_{m} = 0.155$ \citep{planck} }

\label{figure:frac}

\end{minipage}
\end{figure}

In order to make contact with earlier studies, we first investigate the relationships between the total mass, gas mass, and stellar mass in our sample. We fit the relationship between the gas mass within $R_{2500}$ and $M_{2500}$ in HIFLUGCS systems using the bivariate correlated error and intrinsic scatter (BCES) method of \citet{akritas_1996}.  We find a tight correlation between halo mass and gas mass, shown in Figure \ref{figure:frac}, that follows the power law relationship $M_{g,2500}/10^{13}M_{\odot} = 10^{-1.27 \pm0.05} (M_{2500}/10^{13}M_{\odot})^{1.20 \pm 0.04}$.  
The slope of this relationship is consistent with \citet{Gonzalez_2013}.   

This result implies that the gas fraction increases with halo mass.  At the highest masses, the gas fraction is approximately 0.12, comparable to $0.11$ found by 
\citet{Allen_2008} in relaxed clusters with temperatures above 5 keV.  These values lie below 0.155 derived from cosmic microwave background measurements \citep{planck}, in part because we have not included stellar masses.  

\subsection{Cluster Scaling Relations}

Before investigating the effects of AGN feedback on cluster scaling relations, we first investigate how the scaling relations in our sample compare to earlier work. If galaxy cluster
formation was governed solely by gravity, the cluster luminosity, temperature, and mass should be related by simple powerlaw expressions, as $L \propto E(z) T^{2}, M \propto E(z)^{-1} T^{3/2}$, where $E(z) = H(z)/H_{0}$ \citep{Kaiser_1986}. Departures from these relations are likely caused by baryonic processes, such as radiative cooling and heating 
by AGN and supernovae (reviewed by \citealt{Kravtsov_2012}).  Here we search for indications that these scaling relations are affected by ongoing AGN feedback.

We fit the $L-T$ and $M-T$ relations with the bivariate correlated error and intrinsic scatter (BCES) method of \citet{akritas_1996}. Previous studies have found deviations from self-similarity are largest in poor clusters and groups \citep{Arnaud_2005,Sun_2009,Eckmiller_2011}.   We perform fits on both our primary sample and on the systems in the primary sample with temperatures above $3$ keV.  We also fit our sample omitting evolutionary corrections to allow a direct comparison to the results in section \ref{sec:cavscal}.

\subsubsection{L-T relation}

The luminosities calculated  for a flux limited sample are affected by the Malmquist bias, which over-represents the most luminous objects. For a low-redshift sample with a log-normal distribution of luminosities and no redshift cutoff, the bias on $L$ in the $L-T$ relation can be expressed as $\Delta \ln L = \frac{3}{2} \sigma_{i}^{2}$, where $\sigma_{i}$ is the intrinsic log normal scatter in L for a given temperature (described in \citealt{vikhlinin_2009} for the $L-M$ relation). This will be equivalent for any relation between L and a second variable, if they are related by a powerlaw with log-normal scatter. The Malmquist bias in our sample would then simply modify the normalization, but not the exponent of powerlaw relations involving luminosity.

\begin{figure*}
\begin{minipage}{1.0\textwidth}
\centering

\includegraphics[width=0.45\textwidth]{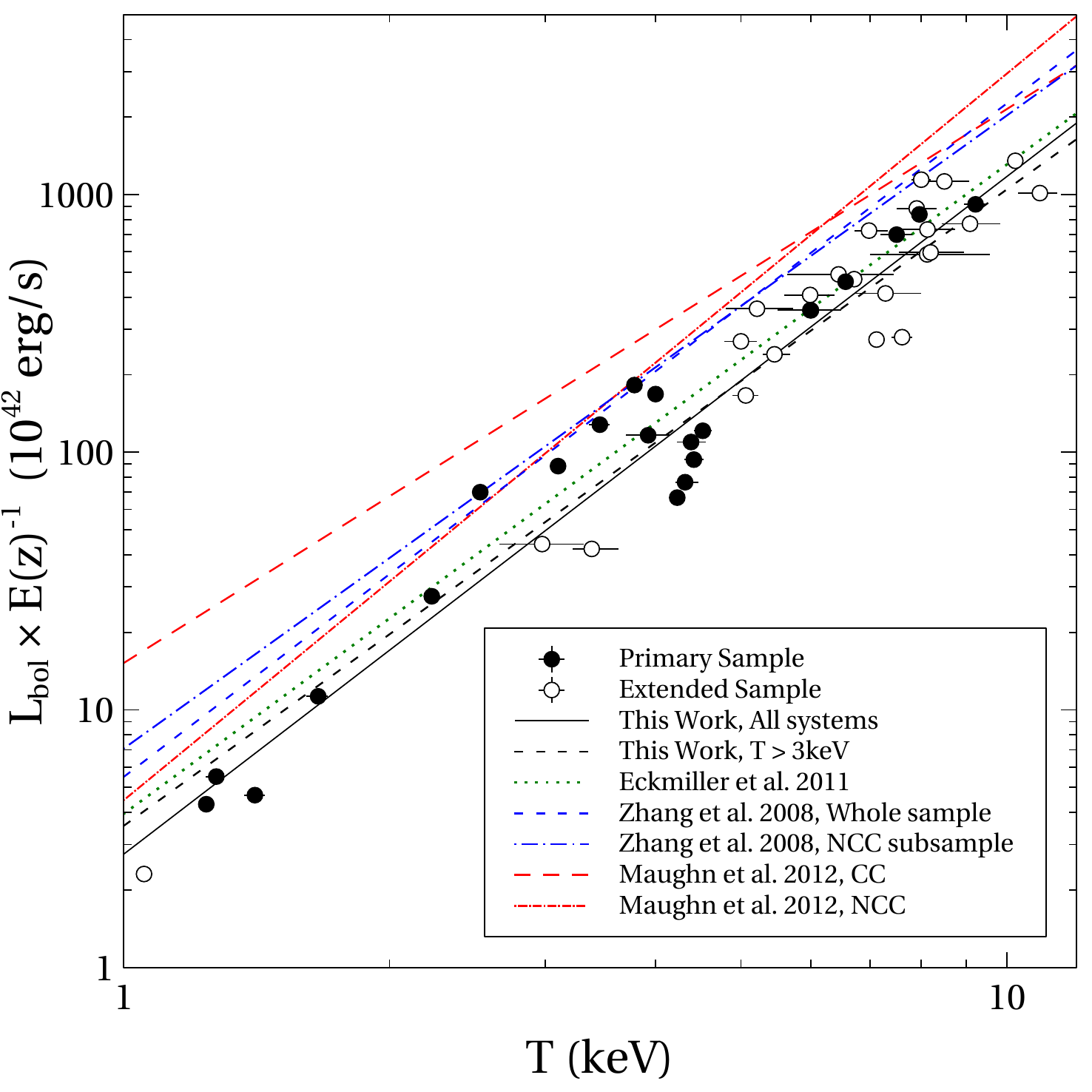} 
\includegraphics[width=0.45\textwidth]{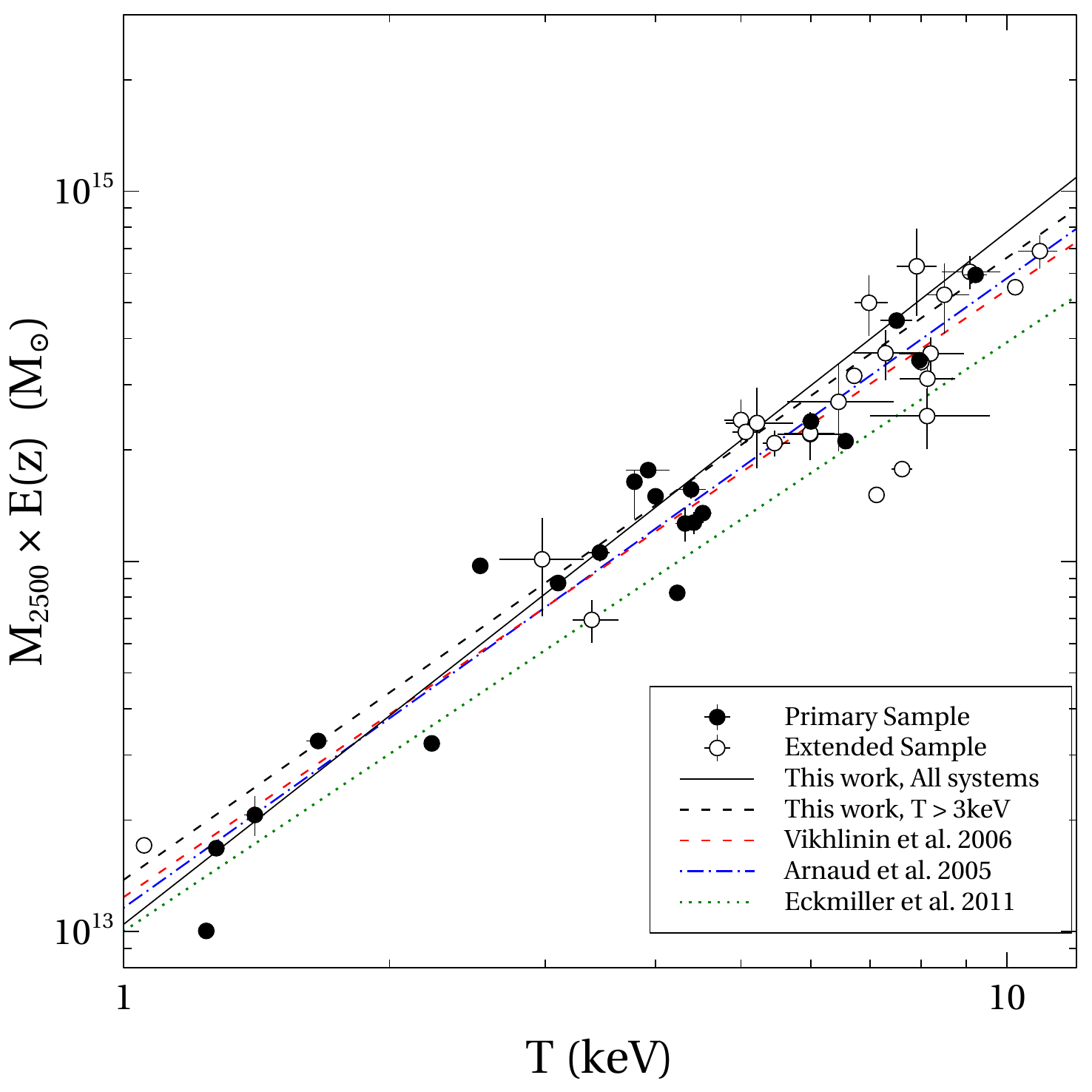}

\caption{\textit{Left:} Bolometric luminosity vs temperature. \textit{Right:} M$_{2500}$ vs. temperature. Luminosity and temperature are measure in the aperture (0.3-1.0)R$_{2500}$. The black lines show the best-fit relation to the primary sample over the full range, and restricted to systems with T $>$ 3 keV. Best-fit relations from the literature for core-excised L-T relations, and the M-T relation within R$_{2500}$ are overlain. The discrepancy between the L-T normalizations is primarily due to differences in the aperture used to determine the luminosity. }

\label{figure:scal}

\end{minipage}
\end{figure*}

The best-fit results to our $L-T$ and $M-T$ relations are given in table \ref{table:scalings}, and are shown in figure \ref{figure:scal}. Our fit to the full sample is $L_{\rm bol} E(z)^{-1} \propto T^{2.63\pm0.10}$. 
This slope is significantly steeper than the self-similar value, but is consistent with the range of slopes of core-excised $L-T$ relations found in previous studies (see \citealt{giodini_2013} for a review, and Figure \ref{figure:scal} for a few comparisons), including $L-T$ relations measured for the full cluster population and in non cool-core clusters \citep{Markevitch_1998,Zhang_2008,Eckmiller_2011}.
Notably, our $L-T$ relation is steeper than the value of $2.15\pm0.17$ of \citet{Maughan_2012} for core-excised cool-core clusters. This may be due to our inclusion of lower temperature systems, which drive the relation away from self-similarity.
The best-fit slope in our systems above 3 keV of $2.47\pm0.34$ is shallower, but is not significantly different from the result for the full sample. Our normalization is lower than previous studies due to a smaller aperture size compared to the typical radius of R$_{500}$.
The raw scatter in our relation is $\sim41\%$, close to other core-excised L-T relations of cool-core systems (eg: $0.103$ dex in \citealt{Markevitch_1998}, $0.242\pm0.110$ dex in \citealt{Pratt_2009}, $33.2\%$ in \citealt{Mittal_2011}).

\subsubsection{M-T relation}

The best-fit $M-T$ relation is given by $(M_{2500}/10^{13}M_{\odot}) E(z) \propto T^{1.87\pm0.12}~\rm keV$ for our primary sample, and $(M_{2500}/10^{13}M_{\odot}) E(z) \propto T^{1.68\pm0.22}~\rm keV$ when fit to the systems in our primary sample with $T > 3$ keV. The $M-T$ relation is steeper than the self-similar expectation for our full sample, and is consistent with $M\propto T^{3/2}$ when only the $T > 3$ keV systems are fit. The normalization of our relation is slightly higher than those of \citet{Arnaud_2005} and \citet{vikh_2006}, but this difference is within the 1-$\sigma$ uncertainties. A slope of $1.87\pm0.12$ is marginally steeper than the slopes of $\sim 1.6-1.7$ which is usually observed over the same mass range (e.g.,\citealt{Arnaud_2005,vikh_2006}). Due to the small number of points at low mass, this could easily be caused by one or two outliers. For example, excluding  NGC 5044,  the best-fit slope is $1.77\pm0.10$. We find no convincing evidence that the $M-T$ relation in our sample is significantly different from previously measured relations.

\subsubsection{Summary of scaling relations results}

We measured the $L-T$ and $M-T$ relations in our primary sample, excluding the inner 0.3 $R_{2500}$. This sample is comprised entirely of cool-core clusters currently experiencing mechanical feedback, as evidenced through X-ray cavities. Both relations are steeper than the self-similar slopes. The $M-T$ relation is consistent with the self-similar scaling, but only in clusters with mean temperatures above 3 keV.  Our measured $L-T$ and $M-T$ relations are consistent with previous studies which included the full cluster population, and studies of non cool-core clusters (see eg: \citealt{Arnaud_2005,vikh_2006,Pratt_2009,Mittal_2011,Maughan_2012} for detailed analyses).  The scaling relations for our sample are generally consistent with the cluster population as a whole, when measured beyond the inner regions of the cluster. This result seems to imply either that mechanical AGN feedback is not strongly affecting the overall temperature and luminosity of clusters beyond the cooling radius (Mantz et al. 2010), or that all clusters have experienced a significant level of heating from AGN or other processes in the past. The above shows that our sample, chosen specifically based on direct evidence of AGN heating, does not appear to be atypical in terms of large scale cluster properties. This helps inform the interpretation of the new correlations which we investigate in further sections.

\begin{table}
\caption{Cluster Scaling Relations.}
\begin{tabular}{c}
Fit Results to $\text{log(y) } = a + b \text{ log(x)}$
\end{tabular}

\bgroup
\setlength{\tabcolsep}{.3em}
\def\arraystretch{1.2}
\begin{tabular}{l l l l l}
\hline
y & x & a & b & $\sigma$\\
\hline
$L_{\rm bol}E(z)^{-1}$ & $T_{x}$ & $0.44 (\pm 0.07)$ & $2.63 (\pm 0.10)$ & 0.15 \\
$L_{\rm bol}E(z)^{-1}$ & $T_{x}(>3 \text{keV})$ & $0.55(\pm 0.26)$ & $2.47 (\pm 0.34)$ & 0.15 \\
$M_{2500}E(z)$ & $T_{x}$ & $ 0.02(\pm 0.07)$ & $1.87 (\pm 0.12)$ & 0.11 \\
$M_{2500}E(z)$ & $T_{x}(>3 \text{keV})$ & $ 0.14(\pm 0.14)$ & $1.68 (\pm 0.22)$ & 0.08 \\
$L_{\rm bol}$& $T_{x}$ & $0.43 (\pm 0.07)$ & $2.76 (\pm 0.09)$ & 0.15 \\
$L_{\rm bol}$ & $M_{2500}$ & $0.40 (\pm 0.11)$ & $1.55 (\pm 0.09)$ & 0.15 \\
$M_{2500}$ & $T_{x}$ & $ 0.02(\pm 0.07)$ & $1.78 (\pm 0.11)$ & 0.11 \\
$M_{g,2500}$ & $M_{2500}$ & $-1.27 (\pm 0.05)$ & $1.20 (\pm0.04)$ & 0.07 \\
\hline
\end{tabular}

\egroup
\label{table:scalings}

Notes: $\sigma$ is the logarithmic RMS residual in dex, $L_{\rm bol}$ is the core-excised luminosity in units of $10^{42}\ergps$, $T_{x}$ is the core-excised temperature in keV, and $M_{2500}$ and $M_{g,2500}$ are in units of $10^{13} M_{\odot}$ .

\end{table}

\subsection{Relationship between Jet Power and Dark Matter Halo Mass}

\label{sec:cavscal}

\newcommand\lcool{L_{\rm cool}}
\newcommand\rcool{R_{\rm cool}}
\newcommand\nelec{n_{\rm e}}
\newcommand\fgas{f_{\rm g}}
\newcommand\rhobar{\overline{\rho}}

\newcommand\rvirial{R_{\rm v}}
\newcommand\tage{t_{\rm age}}
\newcommand\mbh{M_{\rm BH}}

Using scaling arguments \citep{Kaiser_1986}, we now consider how the jet
powers of AGN at the centers of cluster cool-cores are expected to
scale with the properties of their hosts.  In the spirit of scaling
models, we ignore all but the most basic physical processes. Hence,
we assume that the jet mechanical power is the dominant heat source
preventing the hot gas from cooling in every cluster cool-core.
Thus, for the central AGN to regulate cooling and star formation by
feedback, its mean jet power must, at least, match the power radiated
by the gas that could otherwise cool to low temperatures, i.e.~the
cooling power.  The cooling power, $\lcool$, is estimated as the power
radiated by gas within the cooling radius, $\rcool$, where the gas
cooling time equals the age of the cluster.  Although feedback has a
significant impact on the gas distribution within $\rcool$, gas
outside this radius is much less affected (because
the total thermal energy of the gas is a strongly increasing function
of the radius in the region of interest in most clusters).  
\citet{Voit_2005} argue that the entropy index, $\Sigma = kT / \nelec^{2/3}$,
varies with radius as $\Sigma \propto R^\eta$, with $\eta \simeq 1.1$,
in the self-similar regions of clusters, $R > \rcool$.  
Approximating cluster atmospheres as 
constant temperature, which is typically correct to within a factor of a few, 
this implies that the electron density scales as
\begin{equation} \label{eqn:nelec}
\nelec \sim \fgas \rhobar (R / \rvirial)^{-3 \eta / 2},
\end{equation} 
where $\rvirial$ is the virial radius, $\rhobar$ is the mean mass
density of the cluster within the virial radius and $\fgas$ is the gas fraction.  The cooling
time scales as $T / (\nelec \Lambda)$, where we assume that the cooling
function, $\Lambda$, scales with temperature as $\Lambda (T) \propto
T^\alpha$.  Equating the cooling time to the cluster age, $\tage$, then
gives the scaling for $\rcool$,
\begin{equation} \label{eqn:rcscale}
\left(\rcool \over \rvirial\right)^{3\eta/2} \sim {\fgas \rhobar \tage
  \over T^{1 - \alpha}}.
\end{equation}
Observed entropy and gas density profiles are less steep in the
core region that has been affected by AGN feedback than in the self-similar region
beyond $\rcool$, so that the cooling luminosity is dominated by
radiation from around $\rcool$.  Since the thermal energy
within the cooling radius scales as $p(\rcool) \rcool^3$, where $p$
is the pressure, and the cooling time there is $\tage$, the power
radiated from within the cooling radius scales as $\lcool \sim
p(\rcool) \rcool^3 / \tage$.  Noting that $p \sim \nelec T$, we
also have from (\ref{eqn:nelec}) $p(\rcool) \rcool^3 \sim \fgas
\rhobar \rvirial^3  T (\rcool / \rvirial)^{3 - 3\eta/2} \sim \fgas M T (\rcool / \rvirial)^{3 - 3\eta/2}$.  Thus we get
\begin{equation}\label{eqn:paul}
\begin{split}
\lcool & \sim {p(\rcool) \rcool^3 \over \tage}
\sim {\fgas M T \over \tage} \left(\rcool \over \rvirial\right)^{3 - 3\eta/2} \\
& \sim \fgas^{2/\eta} M T^{\{1 - [(2/\eta) - 1] (1 - \alpha)\}}
\rhobar^{[(2 /\eta) - 1]} \tage^{(2/\eta) - 2},
\end{split}
\end{equation}
where the final expression is obtained by using the relation
(\ref{eqn:rcscale}) to eliminate $\rcool / \rvirial$.  Under the usual
self-similar scaling assumptions, $T \sim M^{2/3}
  \rhobar^{1/3}$ and cooling is dominated by thermal 
bremsstrahlung, so that $\alpha = 0.5$.  Using these to eliminate the
temperature from the final expression in (\ref{eqn:paul}) and
retaining only the mass-dependent terms gives the mass dependence of
the cooling power as
$\lcool \sim \fgas^{2/\eta} M^{2 - [2/(3 \eta)]} \sim \fgas^{1.82} M^{1.39}$ for
$\eta = 1.1$.  Finally, using $\fgas \sim M^{0.2}$, as
  measured in section \ref{sec:gasmass}, gives $\lcool \sim
  M^{1.75}$.  For heating to balance cooling, we need $P_{\rm cav} = \lcool$.
Strictly, $\fgas$ is constant for the self-similar model,
so that allowing it to be mass-dependent here highlights a weakness of
scaling models. 
Nevertheless, observed gas fractions increase with cluster mass and we
should expect this to steepen the mass dependence of feedback power.

SMBH co-evolution appears to be tied more closely to the bulge than the halo (Reviewed by \citealt{Kormendy_2013}). In addition, \citet{Bluck_2014} find that quenching in central galaxies is most tightly coupled to the bulge mass, and suggest AGN feedback as the most likely quenching mechanism. In order to examine this, we compare AGN energetics with $M_{2500}$ and with $M_{K}$ of the BCG.

\begin{figure*}
\begin{minipage}{1.0\textwidth}
\centering

\includegraphics[width=0.42\textwidth]{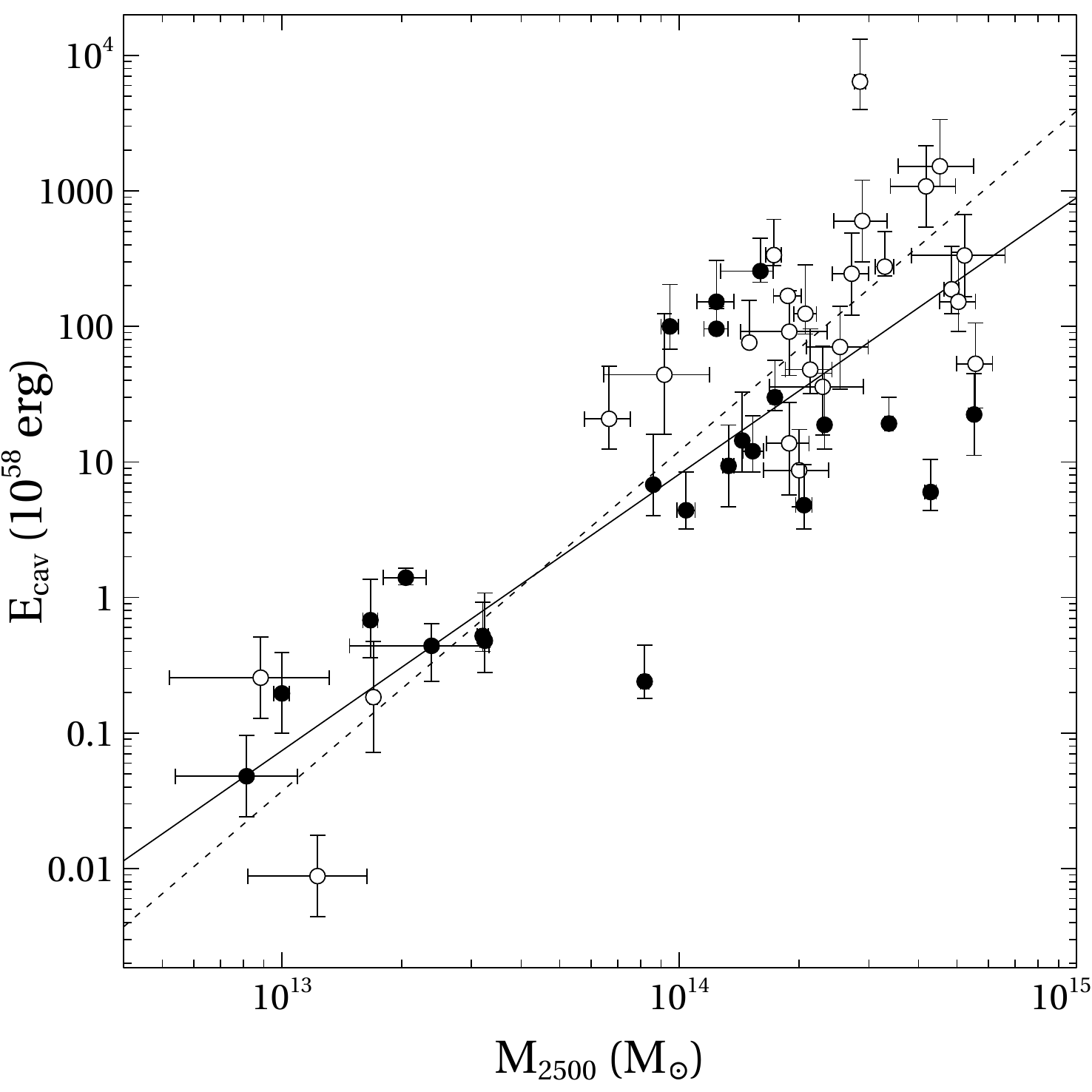} 
\includegraphics[width=0.42\textwidth]{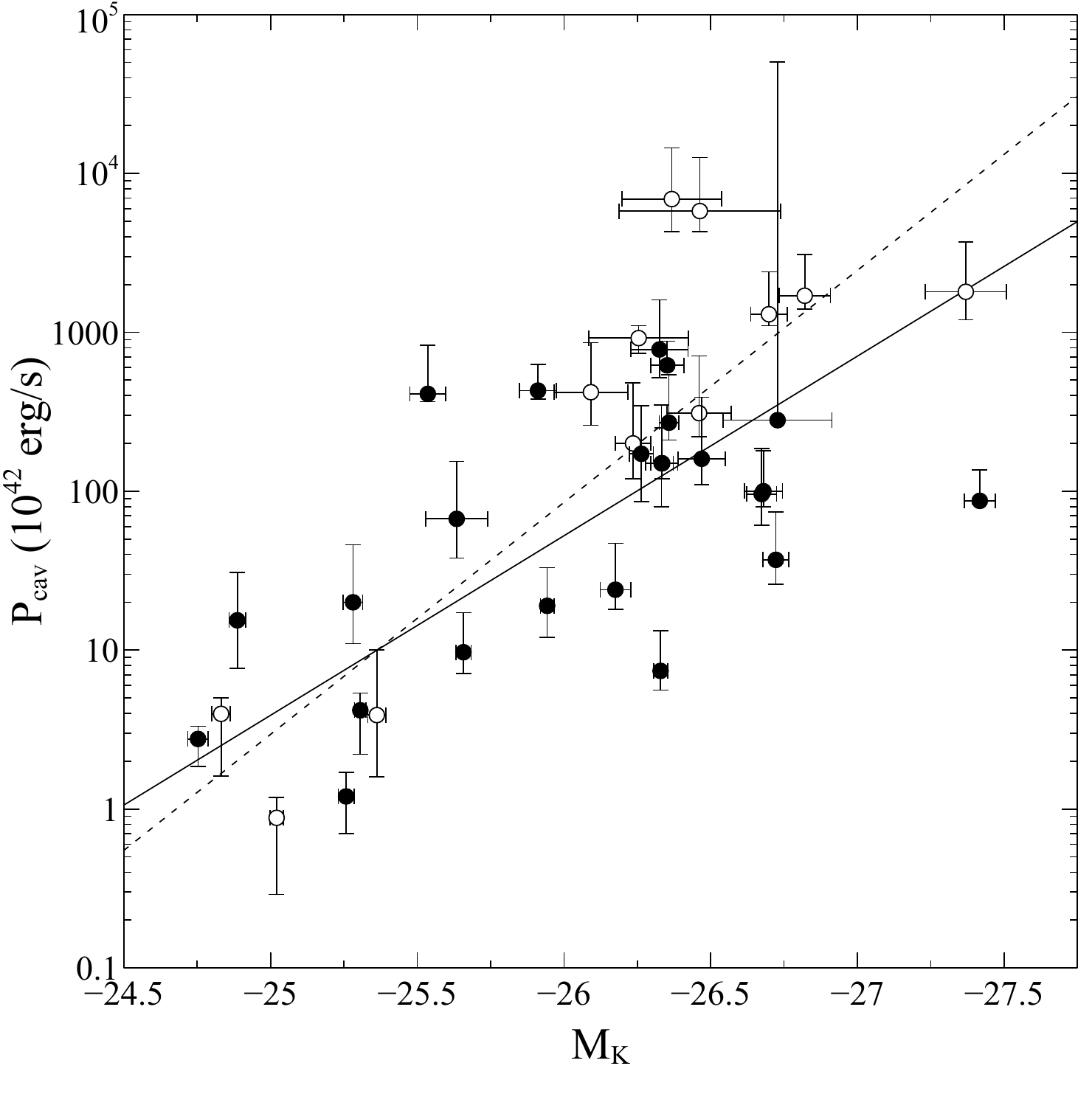} \\
\includegraphics[width=0.42\textwidth]{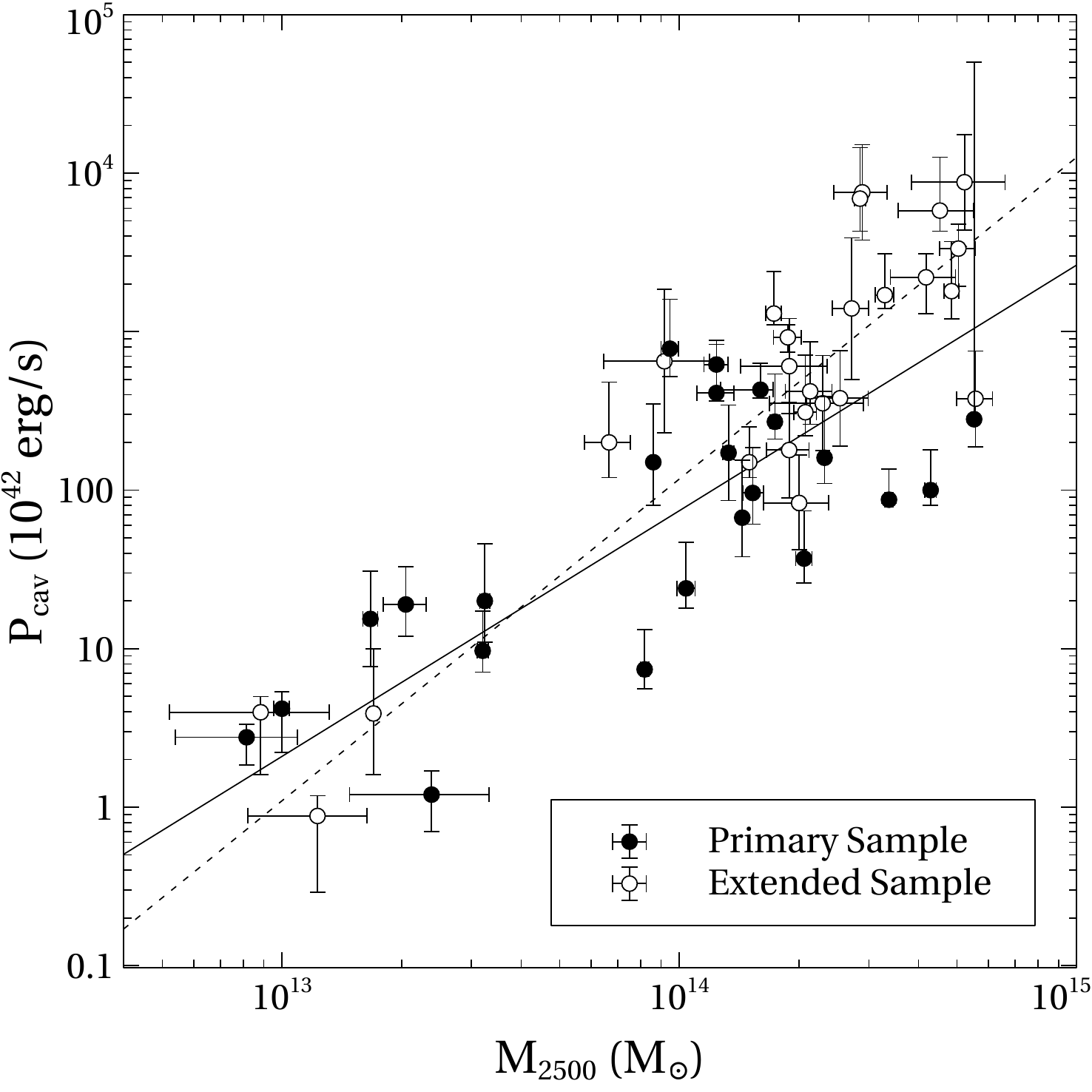} 
\includegraphics[width=0.42\textwidth]{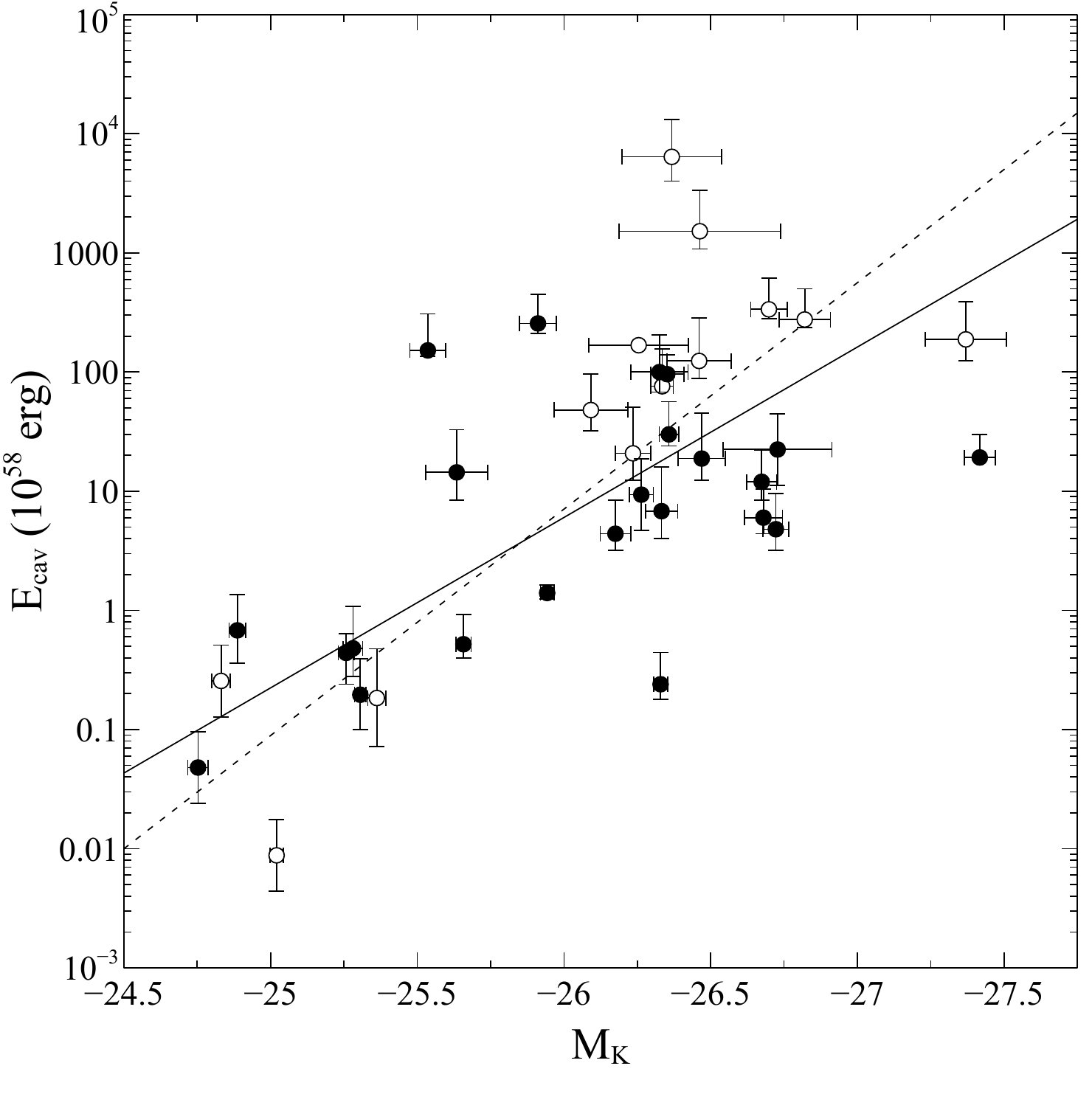} \\
\includegraphics[width=0.42\textwidth]{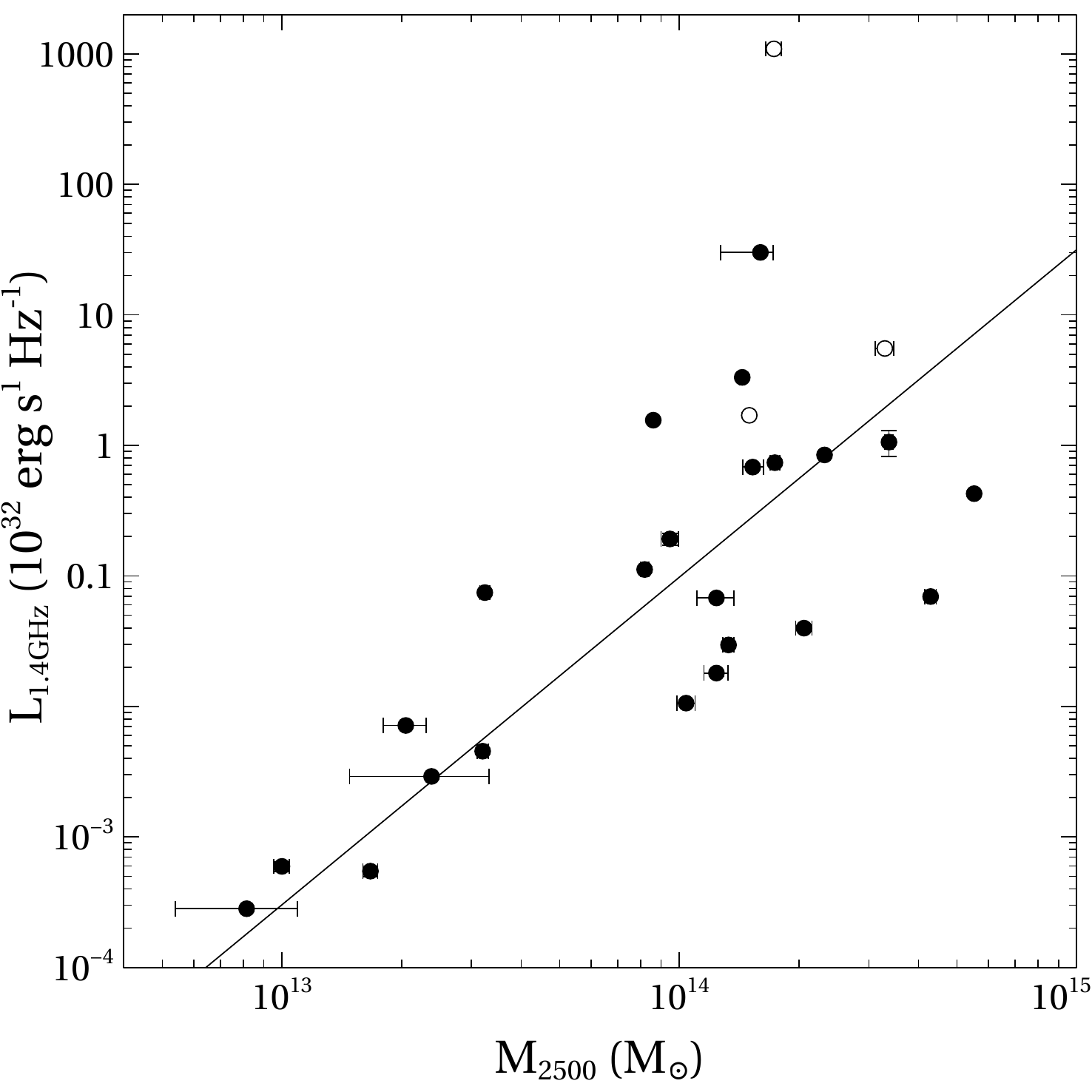} 
\includegraphics[width=0.42\textwidth]{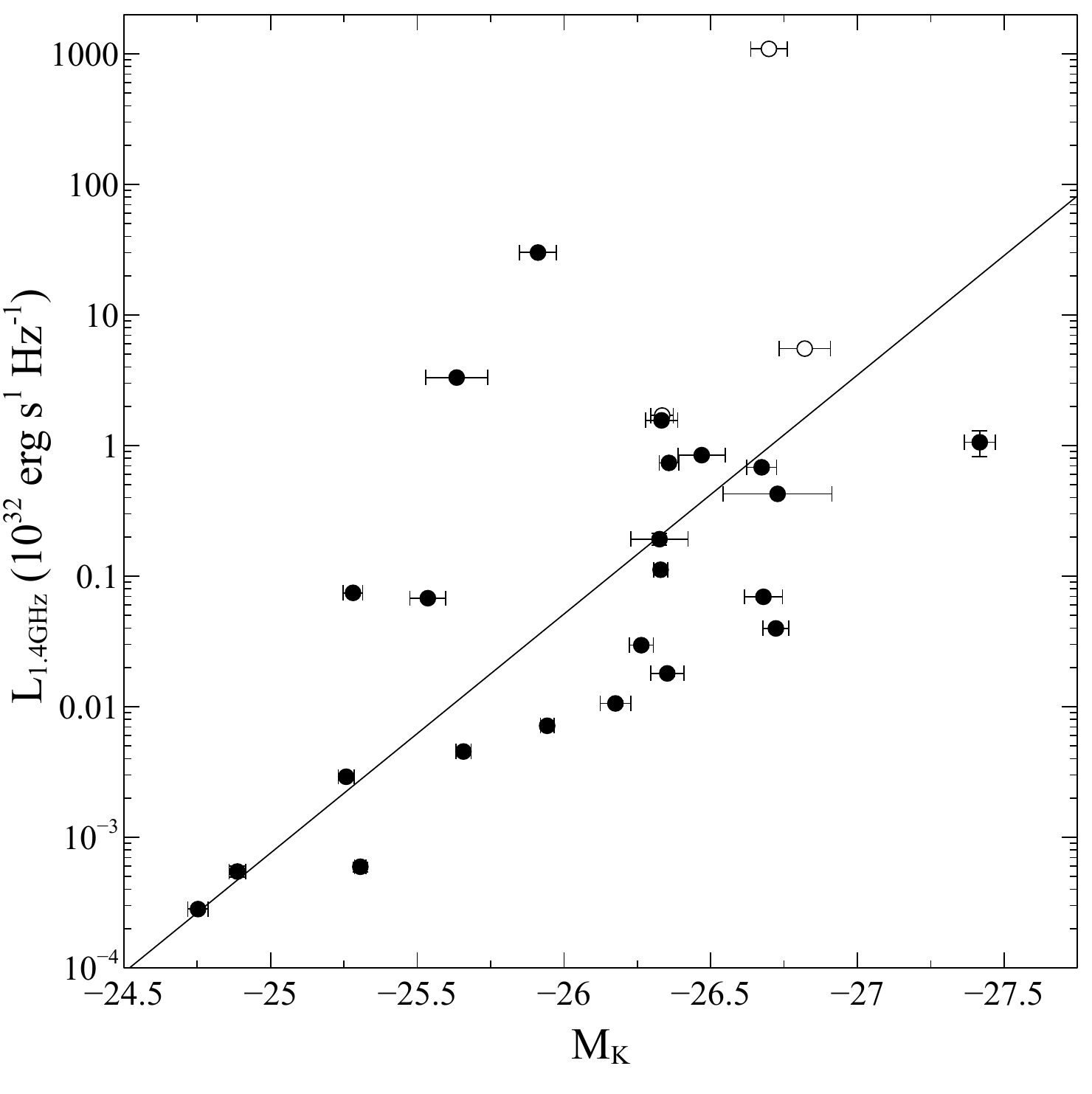}

\caption{Top: Cavity power vs. M$_{2500}$ (\textit{Left}) and vs. M$_{K}$ (\textit{Right}). Middle: Cavity energy vs. M$_{2500}$ (\textit{Left}) and vs. M$_{K}$ (\textit{Right}). Bottom: 1.4GHz luminosity vs. M$_{2500}$ (\textit{Left}) and vs. M$_{K}$ (\textit{Right}). The solid and dashed lines denotes the best BCES bisector fit for the primary and extended samples respectively. Cavity energy calculated as E$_{\rm cav} = 4$pV, and the buoyancy timescale is used in calculating P$_{\rm cav}$. }

\label{figure:mvsp_figure}

\end{minipage}
\end{figure*}

We fit power law relationships using $E_{\rm cav}$, $P_{\rm cav}$, and $L_{\rm 1.4GHz}$.  A2204 is excluded from our fits due to the large systematic uncertainty in the energetics of the outer cavity system \citep{Sanders_2009}. The error on the scatter is calculated through bootstrap resampling with 10000 iterations. The relations for M$_{2500}$ and M$_{K}$ are plotted in Figure \ref{figure:mvsp_figure}, and are listed in Table \ref{table:Pjet_fits} with the 1-$\sigma$ logarithmic vertical scatter. 

\begin{table}
\caption{Mass Dependence of Feedback}
\begin{tabular}{c}
Fit Results to $\text{log}(y) = a + b \text{ log(x)}$
\end{tabular}

\bgroup
\setlength{\tabcolsep}{.3em}
\def\arraystretch{1.2}
\begin{tabular}{l l l l l l}
\hline
x & y & a & b & $\sigma$$^{a}$ & r$^{g}$ \\
\hline
Primary \\
\hline
$M_{2500}$$^{b}$ & $P_{\rm cav}$$^{c}$ & $0.32 (\pm 0.24)$ & $1.55 (\pm 0.26)$ & $0.56 (\pm0.08)$ & 0.73 \\
$M_{2500}$ & $E_{\rm cav}$$^{d}$ & $-1.13 (\pm 0.25)$ & $2.04 (\pm 0.30)$ & $0.70 (\pm0.09)$ & 0.74 \\
$M_{2500}$ & $L_{\rm 1.4}$$^{e}$ & $-3.52 (\pm 0.24)$ & $2.51(\pm 0.33)$ & $0.89 (\pm0.14)$ & 0.75 \\
$M_{\rm K}$$^{f}$ & $P_{\rm cav}$ & $-27.7 (\pm 5.3)$ & $-1.13 (\pm 0.20)$ & $0.72 (\pm0.09)$ & -0.56 \\
$M_{\rm K}$ & $E_{\rm cav}$ & $-36.4 (\pm 7.0)$ & $-1.43 (\pm 0.27)$ & $0.93 (\pm0.15)$ & -0.54 \\
$M_{\rm K}$ & $L_{\rm 1.4}$ & $-48.5 (\pm 7.1)$ & $-1.83 (\pm 0.28)$ & $1.11 (\pm0.20)$ & -0.60 \\
\hline
\end{tabular}
\begin{tabular}{l l l l l}
Extended \\
\hline
$M_{2500}$ & $P_{\rm cav}$ & $0.04 (\pm 0.20)$ & $2.03(\pm 0.17)$ & $0.60 (\pm0.06)$ \\
$M_{2500}$ & $E_{\rm cav}$ & $-1.43 (\pm 0.25)$ & $2.51 (\pm 0.22)$ & $0.73 (\pm0.08)$ \\
$M_{\rm K}$ & $P_{\rm cav}$ & $-36.0 (\pm 4.6)$ & $-1.46 (\pm 0.18)$ & $0.78 (\pm0.09)$ \\
$M_{\rm K}$ & $E_{\rm cav}$ & $-48.6 (\pm 6.3)$ & $-1.90 (\pm 0.24)$ & $1.06 (\pm0.13)$ \\
\hline
\end{tabular}

\egroup
\label{table:Pjet_fits}

Notes: $^{a}$$\sigma$ is the RMS residual. $^{b}$$M_{2500}$ is in units of $10^{13} M_{\odot}$. $^{c}$$P_{\rm cav}$ is given in units of $10^{42} \ergps$. $^{d}$$E_{\rm cav}$ is in units of $10^{58}$erg. $^{e}$$L_{\rm 1.4}$ is the central 1.4GHz luminosity in units of $10^{40}$Hz. $^{f}$The K-band luminosity of the BCG from 2MASS. $^{g}$ Correlation coefficient between the parameters.

\end{table}

In the primary sample, we find best-fitting relation between cavity energy and halo mass, $E_{\rm cav} \propto M_{2500}^{2.04\pm0.30}$. 
The best-fitting relationship between jet power and halo mass, $P_{\rm cav} \propto M_{2500}^{1.55\pm0.26}$, is 
consistent with our self-similar model scaling of (\ref{eqn:paul}), and with the \citet{Somerville_2008} semi-analytic model that assumes jet power scaling
with black hole mass as $M_{\rm BH}^1.6$.
The relation between $P_{\rm cav}$ and $M_{2500}$ is further consistent with the model detailed in \citet{Sijacki_2006} with $P_{\rm cav} \propto M_{\rm BH}$ and an M-$\sigma$ relation of $M_{\rm BH} \propto \sigma^{4}$ or $M_{\rm BH} \propto \sigma^{5}$. The relation between $E_{\rm cav}$ and $M_{2500}$ is steeper than either of these models, but favours co-evolution of the forms $P_{\rm cav} \propto M_{\rm BH}$, and $M_{\rm BH} \propto \sigma^{5}$. The assumption that $E_{\rm cav} \propto M_{\rm BH}$ is equivalent to fixing accretion to be proportional to the Eddington rate (e.g., \citealt{Somerville_2008}, which we investigate further in Section \ref{sec:accretion}.

The extended sample yields steeper scalings of the forms $P_{\rm cav} \propto M_{2500}^{2.03\pm0.17}$ and $E_{\rm cav} \propto M_{2500}^{2.51\pm0.22}$,  both of which are steeper than the primary sample and  the simple analytic model of \citep{Sijacki_2006}.  As the extended sample includes higher redshift clusters, its steepening could be either a real evolutionary effect or selection bias introduced by our insensitivity to low power AGN at higher redshifts that would lie below the relation. We cannot distinguish between these
possibilities.  Nevertheless, having been drawn from a complete, flux-limited sample, the results for the primary sample are reliable. 

\subsubsection{Scaling between AGN Power, Gas, and Stellar Mass}

The dependence of $P_{\rm cav}$ on gas mass is similar to that on total mass.
The RMS residual of $P_{\rm cav}$ on gas mass is $0.54\pm0.08$, and the fits are equally good for gas and total mass. The situation is much the same for  $T$, $L_{\rm bol, total}$, $L_{\rm bol, exc}$, due to the close relationship
between the gas and total masses.   Assuming the black hole is fuelled by gas, jet power would more naturally be expected
to correlate with gas mass.  However, we find no quantitative evidence that gas mass is preferred over total mass. 

The scatter in the trend between jet power and the BCG's K-band magnitude (a proxy for stellar mass) in Figure \ref{figure:mvsp_figure} is not significantly larger than that for the trend with $M_{2500}$ for both the primary and extended samples.  

We also used the 1.4 GHz luminosity as an independent proxy for mechanical power. We find that the scatter in the relationship between $L_{\rm 1.4 GHz}$ and $M_{2500}$ 
is $0.89\pm0.14$ dex and the scatter for $M_{K}$ is $1.11\pm0.20$ dex.  The difference is insignificant.   

\subsubsection{Correlation Bias}

A biased cavity detection procedure could potentially affect the scaling relations involving cavity power. One possible source of bias would be a preferred angular size
for detectable cavities due to Chandra's spatial resolution. With $0.492"$ pixels, cavities need to be a few arcseconds across to be resolved by Chandra. The most obvious manifestation of a preferred angular size would be a spurious correlation with redshift. Cavity power would scale as $D_{A}^{2}$, since $V \propto D_{A}^{3}$, and $t_{\rm buoy} \propto D_{A}$. In a flux-limited sample, this would cause a spurious correlation between cavity power and X-ray luminosity. 

\begin{figure}
\begin{minipage}{1.0\columnwidth}
\centering

\includegraphics[width=1.0\columnwidth]{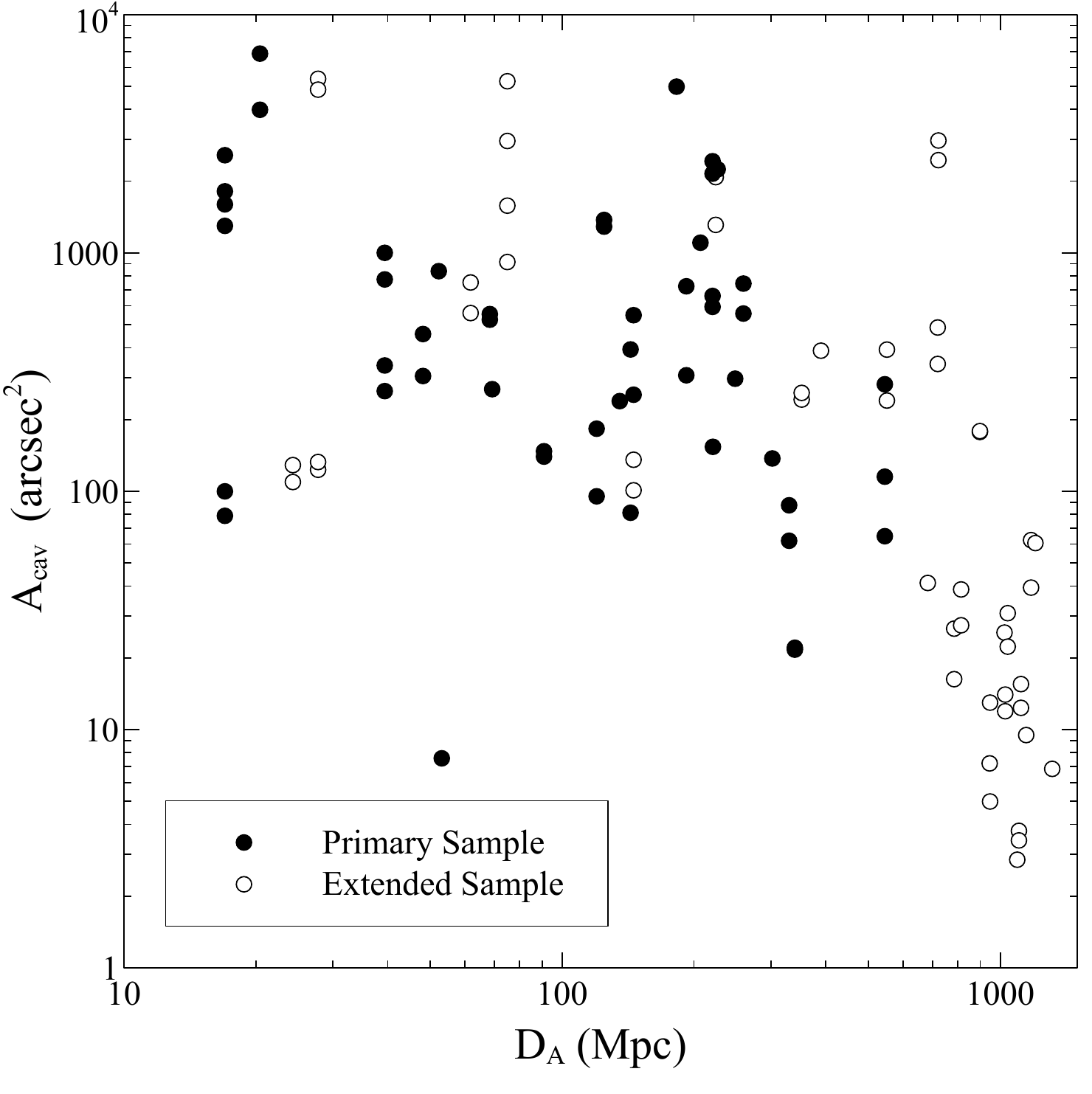}

\caption[Angular area of cavities vs. angular diameter distance]{Angular area of all detected cavities in our sample vs. angular diameter distance. If there is a bias towards detecting a specific angular size of cavities, the relation should appear flat with small scatter.}

\label{figure:cavsizes}
\end{minipage}
\end{figure}

The angular areas of the cavities in our systems are plotted against angular diameter distance in Figure \ref{figure:cavsizes}. The areas span 3 decades and, on average, decrease with redshift. This trend is inconsistent with a preferred angular size, and implies no spurious redshift dependence.  Thus our correlations of cavity power with redshift dependent quantities are likely physical trends.

Several other factors suggest a genuine relationship between halo mass and $P_{\rm cav}$. It is unlikely that much more powerful outbursts are missed in the nearby, low-mass systems as they would be easily detected. In the full HIFLUGCS sample, we also find a strong correlation in clusters with short central cooling times between jet power 
and halo mass when using 1.4 GHz luminosity as a proxy for jet power (see Figure \ref{fig:MP_HIFLUGCS} and Section \ref{sec:bs}). All HIFLUCGS clusters with central cooling times 
below 1 Gyr are radio audible. Therefore, we are missing few if any objects to the bottom-right or top-left of the relationship between $P_{\rm cav}$ and $M_{2500}$. 
Therefore, the correlation is likely real. 

\subsubsection{Integrated star formation efficiency}
\label{sec:sfe}

We now investigate the potential impact of AGN feedback on the efficiency of star formation, adopting the ratio of the BCG's stellar mass to the total baryon mass 
as a probe  i.e. $M_{*} / M_{\rm baryon}$ where $M_{\rm baryon} \equiv M_{g,2500} + M_{*,\rm BCG}$. We do not consider the masses of satellite galaxies in the core , but note that the BCG dominates the stellar mass there  \citep{Burke_2015}.  

\begin{figure*}
\begin{minipage}{1.0\textwidth}
\centering

\includegraphics[width=0.42\textwidth]{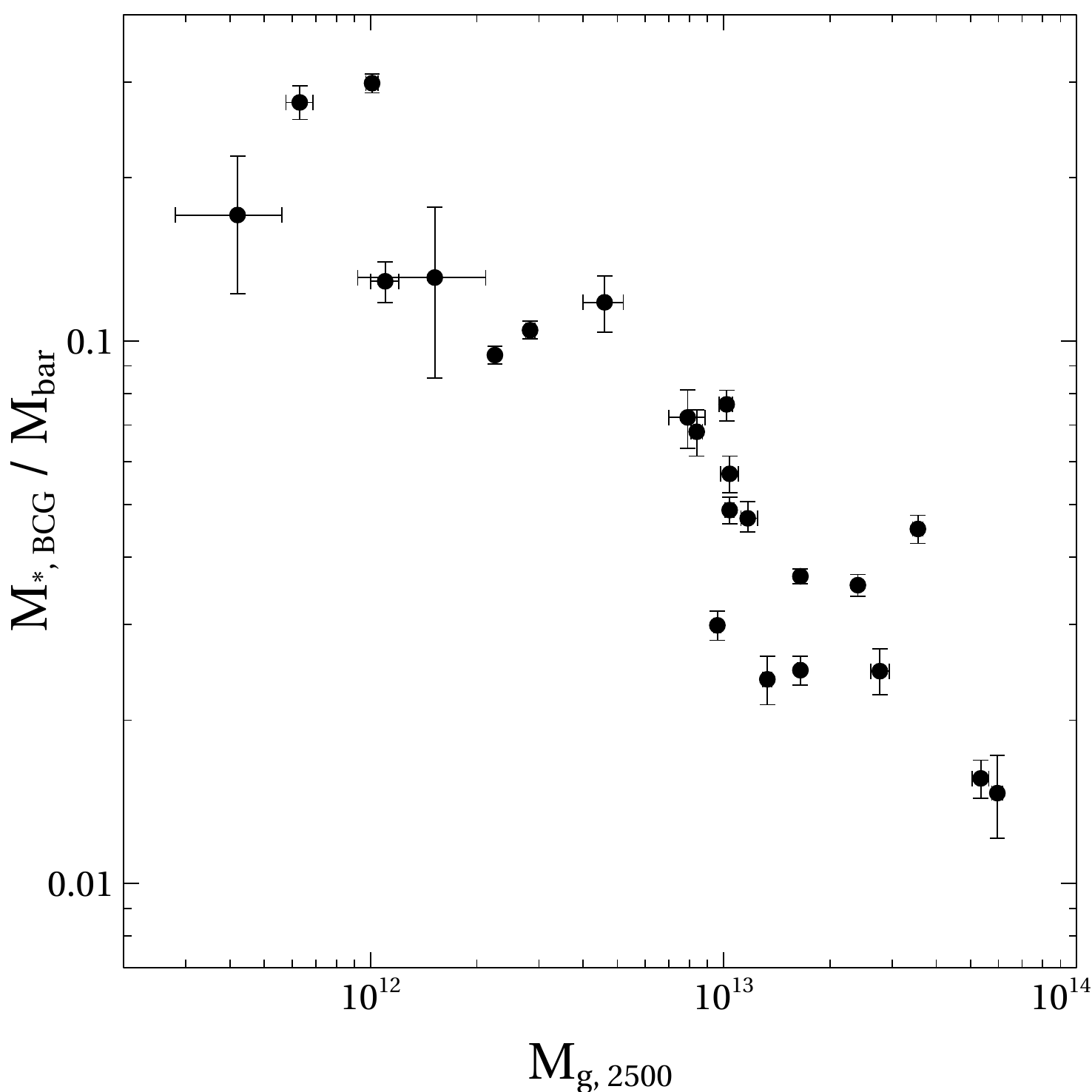} 
\includegraphics[width=0.42\textwidth]{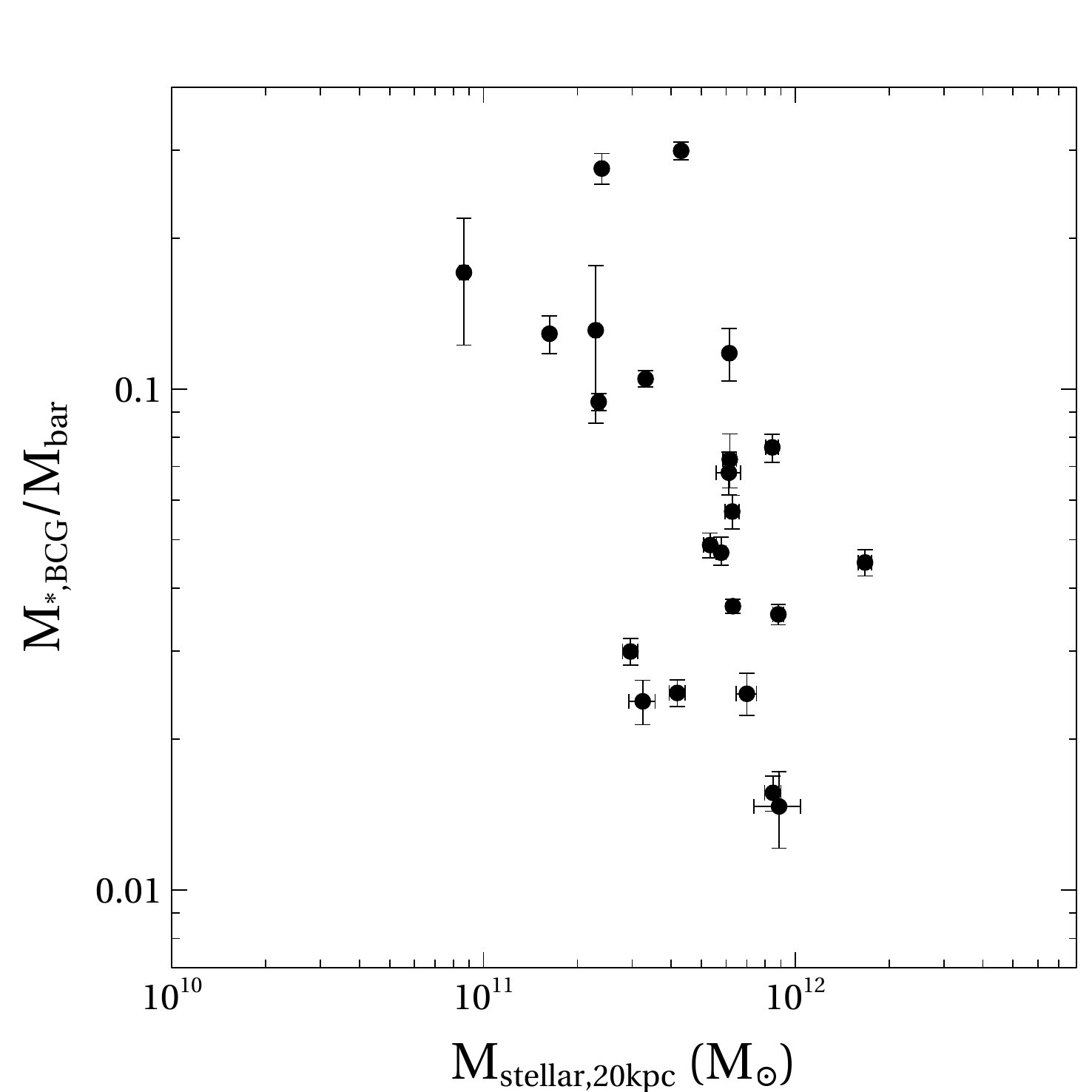} \\
\includegraphics[width=0.42\textwidth]{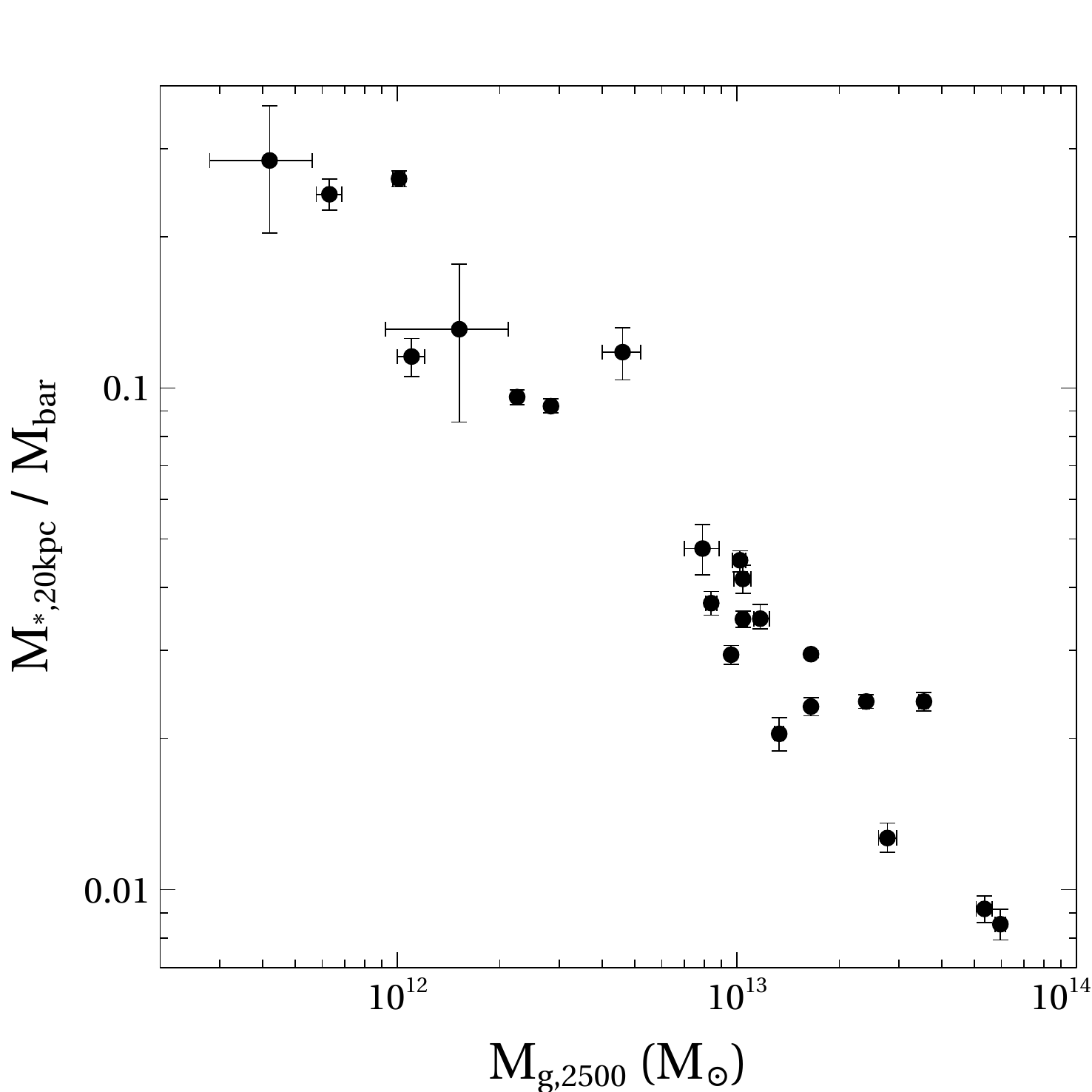} 
\includegraphics[width=0.42\textwidth]{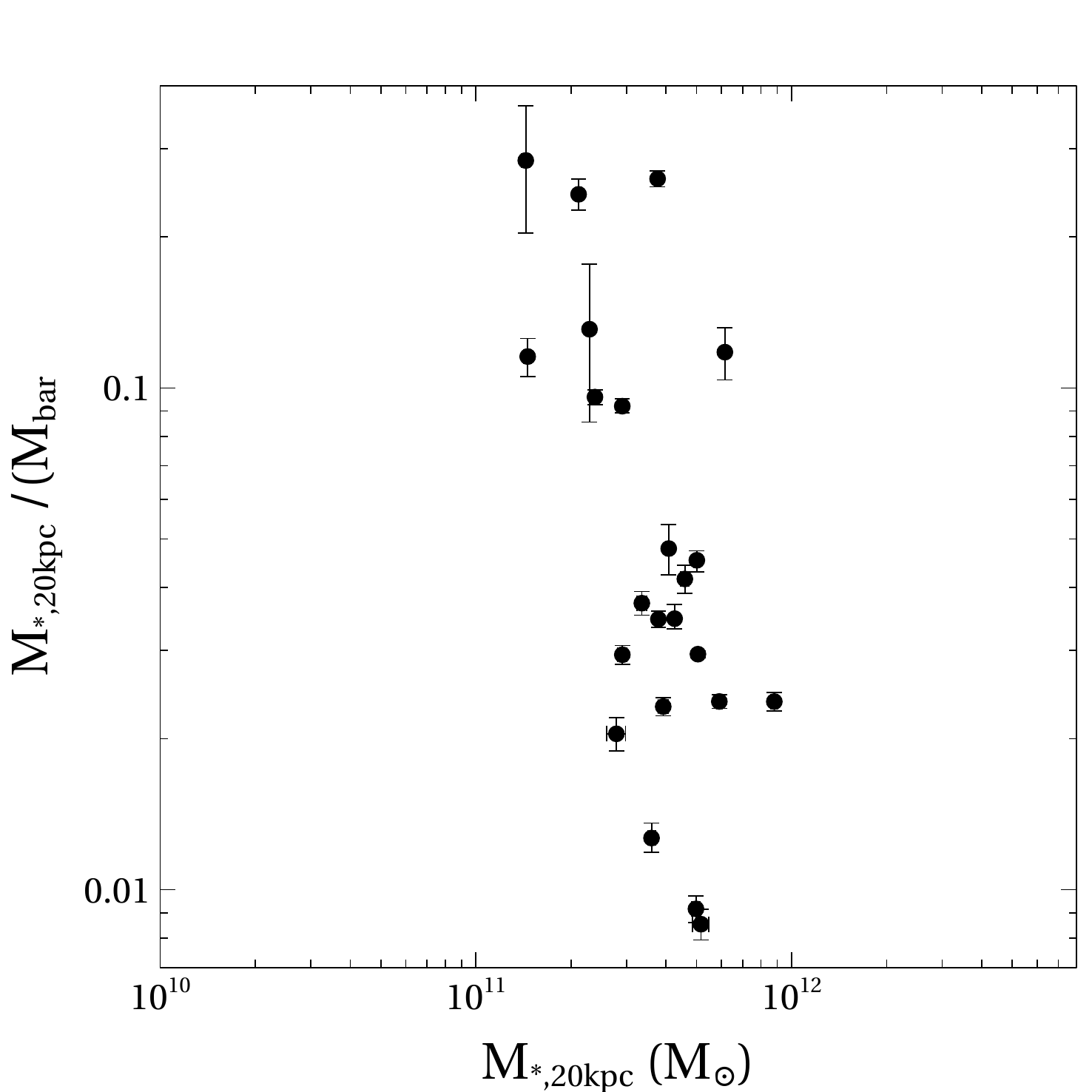}

\caption{Top: $M_{*,\rm BCG}/(M_{g,2500}+M_{*,\rm BCG})$ vs. M$_{g,2500}$ (\textit{left}), $M_{*,\rm BCG}/(M_{g,2500}+M_{*,\rm BCG})$ vs. M$_{*,\rm BCG}$ (\textit{right}). Bottom: Same as above, but with stellar mass calculated at 20kpc.}

\label{figure:BCG_frac}

\end{minipage}
\end{figure*}

We plot the fraction of baryons in stars against both gas mass and stellar mass in Figure \ref{figure:BCG_frac}. These diagrams show that
that the gas fraction locked-up in stars declines with total gas mass from a value of about 20\% in groups with baryon masses of $\sim 10^{12}~\rm M_\odot$ to
roughly $1\%$ or less in the centres of rich clusters with gas masses upward of $10^{14}~\rm M_\odot$.  Interpreted as the efficiency of conversion of gas into stars, the integrated star formation efficiency of BCGs decreases dramatically with the total gas and halo masses, as previously observed (e.g. \citealt{David_1995,Behroozi_2013,Gonzalez_2013}). This indicates that the central stellar mass is remarkably consistent from galaxy to galaxy, and the trend is sharper when restricting the measurement to stellar mass within a 20 kpc aperture.

The right hand panels of Figure \ref{figure:BCG_frac} show that the stellar mass range is much smaller than the range of halo mass.
The stellar mass increases with halo mass with an amplitude no larger than the scatter in the trend.  The data show that BCGs have grown to a relatively narrow range of masses despite the
enormous fuel supply available to continue fuelling their growth.  BCGs residing in low mass halos containing almost $30\%$ of the baryonic mass within $R_{2500}$,  consumed
their fuel supplies more efficiently than BCGs in the highest mass systems, which consumed only $1-2\%$ of the baryonic mass within $R_{2500}$.  
 
\subsubsection{Effects of Cooling Atmospheres}
\label{sec:sfecf}
In order to convert the fuel reserves locked-up in hot atmospheres into stars, the gas must cool and accrete into the centre of the halo.  
In order to investigate the potential effects of cooling atmospheres on the integrated star formation efficiency in BCGs,  we compare the stellar mass to the halo mass in the HIFLUGCS sample. We examine the relationship between the ratio of stellar to halo mass $M_{\rm *,BCG} / M_{2500}$ and halo mass $M_{2500}$ in clusters with and without 
cooling atmospheres.  To isolate the cooling and non-cooling atmospheres, we have divided the sample by central cooling time, indicating those clusters whose 
central cooling times lie above and below 1 Gyr, which corresponds approximately to the cooling time threshold above which cold clouds and star formation
are observed in BCGs \citep{Rafferty_2008}.  

\begin{figure*}
\begin{minipage}{1.0\textwidth}
\centering

\includegraphics[width=0.45\textwidth]{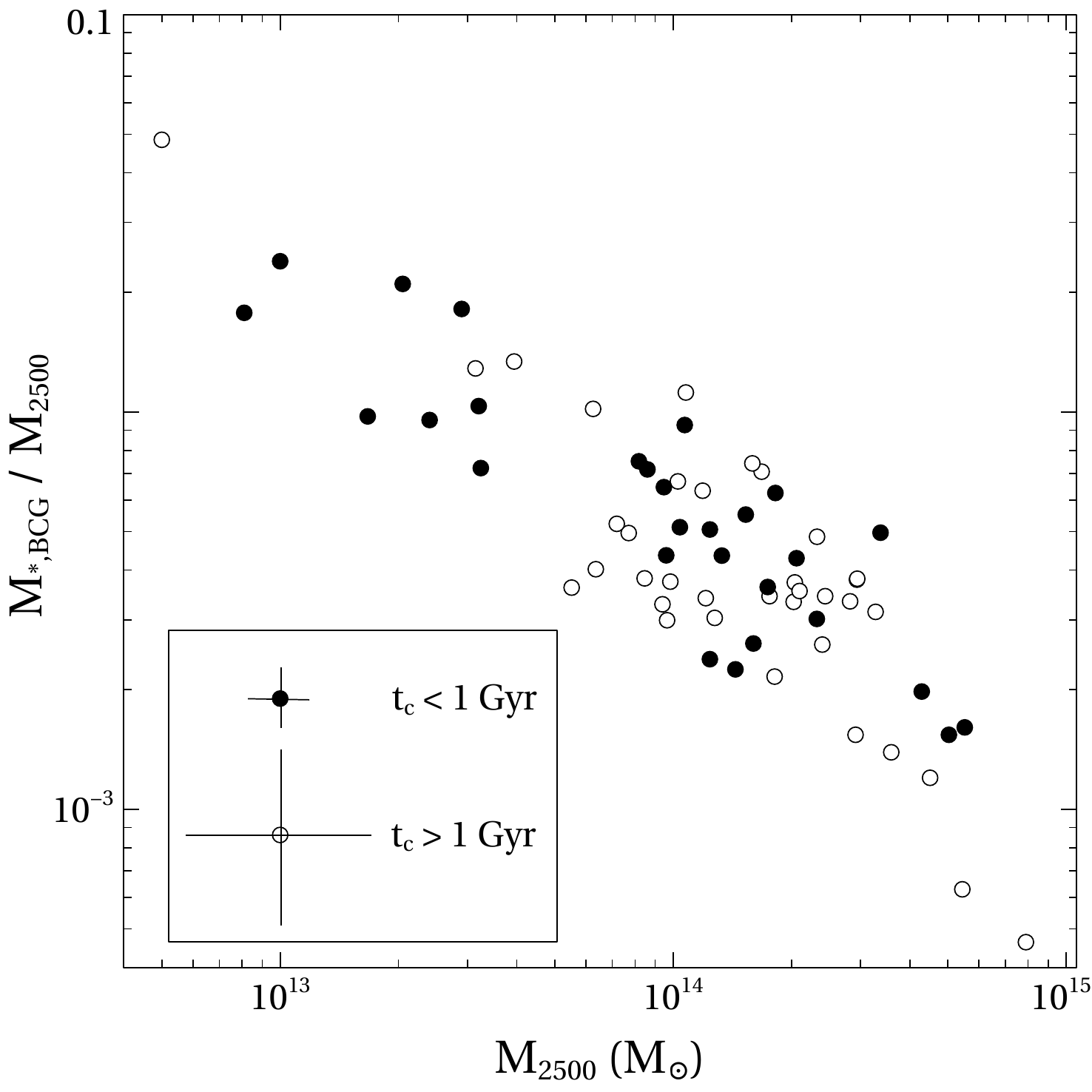} 
\includegraphics[width=0.45\textwidth]{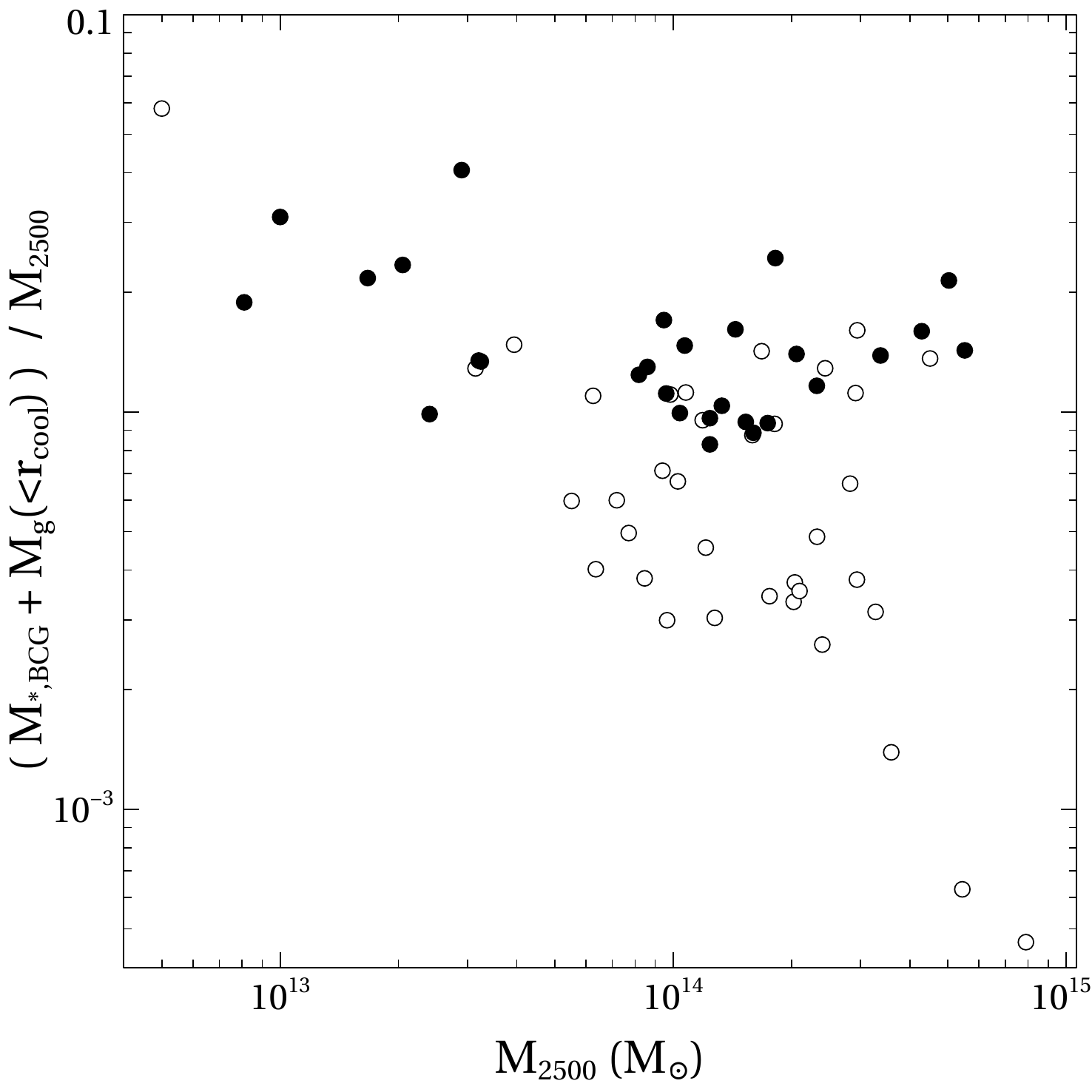}

\caption{Left: M$_{\rm *,BCG}$/M$_{2500}$ vs. M$_{2500}$ in the full HIFLUGCS sample. Right: (M$_{\rm *,BCG}$ + M$_{g}(<r_{\rm cool}) )/M_{2500}$ vs. M$_{2500}$, representing the total possible stellar mass in the BCGs assuming cooling proceeds unimpeded. Systems are shown as either open or closed symbols based on the central cooling times from \citet{Hudson_2010}. The average size of the error bars is shown on the bottow left.}

\label{fig:Mfrac_HIFLUGCS}

\end{minipage}
\end{figure*}

We find no discernible separation between BCGs with long and short central cooling times in the left panel of Figure \ref{fig:Mfrac_HIFLUGCS}.  Although most BCGs with short central cooling time are forming stars,
recent star formation has apparently contributed little to the total stellar mass of the galaxies.  
If cooling and star formation were able to ensue unimpeded over the past several Gyr, the gas available within the cooling radius should fuel enough star formation to significantly increase the stellar mass.  

In order to examine this question quantitatively, we calculated the gas mass within the cooling region and added it to the observed stellar mass to estimate
how much the galaxies would have grown had this gas formed stars.  The values of $r_{\rm cool}$ represent the radius within which the cooling time of the ICM is less than 7.7 Gyr, taken from \citet{birzan_2012}. In those systems where we do not have the full gas mass profile, the gas mass within r$_{\rm cool}$ was estimated assuming the gas density follows an NFW profile with $c_{500}=3$, a typical value for halos in our mass range \citep{vikh_2006}. M$_{g,2500}$ is calculated using $M_{g}/10^{13}M_{\odot} = 10^{-1.27} (M_{2500}/10^{13}M_{\odot})^{1.20}$, derived in section \ref{sec:gasmass}, which with the NFW assumption gives a value of $M_{g}( <r_{\rm cool})$. The value $(M_{\rm *,BCG} + M_{g}( <r_{\rm cool}) ) / M_{2500}$ then indicates how much the BCG would have
grown if the hot atmosphere had cooled efficiently and formed stars shown in the right panel of Figure \ref{fig:Mfrac_HIFLUGCS}.  A comparison between the left
panel of Figure \ref{fig:Mfrac_HIFLUGCS}, showing BCGs as they are, and the right panel which shows their mass had all the gas that could have cooled over the last 7.7Gyr formed stars, indicates the
dramatic difference in their masses by factors of several to nearly a factor of 10.  That BCGs do not segregate in this diagram is a graphic indication of how
efficiently cooling has been suppressed.  Virial shocks in groups and clusters heat accreting gas, which limits the cooling gas that can reach the BCG in halos above $\sim 10^{12} M_{\odot}$ \citep{Dekel_2006}.  However, one or more additional mechanisms are need to suppress cooling, possibly including thermal conduction  \citep{Voigt_2004} and AGN feedback
\citep{Mcnamara_2007}.  Figure \ref{fig:Mfrac_HIFLUGCS} shows that heating has been remarkably efficient over the half the age of the universe (i.e. since z=1) 
\citep{Ma_2011,Ma_2013,McDonald_2013}. Studies have shown \citep{Lidman_2012,Lidman_2013} that BCGs have grown by less than a factor of two in stellar mass since $z \sim 1$, driven primarily by mergers rather than cooling. 

\subsection{Accretion Rate} \label{sec:accretion}

The analytic scaling between cavity energy and halo mass discussed above in section \ref{sec:cavscal} assumes the accretion rate is a low, fixed fraction of the Eddington rate. In this section, we investigate whether accretion rate is a constant function of mass in our sample.

As matter accretes onto the central SMBH, the binding energy released by the accreting mass will drive an outburst with some efficiency $\epsilon$. Assuming that the X-ray cavities dominate the total energy output of the AGN, the accretion rate can be estimated as
\begin{equation} \label{eq:accretion}
\dot{M}_{\rm BH} = \frac{P_{\rm cav}}{\epsilon c^{2}}.
\end{equation}
The maximum accretion rate in which gravity is balanced by radiation pressure from accretion is given by the Eddington accretion rate, which for a fully ionized plasma can be calculated as
\begin{equation} \label{eq:eddington}
\frac{\dot{M}_{\rm Edd}}{M_{\odot} \text{yr}^{-1}} \approx \frac{2.2}{\epsilon} \left( \frac{M_{\rm BH}}{10^{9}M_{\odot}} \right).
\end{equation}

\begin{figure}
\begin{minipage}{1.0\columnwidth}
\centering

\includegraphics[width=1.0\textwidth]{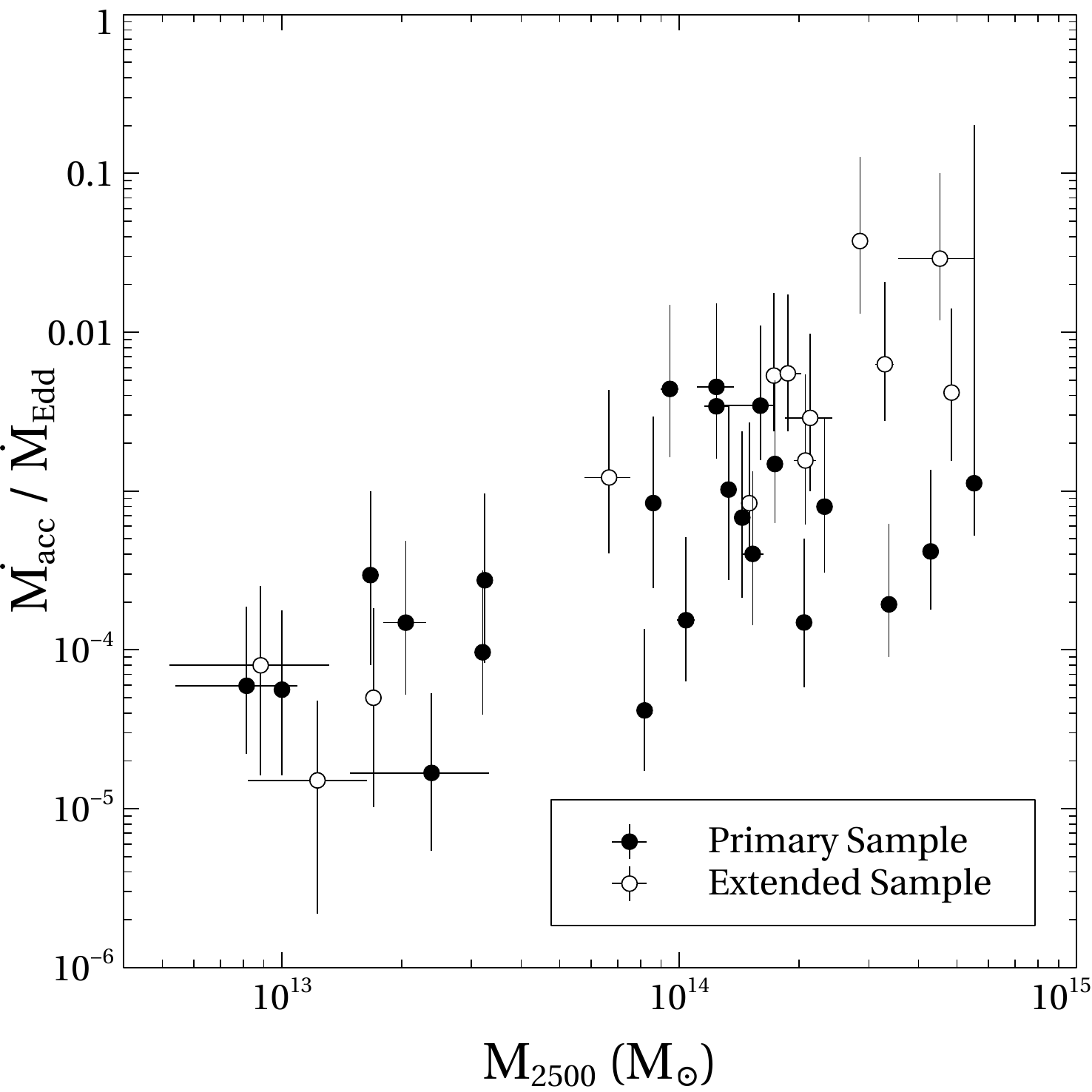} 

\caption{Accretion rate is compared to halo mass in our primary sample. Black hole masses are estimated from \citet{Graham_2007} using K-band luminosities derived from 2MASS, and $\dot{M}_{\rm acc}$ is estimated as $\dot{M}_{\rm acc} = P_{\rm cav}  /  \epsilon c^{2}$.}

\label{figure:Efficiency}

\end{minipage}
\end{figure}

In Figure \ref{figure:Efficiency} we plot the accretion rate $\dot{M}_{\rm acc} / \dot{M}_{\rm Edd}$ against halo mass, using the black hole masses derived in section \ref{sec:bhmass}.
The accretion rates in our systems range between $10^{-5}\dot{M}_{\rm Edd} $ and $ 0.04 \dot{M}_{\rm Edd}$, all well below the Eddington rate, as expected for radiatively inefficient accretion. The weak positive trend shown in Figure \ref{figure:Efficiency} is
harder to interpret.  Taken at face value, it implies that higher mass halos accrete at higher specific accretion rates.  However, the trend may be the result of incorrect underlying assumptions, such as misestimation of black hole masses.  
The estimates of $\dot{M}_{\rm Edd}$ rely on the scaling relation between $M_{\rm BH}$ and $L_{K}$. The black hole mass scaling relation is poorly constrained above $M_{\rm BH} \sim 10^{9}M_{\odot}$ and data suggest an upturn to larger black hole masses \citep{Lauer_2007} in
more massive galaxies \citep{Graham_2015}. Assuming accretion rate is indeed proportional to mass, in our sample, with $\dot{M}_{\rm acc} / \dot{M}_{\rm Edd} = 10^{-4}$,  black hole masses exceeding $10^{10} M_{\odot}$ would be implied in the most massive clusters.  Given M87's black hole mass of several $10^9~\rm M_\odot$ lies in a less luminous central galaxy, the existence of ultramassive black holes is plausible.

\section{AGN in Cooling and non-Cooling Atmospheres}
\label{sec:ncc}

\subsection{Jet Power as a Function of Mass \& Central Cooling Time}
\label{sec:bs} 

Earlier studies have suggested that AGN power may scale differently in cool and non cool-core systems. \citet{Mittal_2009} found a correlation between the radio luminosity and the bolometric X-ray luminosity in cool-core systems, but no evidence for a correlation in weak or non cool-cores. In addition, \citet{Ma_2013} found only a weak trend between $ROSAT$ X-ray luminosity and NVSS radio luminosity, presumably including both cool-core and non cool-core clusters. In this section we investigate the effects of AGN heating using mechanical power in the complete HIFLUGCS sample, which includes clusters with and without cool-cores. 

In order to examine the relationship between feedback in cooling and non-cooling atmospheres, we have separated the HIFLUGCS sample into systems with central cooling times above and below 1 Gyr.    
In all systems, we estimate the mechanical power from the central 1.4 GHz radio luminosity, using the relation from \citet{Cavagnolo_2010}
\begin{equation} \label{eq:jetrad}
\text{log} P_{\rm mech} = 0.75~(\pm0.14) \text{log} P_{1.4} + 1.91~(\pm0.18),
\end{equation}
where $P_{\rm mech}$ is in units of $10^{42}\ergps$ and $P_{1.4}$ is in units of $10^{40}\ergps$. This relation has been measured over the full range of radio luminosities in our sample, and has a scatter of 0.78 dex.  Because many of our systems do not have well defined X-ray cavities and/or uniformly deep Chandra observations to
detect them, applying the scaling relation to all systems yields uniform but somewhat higher variance over the sample.  We refer to the power estimated using this relation as the ``mechanical power'', or $P_{\rm mech}$, throughout this section. 

\begin{figure}
\begin{minipage}{1.0\columnwidth}
\centering

\includegraphics[width=1.0\textwidth]{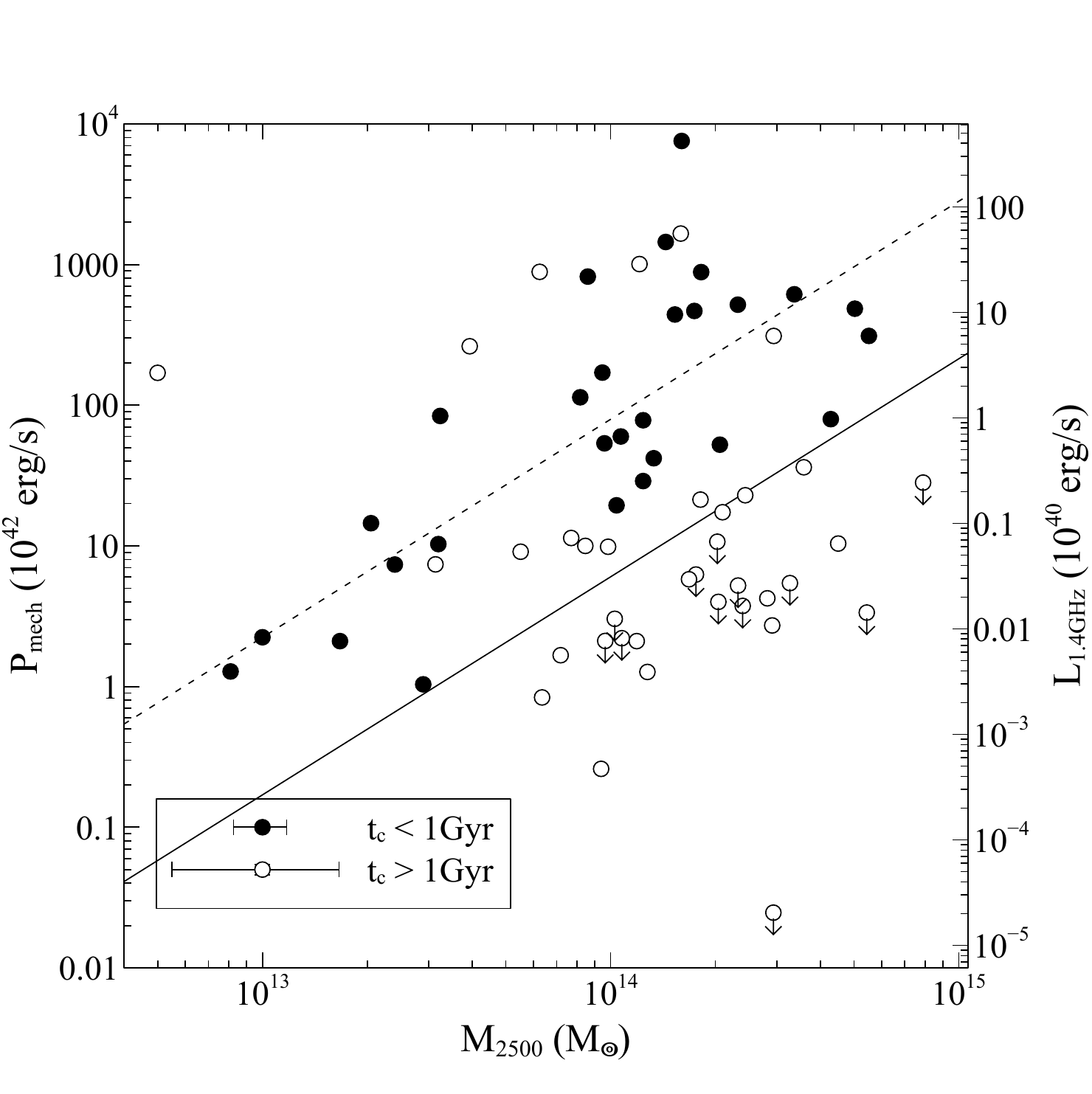} 

\caption{Mechanical Power (derived from 1.4 GHz luminosity) vs. M$_{2500}$ in the full HIFLUGCS sample, with mechanical powers  estimated using equation \ref{eq:jetrad}. Systems are shown as either open or closed symbols based on the central cooling times from \citet{Hudson_2010}. The dotted line is our best-fit P$_{\rm mech}$-M$_{2500}$ relation from section \ref{sec:cavscal}, and the solid line is the 2 $\sigma$ lower limit on the relation. The average size of the error bars is shown at the bottom left. The average error on the 1.4GHz is smaller than the symbol size. }
 
\label{fig:MP_HIFLUGCS}

\end{minipage}
\end{figure}

We plot the mechanical power against $M_{2500}$ for the full HIFLUGCS sample in Figure \ref{fig:MP_HIFLUGCS}. 
The distributions of $P_{\rm mech}$ for systems with short and long central atmospheric cooling times are segregated in Figure \ref{fig:MP_HIFLUGCS}.  Systems with central cooling
times exceeding 1 Gyr  experience AGN outbursts that are typically two orders of magnitude less powerful than systems with central cooling times below 1 Gyr.  Systems with central cooling times less than 1 Gyr are not only more powerful, their AGN powers also correlate with halo mass.
The correlation with halo mass is consistent with our earlier result using direct measurements of cavity power.   In contrast,  the AGN powers of systems with long central cooling times
are not correlated with halo mass.   Most are relatively weak or dormant.  Nevertheless, some are experiencing powerful AGN outbursts that rival the most powerful in the entire sample. 

We have superposed in Figure \ref{fig:MP_HIFLUGCS},  our $P_{\rm cav}$-$M_{2500}$ relation from section \ref{sec:cavscal} and its 2$\sigma$ lower limit to distinguish systems with high mechanical power for their halo mass. We find $11/36~ (\sim 30\%)$ of the clusters with long cooling times lie above this limit, while all 28 of the clusters with short cooling times harbour powerful radio outbursts.   In the latter, the AGN are presumably being fueled by cooling atmospheres, while the fueling mechanism in the clusters with long cooling times is unknown. Unresolved coronae \citep{Sun_2009} or cold gas accreted from interloping galaxies \citep{Hardcastle_2007} are candidates. This result shows heating from AGN is significant in all clusters, but systems with long central cooling times are not necessarily fueled by a feedback cycle.

\section{Conclusions}

We derive X-ray mass, luminosity, and temperature profiles for 45 galaxy clusters to explore relationships between halo mass,  AGN feedback, and central cooling time using
the HIFLUGCS sample.  We find that radio--mechanical feedback power, determined from X-ray cavities, correlates with halo mass, but only in halos with central  atmospheric cooling times shorter than 1 Gyr.  The 1 Gyr timescale corresponds approximately to the cooling time threshold for the onset of star formation \citep{Rafferty_2008} and the entropy threshold for H$\alpha$ emission \citep{Cavagnolo_2008}.    For the HIFLUCGS sample, the relationship between cavity power and mass is of the form $P_{\rm cav} \propto M_{2500}^{1.55\pm0.26}$.  No correlation is found in systems with central cooling times greater than 1 Gyr.   

If cooling is regulated by AGN feedback fuelled by accretion at a fixed fraction of the Eddington rate, the trend of jet power with halo mass is expected to be $P_{\rm cav}\simeq \lcool \sim
M^{1.75}$, where the value of the exponent depends on the radial entropy profile and the mass dependence of the gas mass fraction, both of which are uncertain. 
This scaling is consistent with the observed trend between radio-mechanical power and halo mass.  

The observed trend is also consistent with the assumption that jet power is proportional to black hole mass \citep{Sijacki_2006, Somerville_2008}.   This implies analytic scalings of $P_{\rm cav} \propto M^{4/3}$, for $M_{\rm BH} \propto \sigma^{4}$, or $P_{\rm cav} \propto M^{5/3}$, 
for $M_{\rm BH} \propto \sigma^{5}$.  AGN power is correlated equally well with X-ray gas mass and total mass, but less well with the K-band luminosity of the central galaxy.  
Radio--mechanical powers in clusters with central atmospheric cooling times longer than $\sim 1$ Gyr typically lie two orders of magnitude below those with shorter central cooling times.   We find that $\sim 30\%$ of these clusters are experiencing radio outbursts comparable in power to the clusters with short central cooling times.  This indicates that central galaxies undergo powerful radio outbursts regardless of their halo mass or central cooling time on the scales resolved in this study.  Rare but powerful radio outbursts may well be
the primary source of excess energy injected into the ICM. 

We further investigate the impact of  feedback on cluster scaling relations.  
We find $L-T$, and $M-T$ relations, excluding regions directly affected by AGN,
with the forms $L \propto T^{2.63\pm0.10}$, $M\propto T^{1.87\pm0.12}$.
These forms are generally consistent with the cluster population when measured in our temperature range. 

We examined the star formation history of central galaxies using the ratio of stellar mass (i.e., K-band luminosity) to total halo mass within $R_{2500}$.   The fraction of the mass in the stars of the central galaxy decreases sharply with halo mass, consistent with \citet{David_1995} and \citet{Gonzalez_2013}. 
While the gas mass rises, the stellar mass remains nearly constant with rising total mass.
This trend is seen in all clusters regardless of central cooling time, which implies that star formation in central galaxies is tightly regulated throughout 
the buildup of their halos.  It further implies a long-term balance between AGN heating and atmospheric cooling.

\section*{Acknowledgements}

RAM, BRM and ANV acknowledge generous support from the Natural Sciences and Engineering Council of Canada. RAM and BRM acknowledge generous support from the Canadian Space Agency associated with the Astro-H project. PEJN was supported by NASA contract NAS8-03060. HRR acknowledges support from ERC Advanced Grant Feedback. RAM acknowledges the anonymous referee, whose feedback greatly improved this manuscript.

The plots in this paper were created using \textsc{veusz}\footnote{http://home.gna.org/veusz/}. This research has made use of data obtained from the Chandra Data Archive and the Chandra Source Catalog, and software provided by the Chandra X-ray Center (CXC) in the application packages CIAO, ChIPS, and Sherpa. This publication makes use of data products from the Two Micron All Sky Survey, which is a joint project of the University of Massachusetts and the Infrared Processing and Analysis Center/California Institute of Technology, funded by the National Aeronautics and Space Administration and the National Science Foundation. This research has made use of the NASA/IPAC Extragalactic Database (NED), which is operated by the Jet Propulsion Laboratory, California Institute of Technology, under contract with the National Aeronautics and Space Administration.

\bibliographystyle{mn2e}
\bibliography{writeupBRM5}

\clearpage

\appendix
\section{Mass Profiles}

Here we present our derived mass profiles:

\begin{figure*}
\begin{minipage}{1.0\textwidth}

\includegraphics[width=0.32\textwidth]{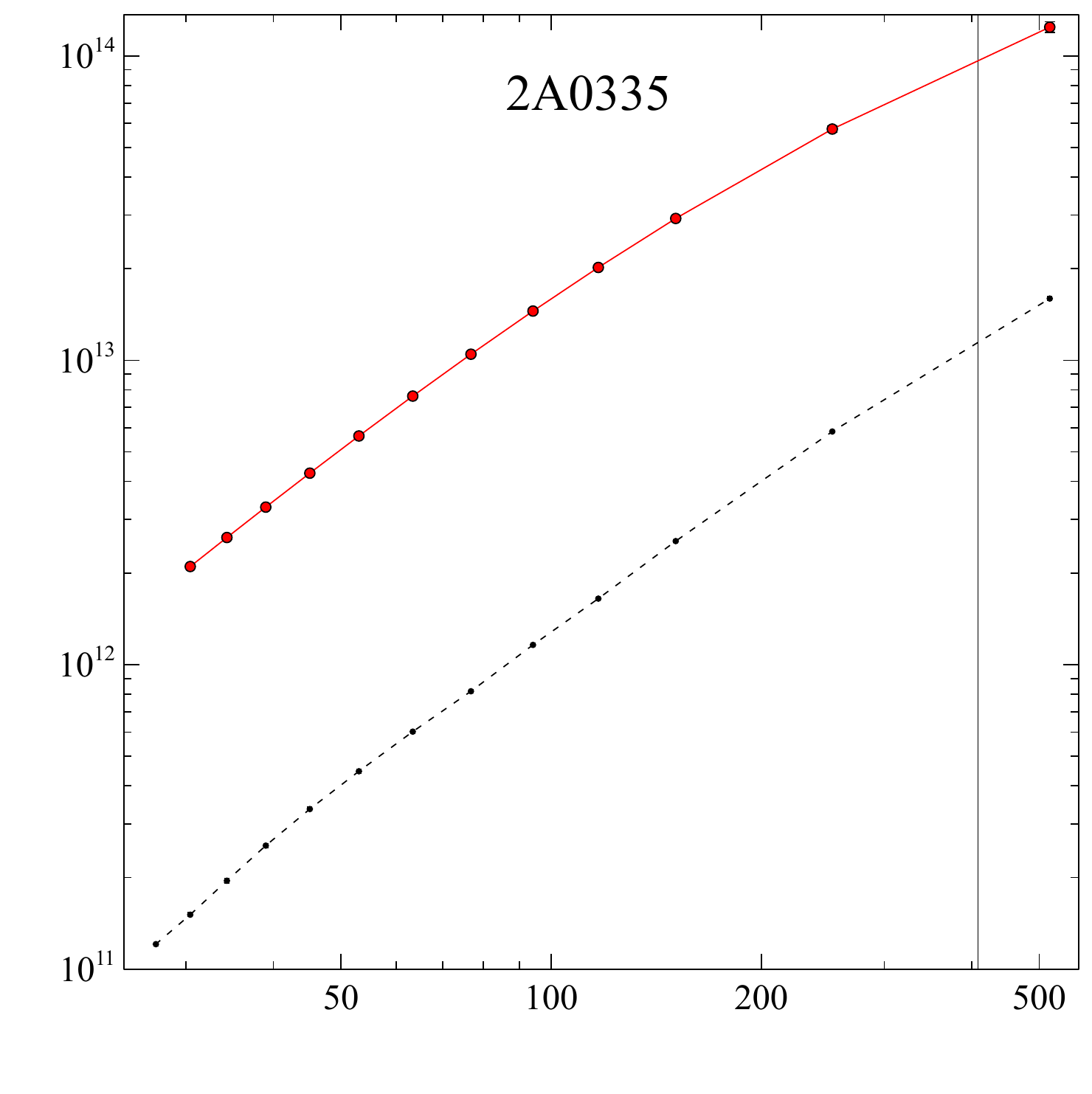} 
\includegraphics[width=0.32\textwidth]{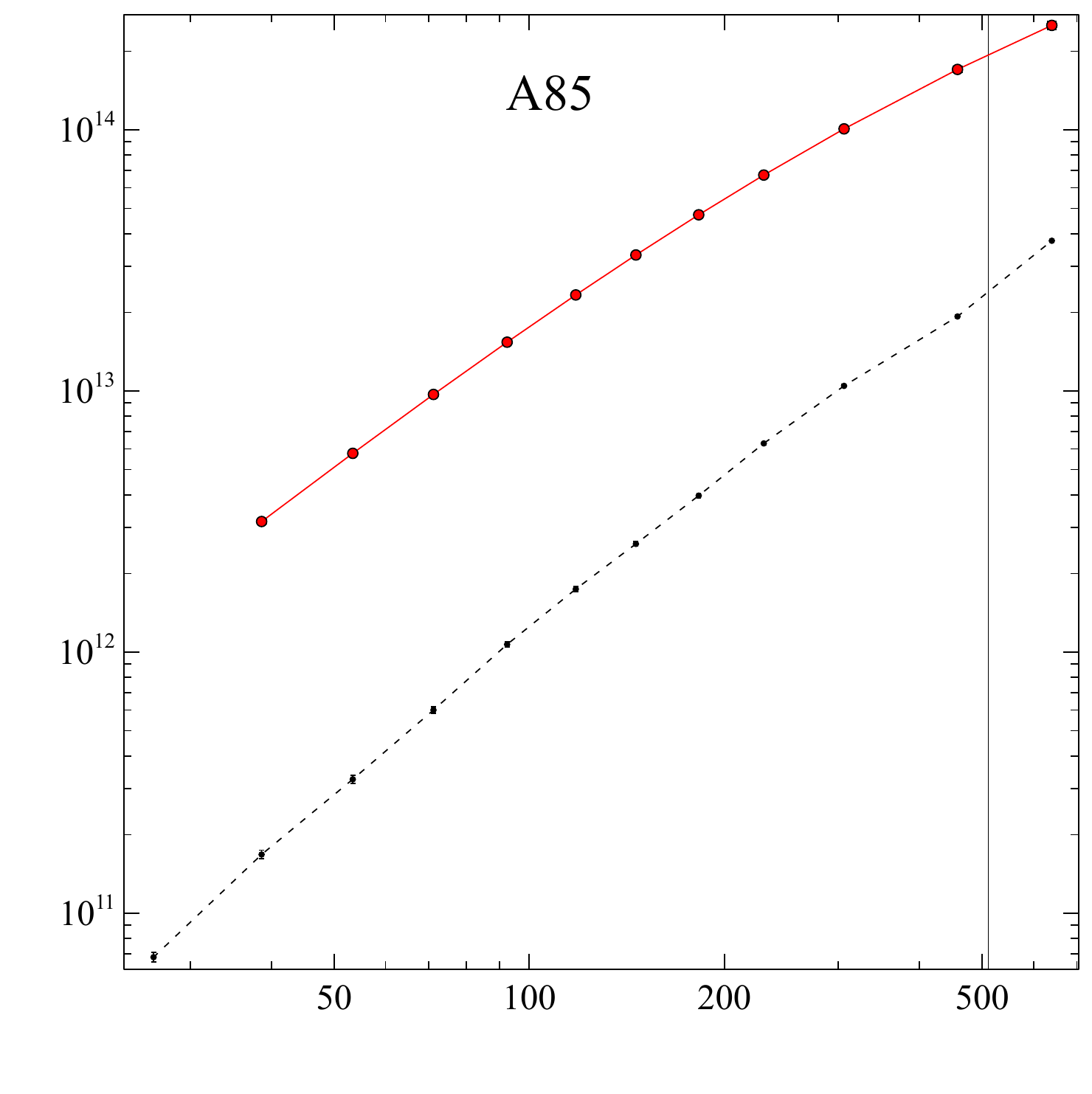}
\includegraphics[width=0.32\textwidth]{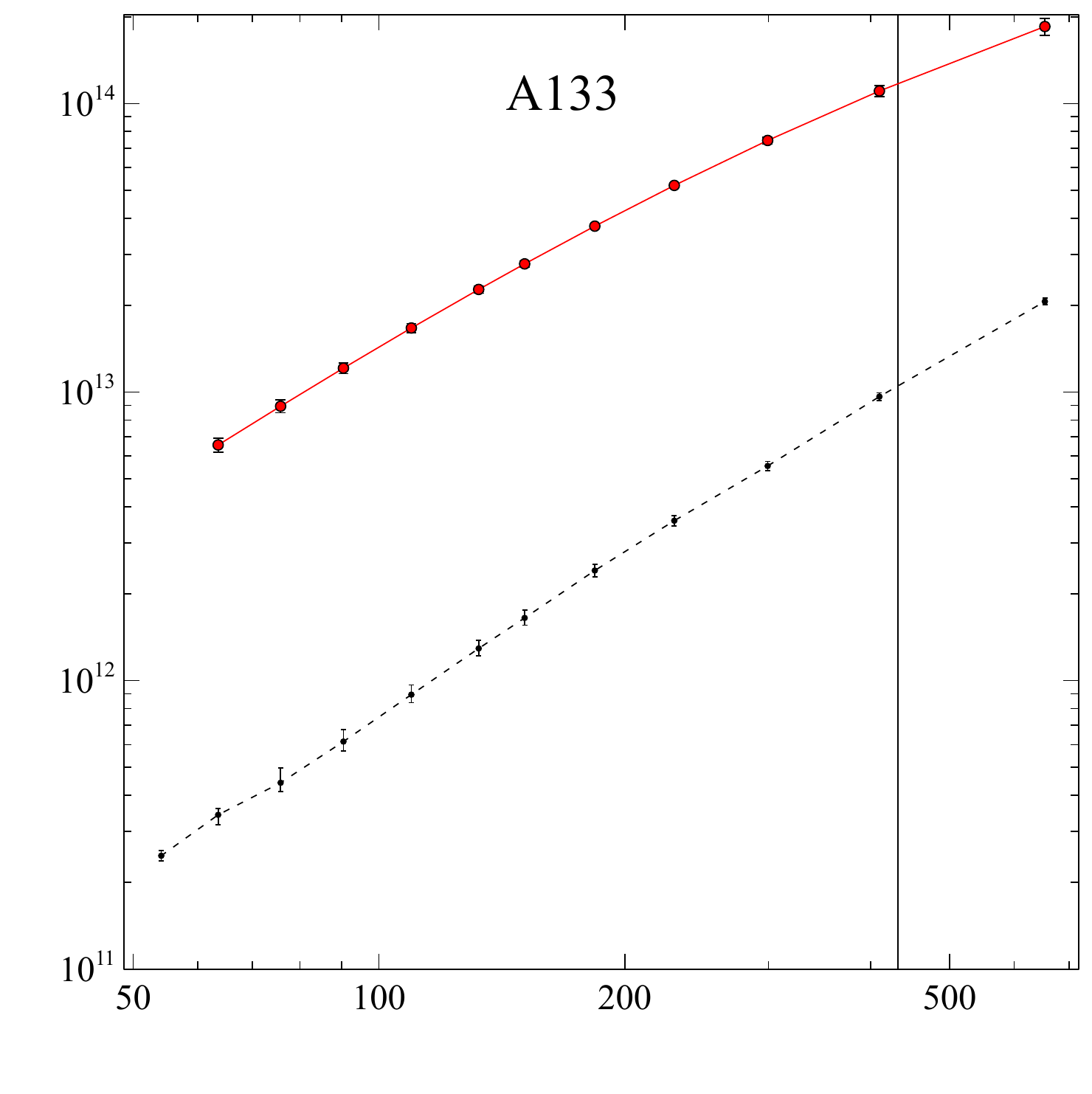} \\
\includegraphics[width=0.32\textwidth]{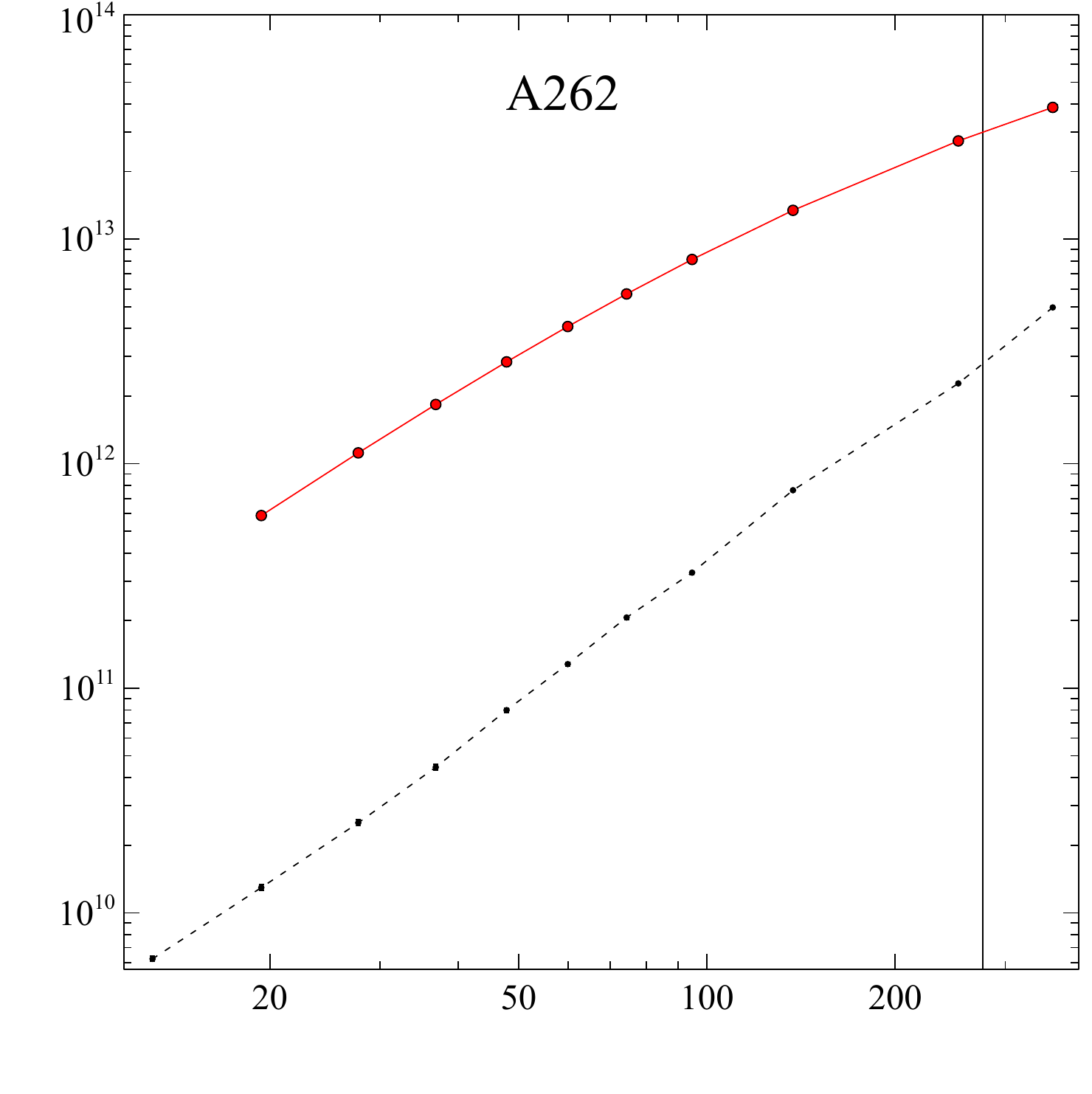}
\includegraphics[width=0.32\textwidth]{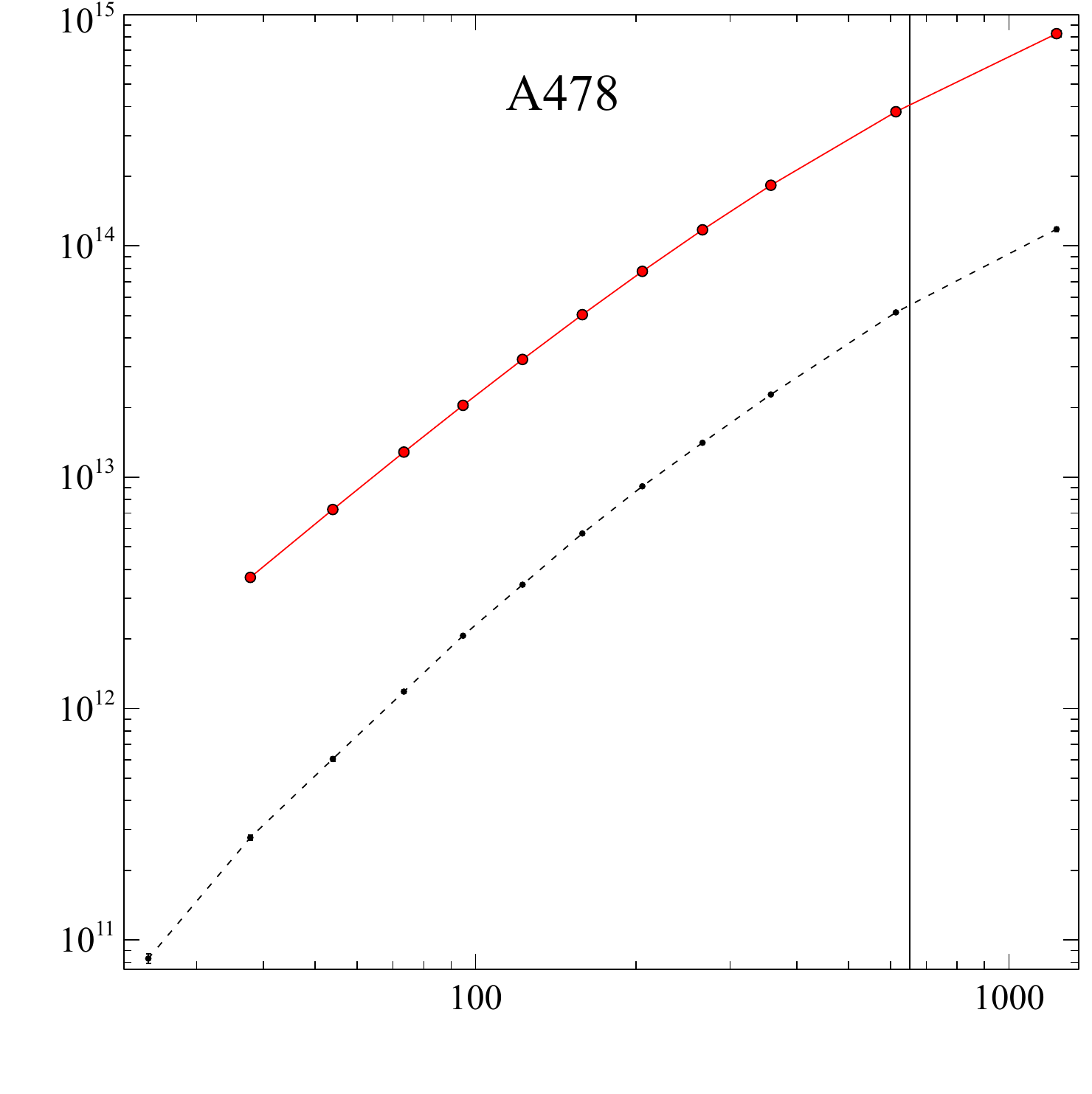}
\includegraphics[width=0.32\textwidth]{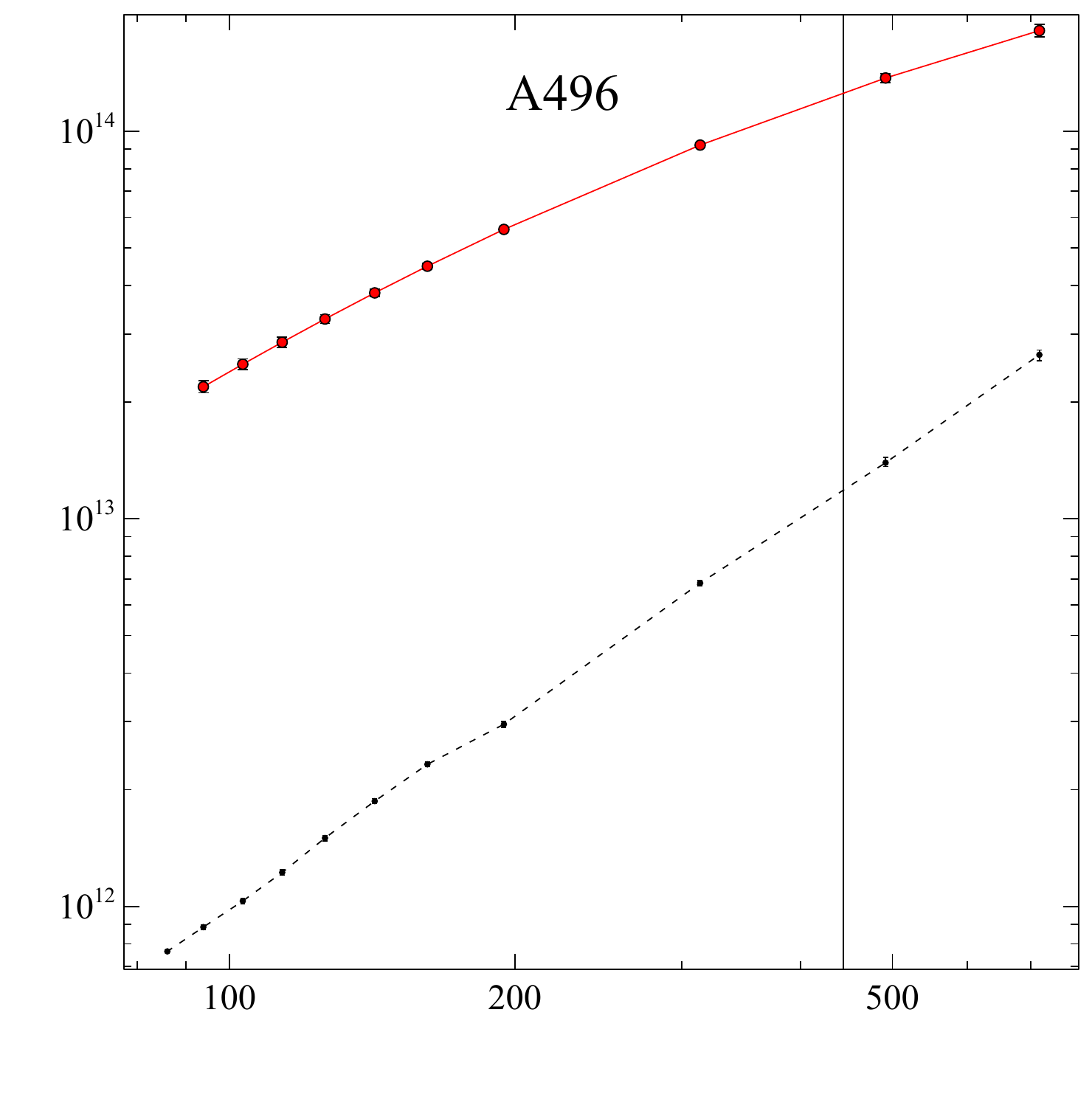} \\
\includegraphics[width=0.32\textwidth]{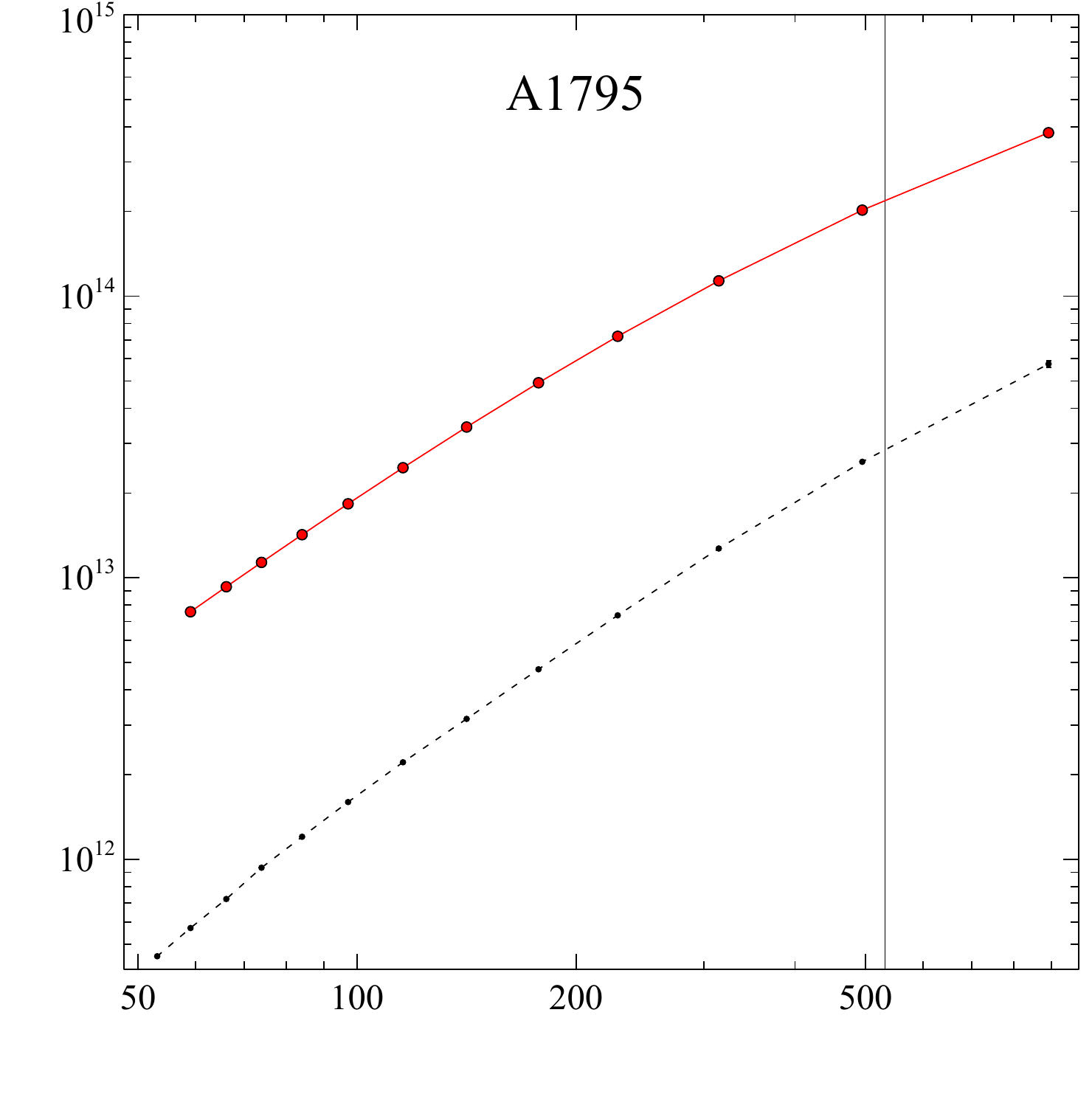} 
\includegraphics[width=0.32\textwidth]{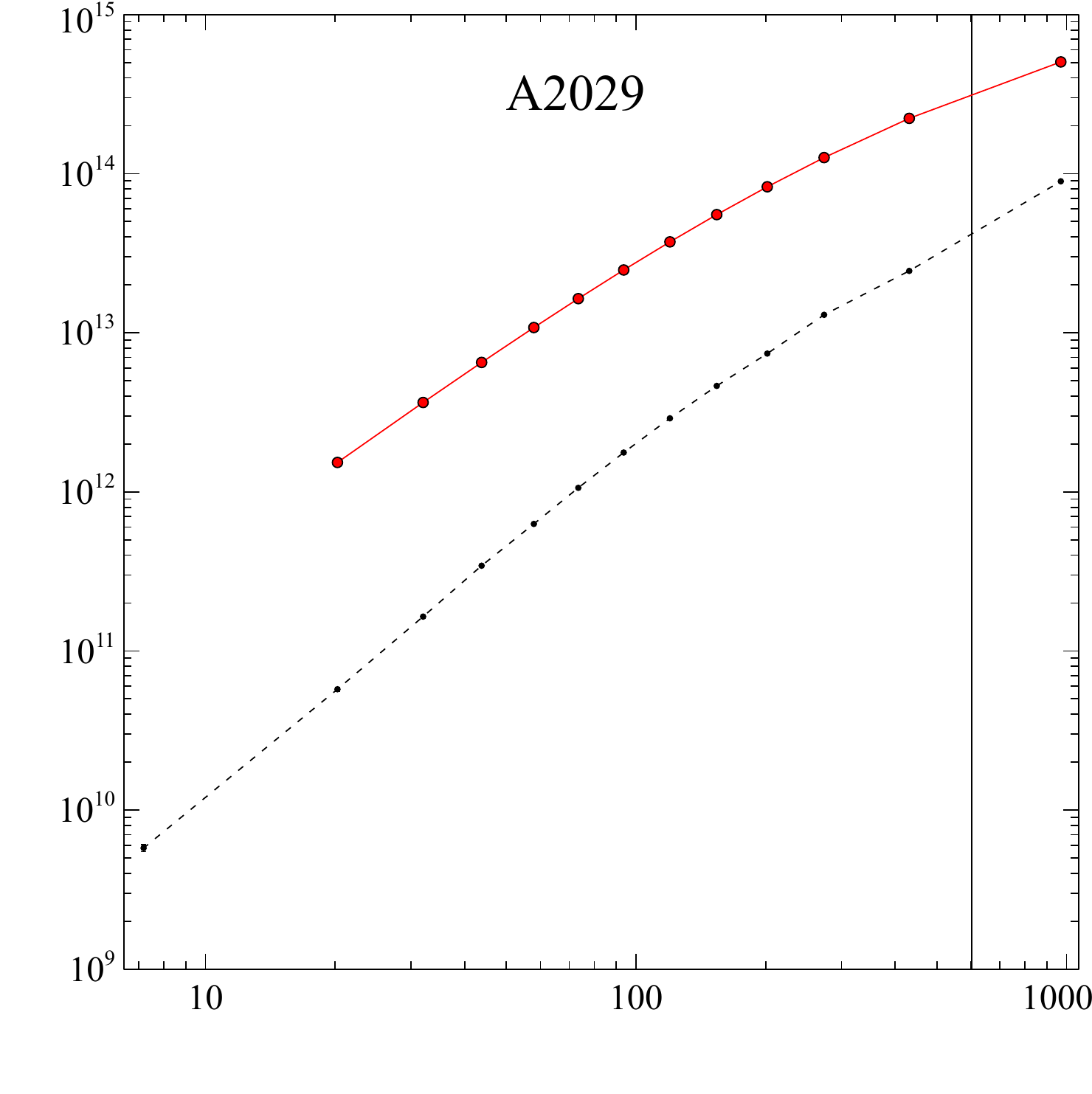}
\includegraphics[width=0.32\textwidth]{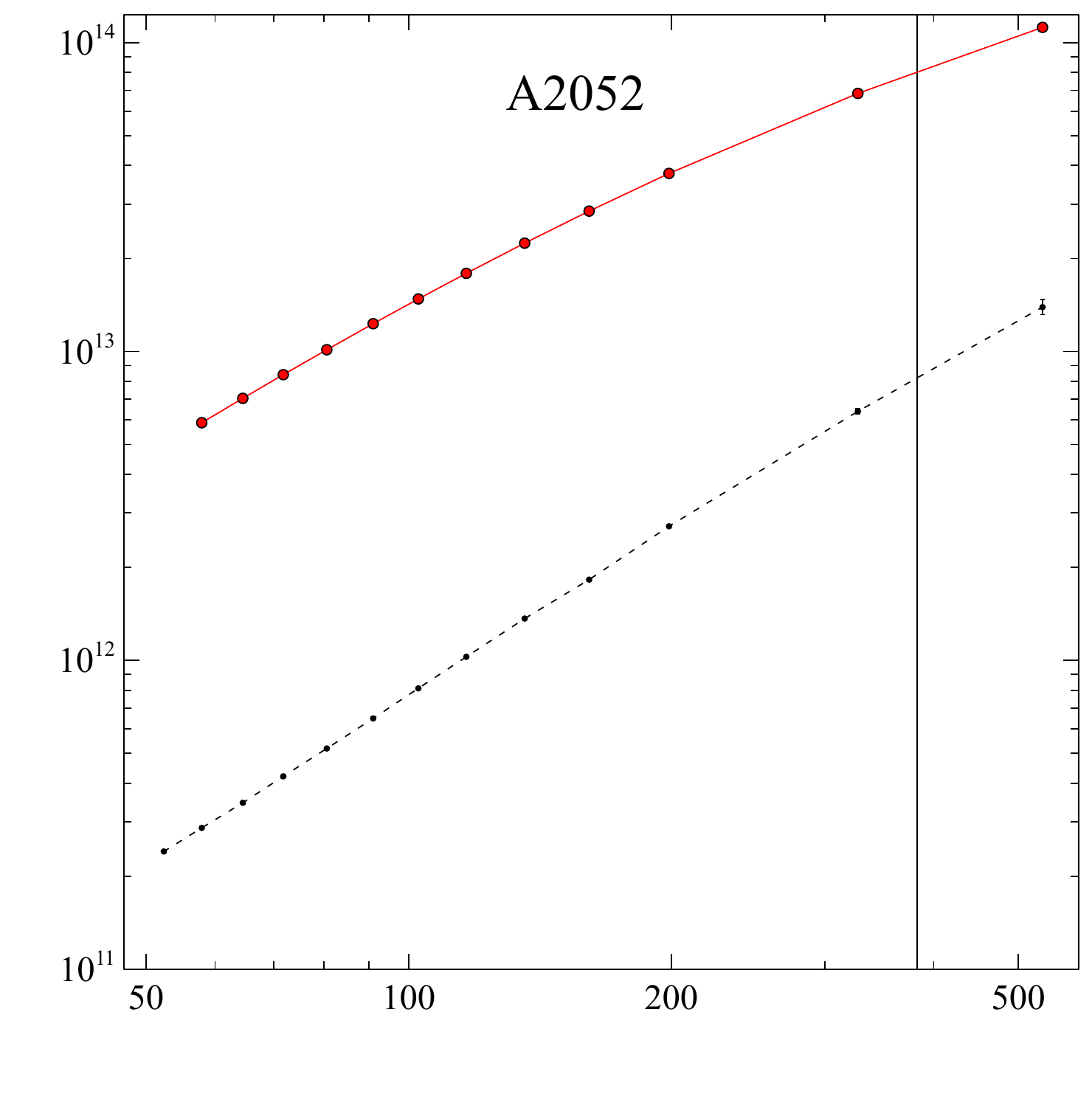}

\caption{Mass profiles of total and gas mass. The y axis gives the enclosed mass in $M_{\odot}$, the x-axis gives the radius from the cluster centre in kpc. The red curve denotes the total mass profile, and the dashed black curve is the gas mass profile. The solid vertical line denotes R$_{2500}$. Note that the error bars in each profile are correlated.}

\label{figure:massprofs1}

\end{minipage}
\end{figure*}

\clearpage

\begin{figure*}
\begin{minipage}{1.0\textwidth}
\centering

\includegraphics[width=0.32\textwidth]{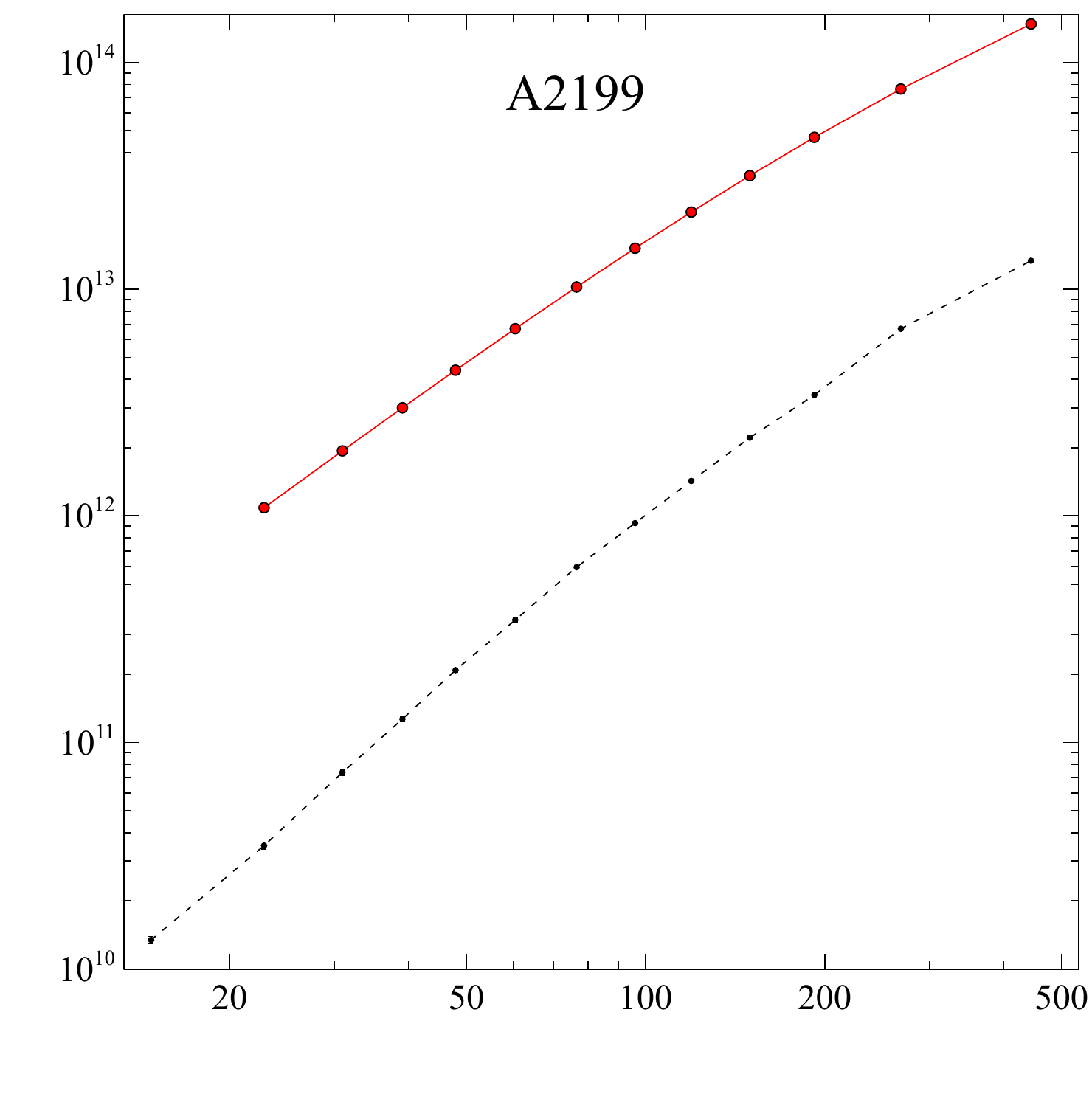}
\includegraphics[width=0.32\textwidth]{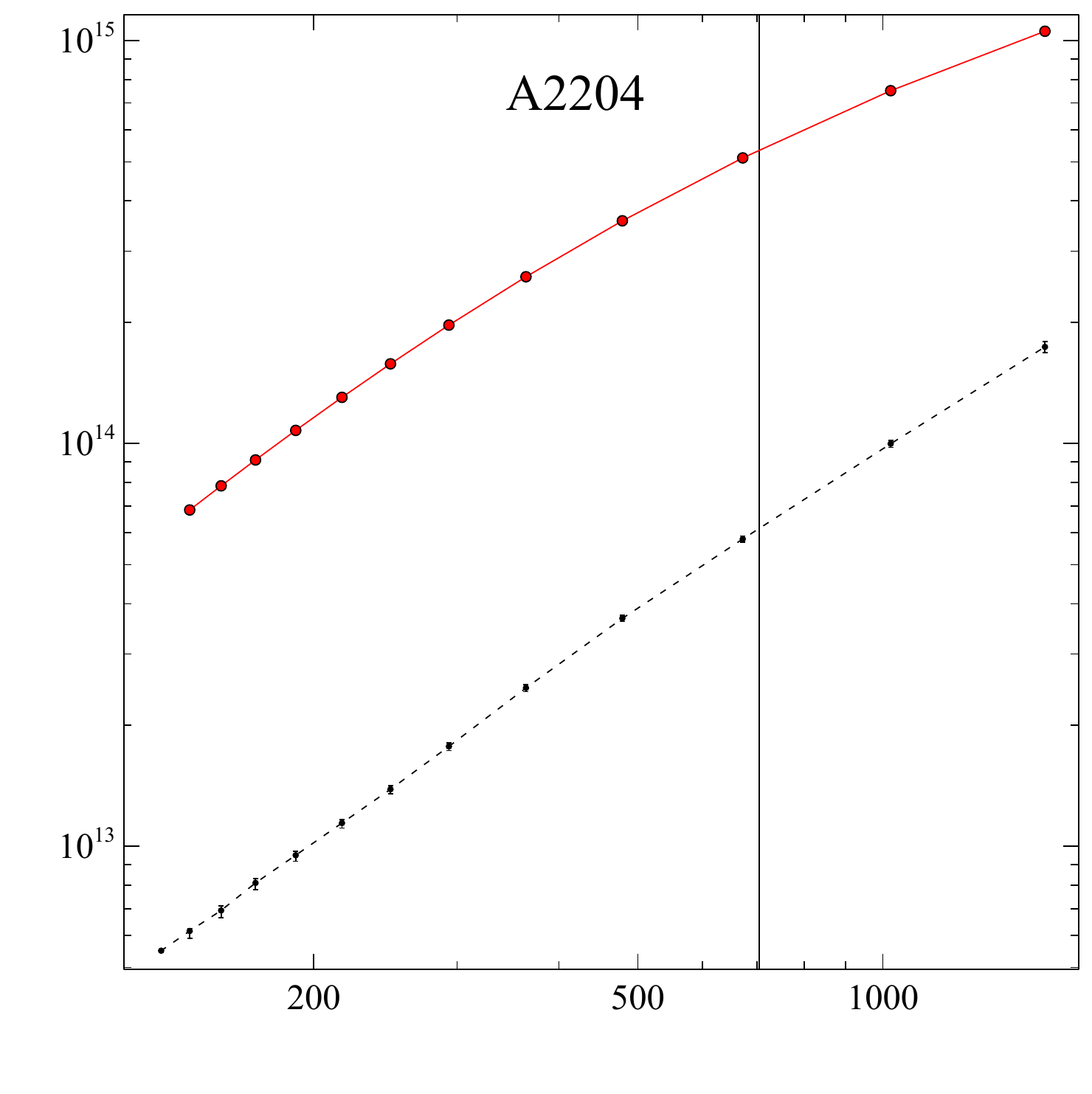}
\includegraphics[width=0.32\textwidth]{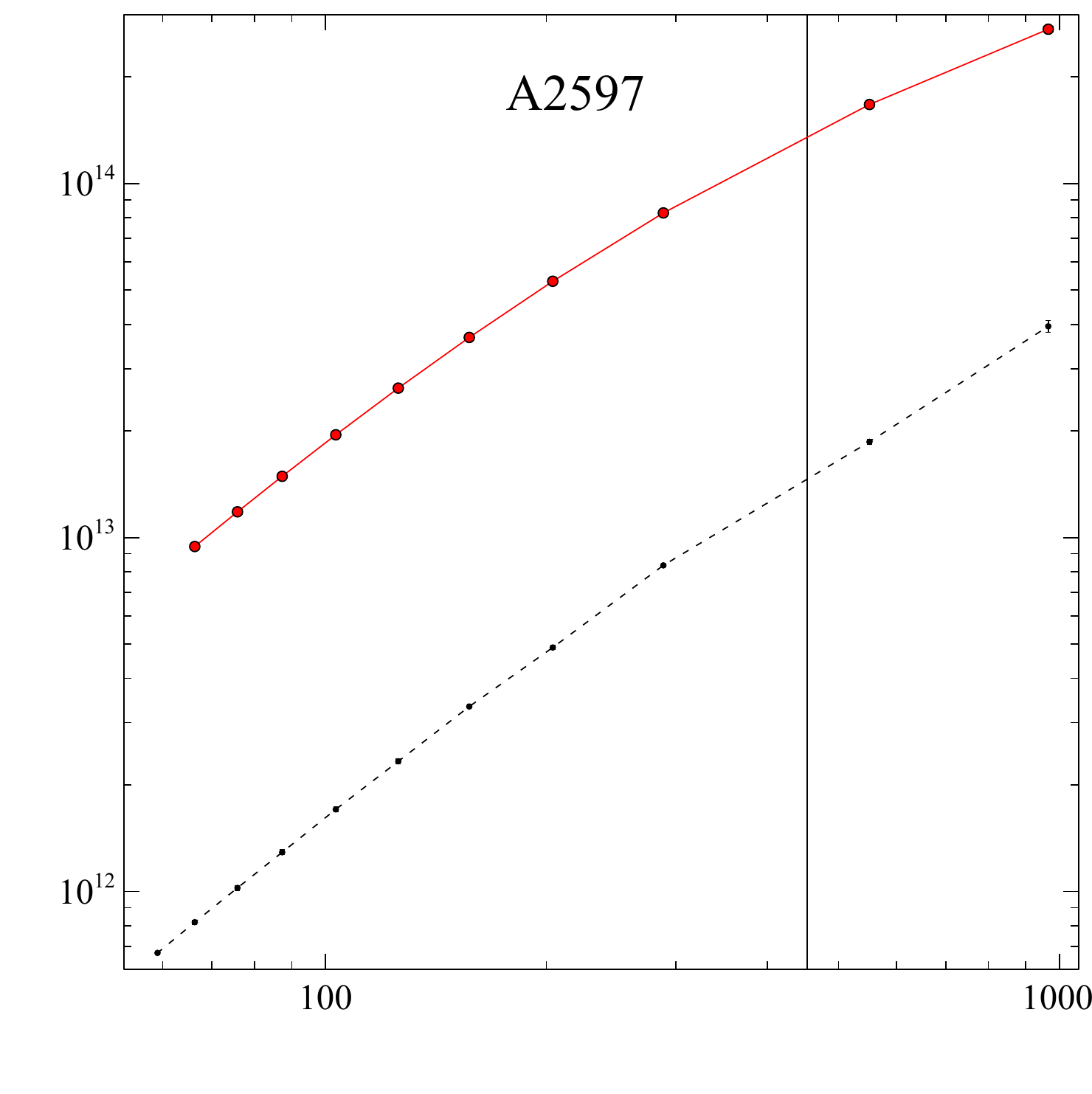} \\
\includegraphics[width=0.32\textwidth]{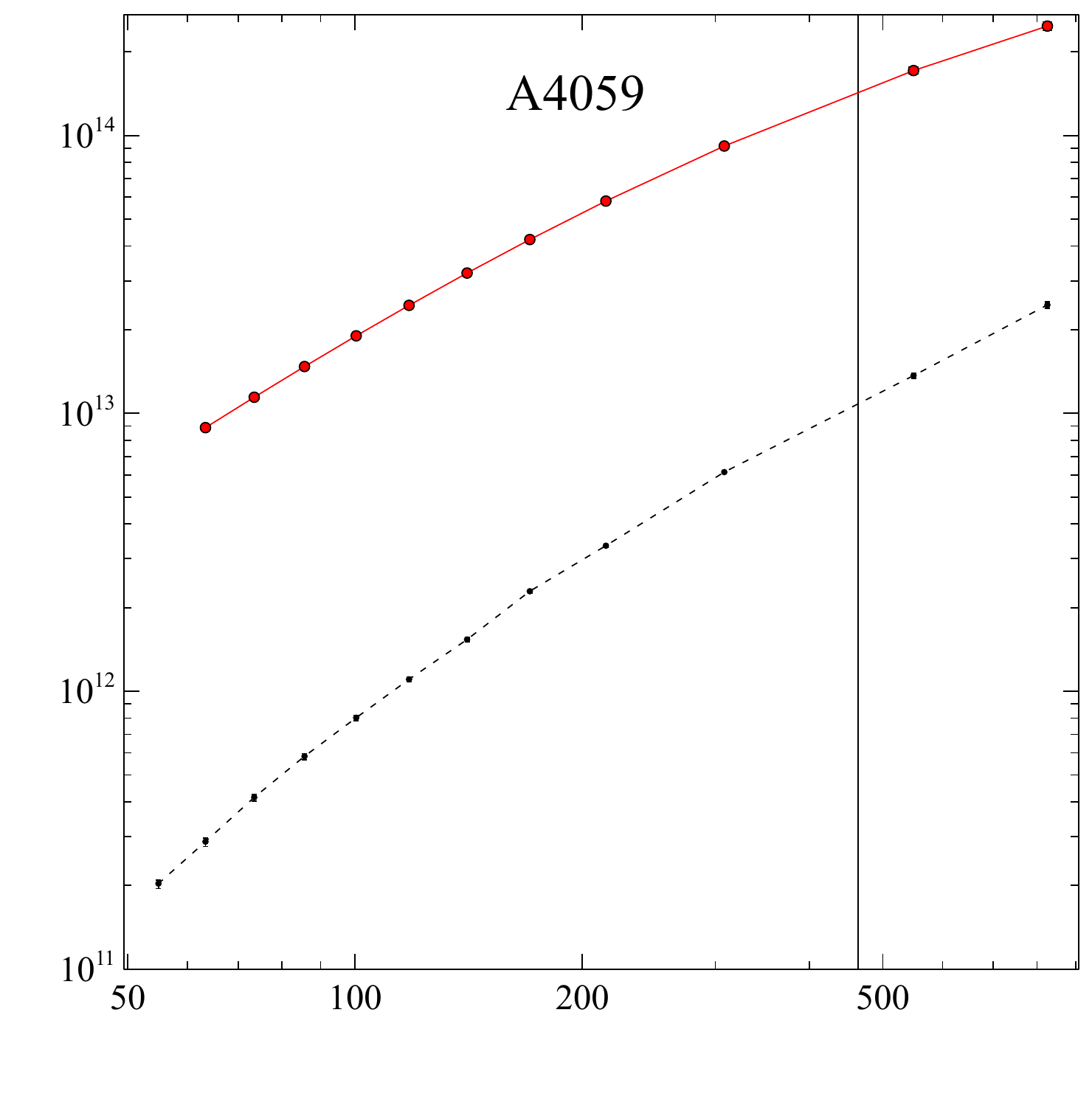} 
\includegraphics[width=0.32\textwidth]{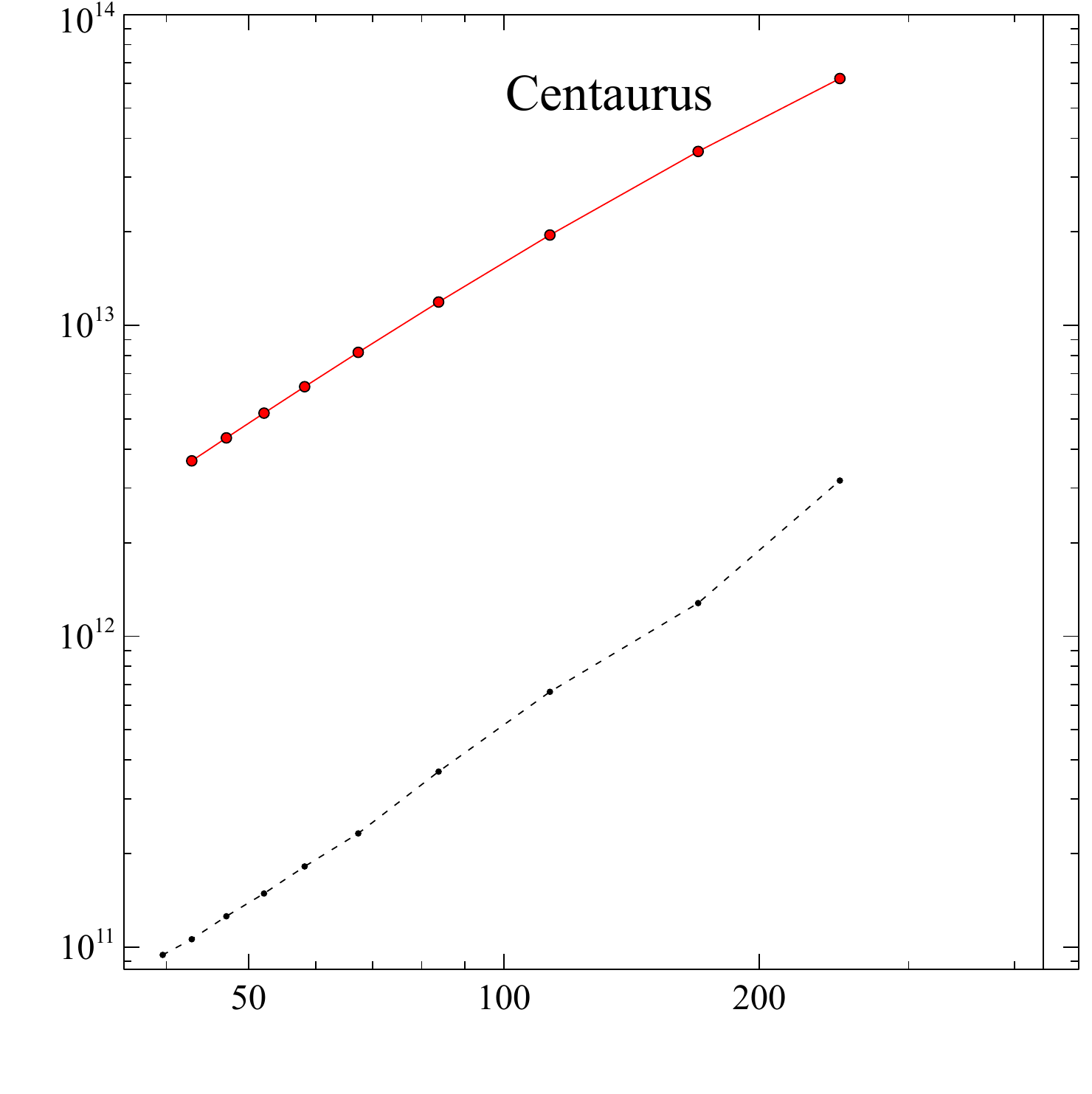}
\includegraphics[width=0.32\textwidth]{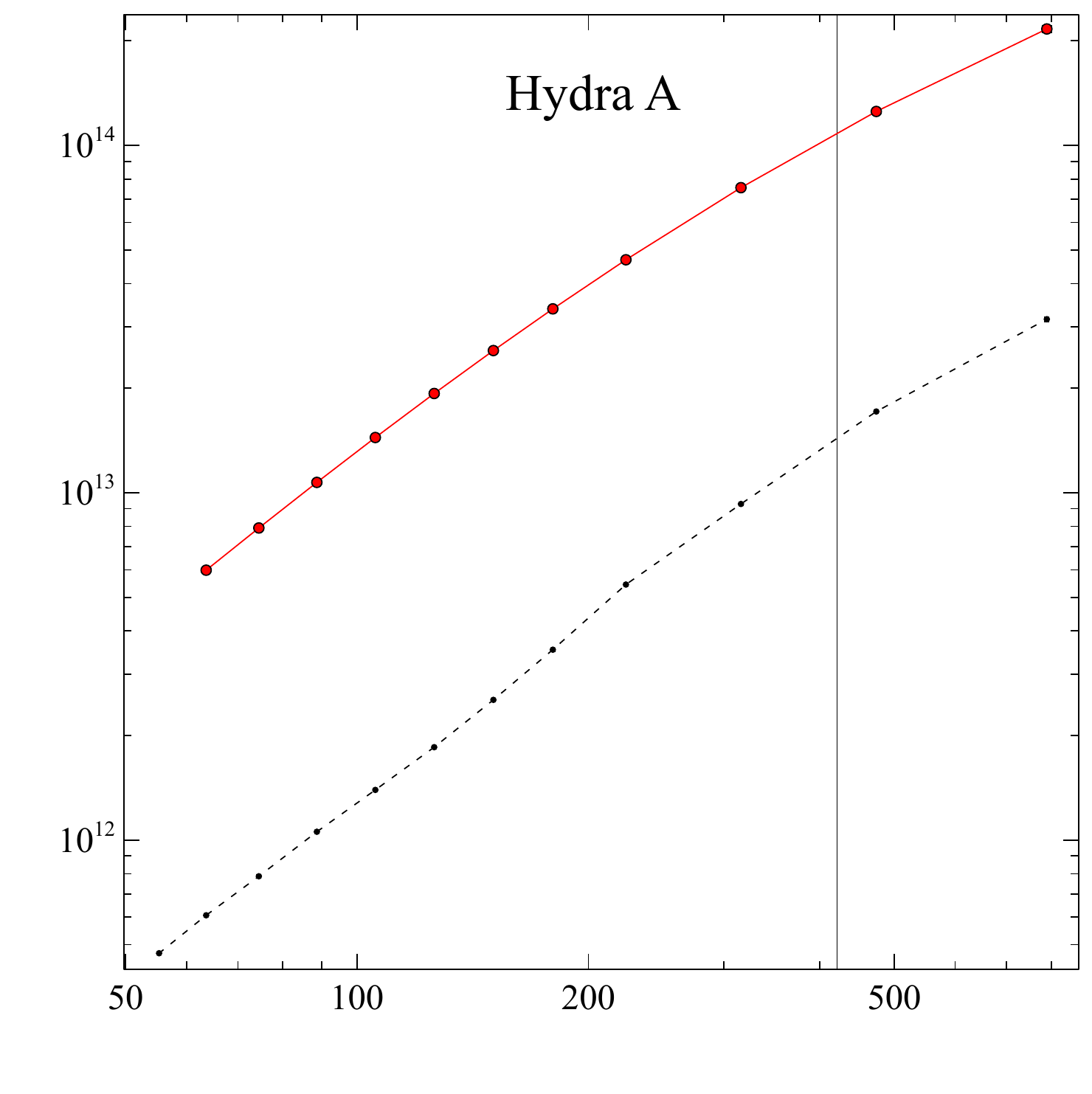} \\
\includegraphics[width=0.32\textwidth]{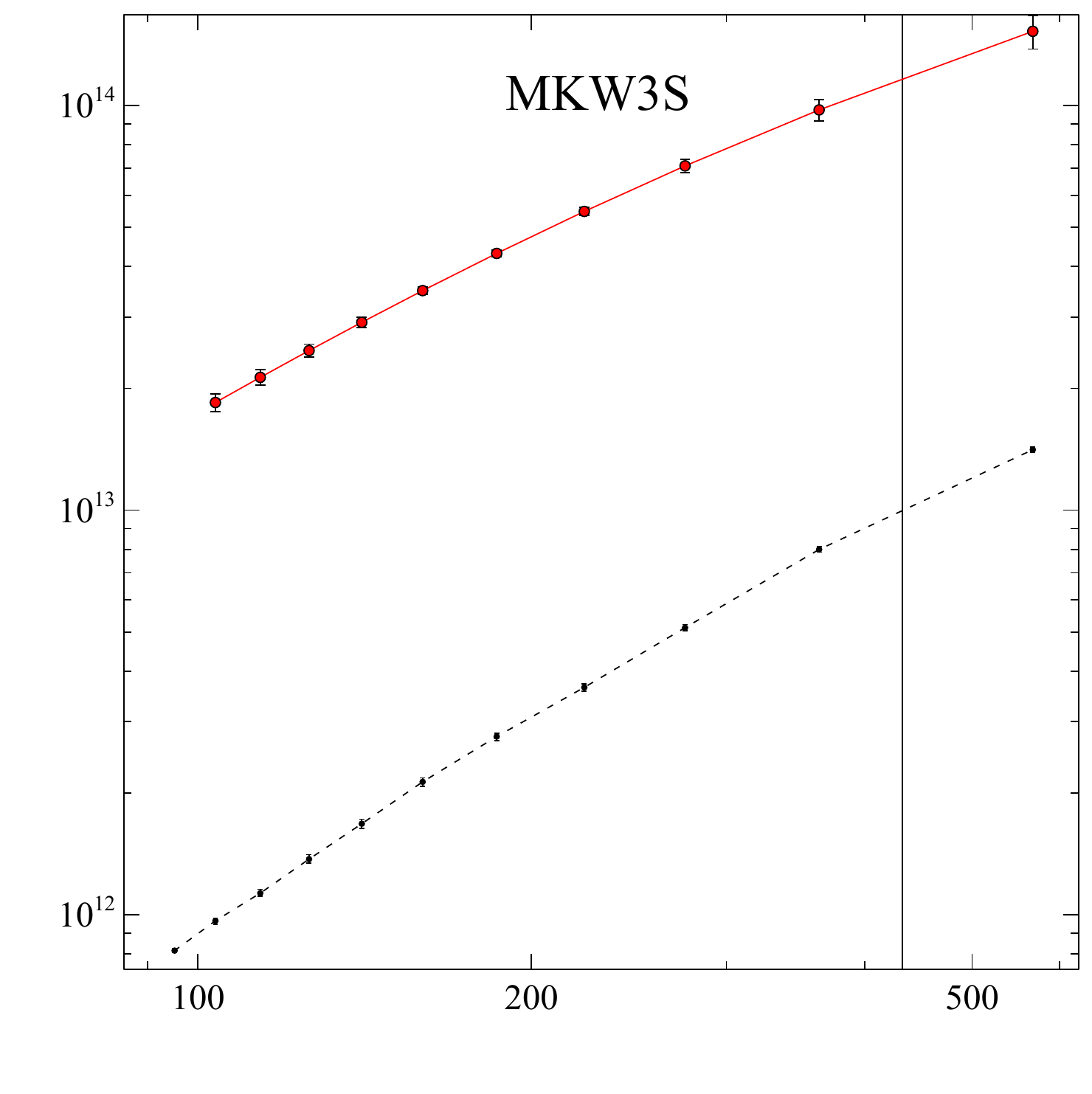}
\includegraphics[width=0.32\textwidth]{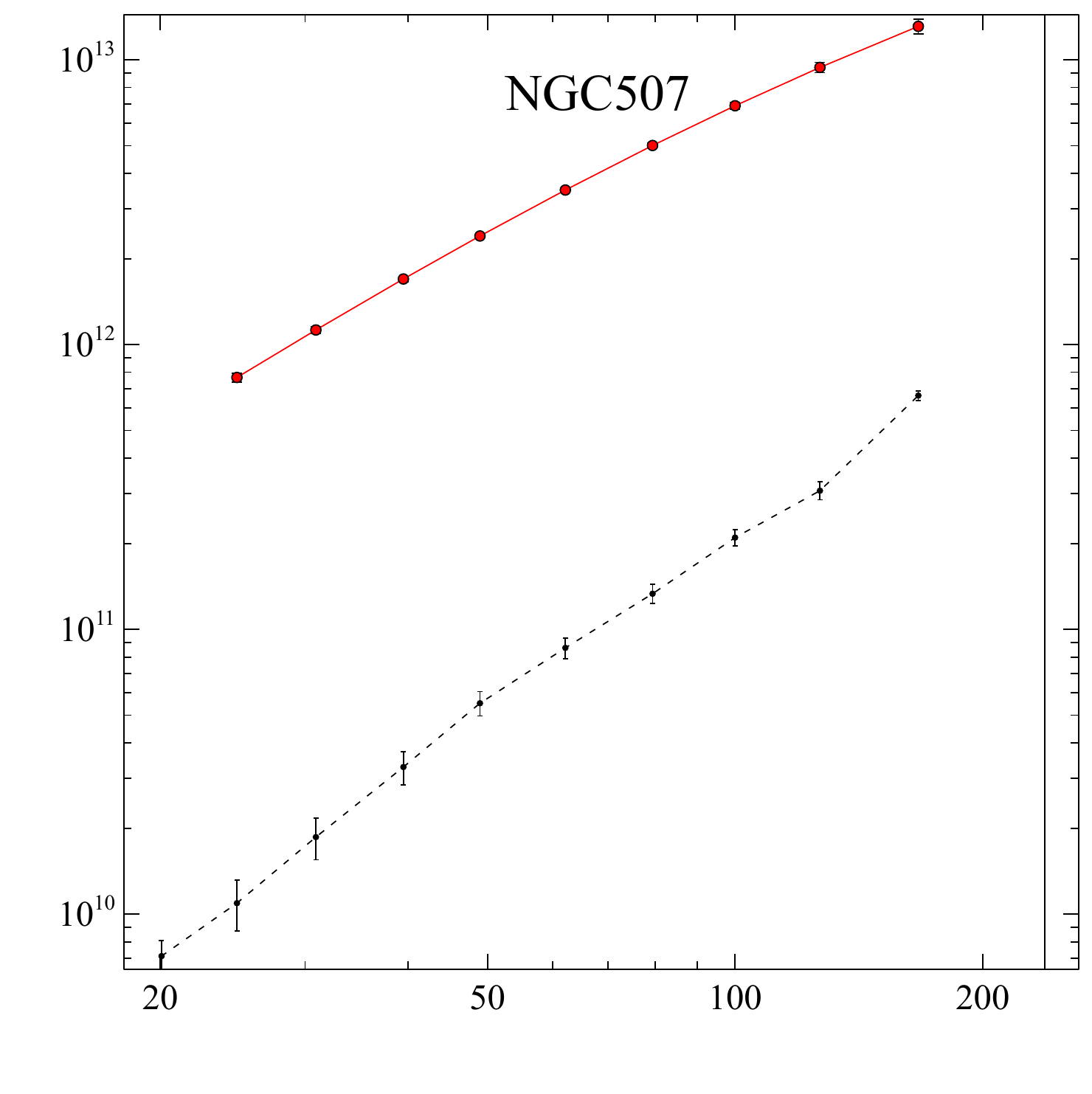} 
\includegraphics[width=0.32\textwidth]{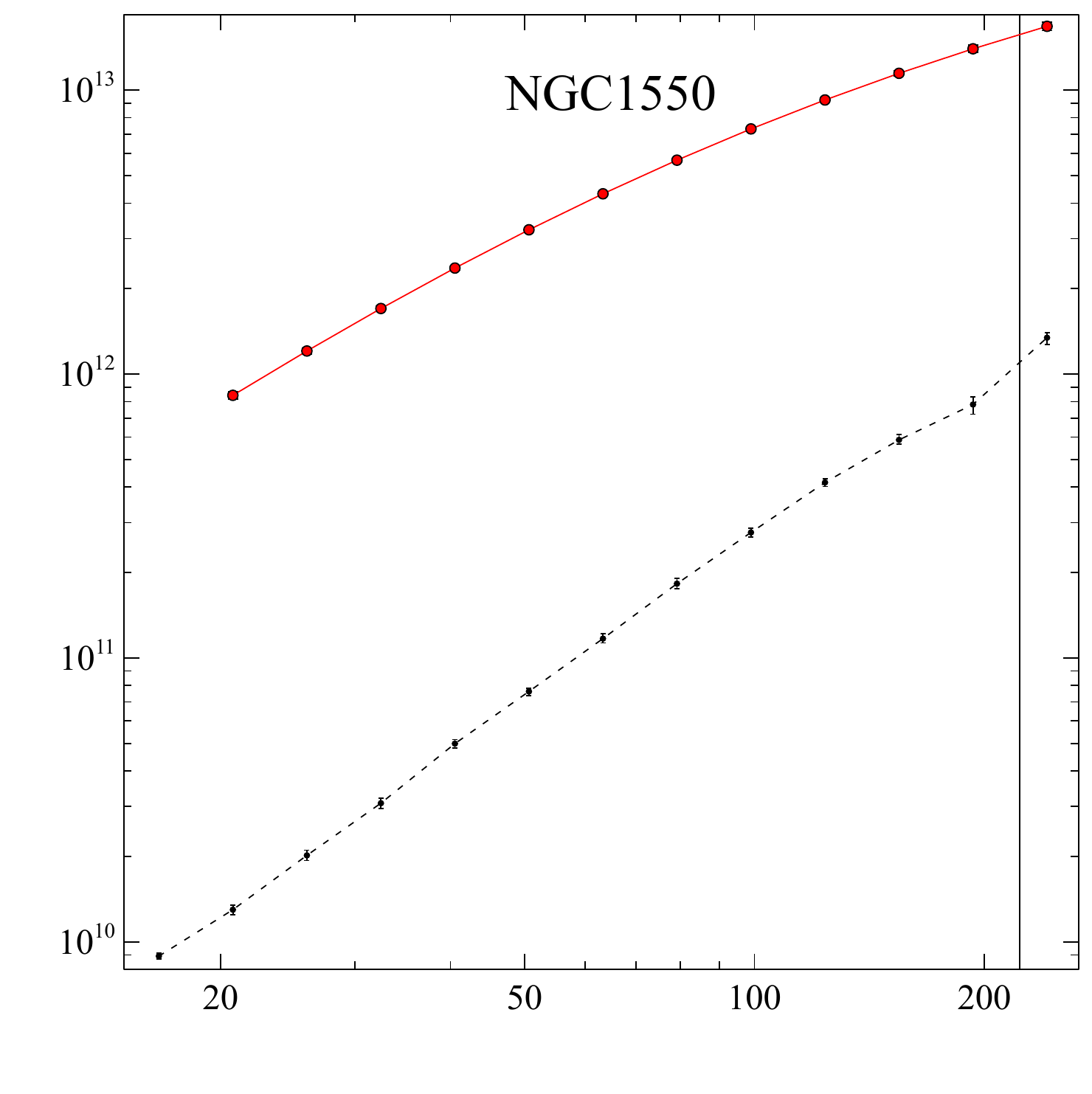} \\ 
\includegraphics[width=0.32\textwidth]{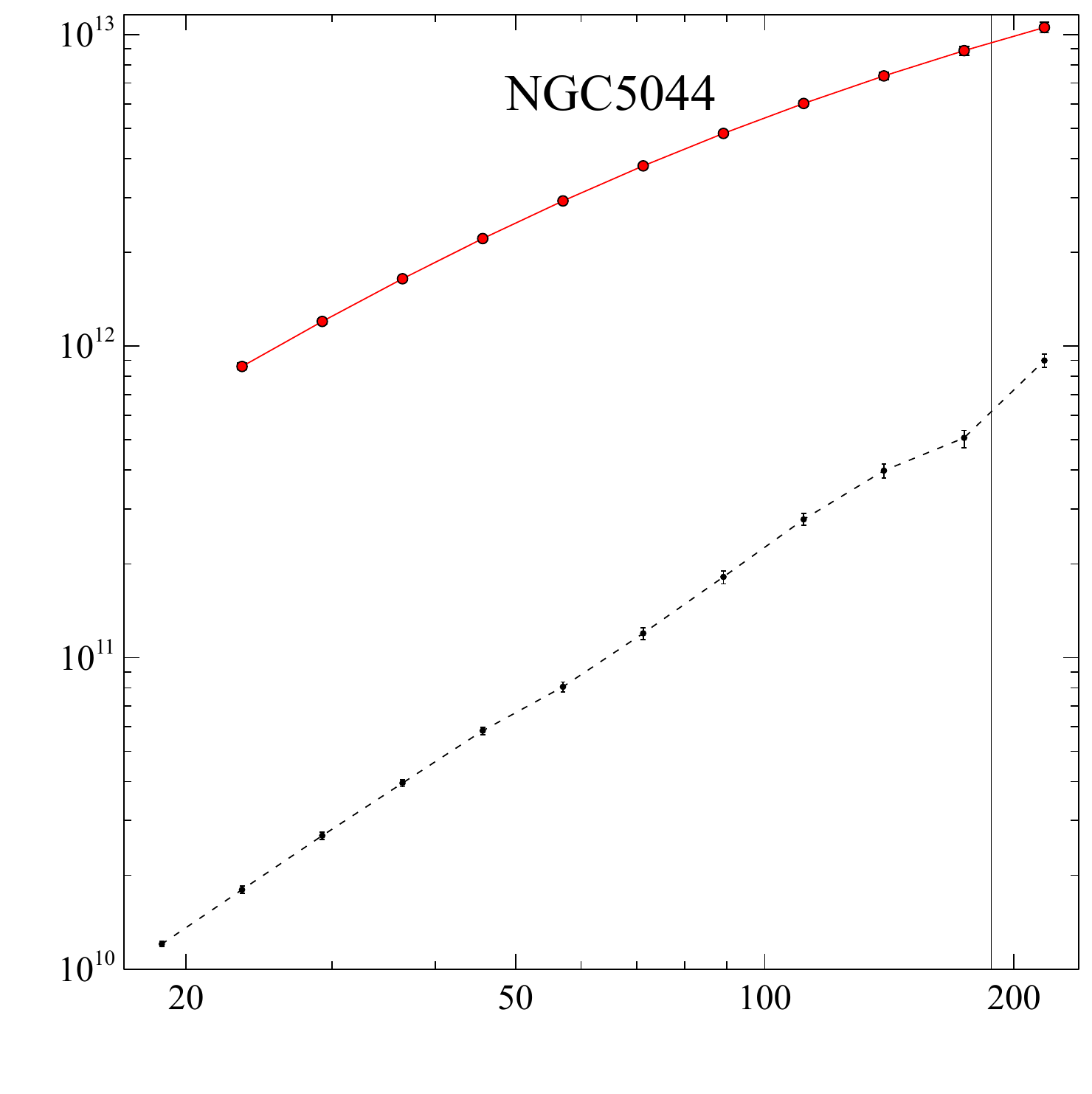}
\includegraphics[width=0.32\textwidth]{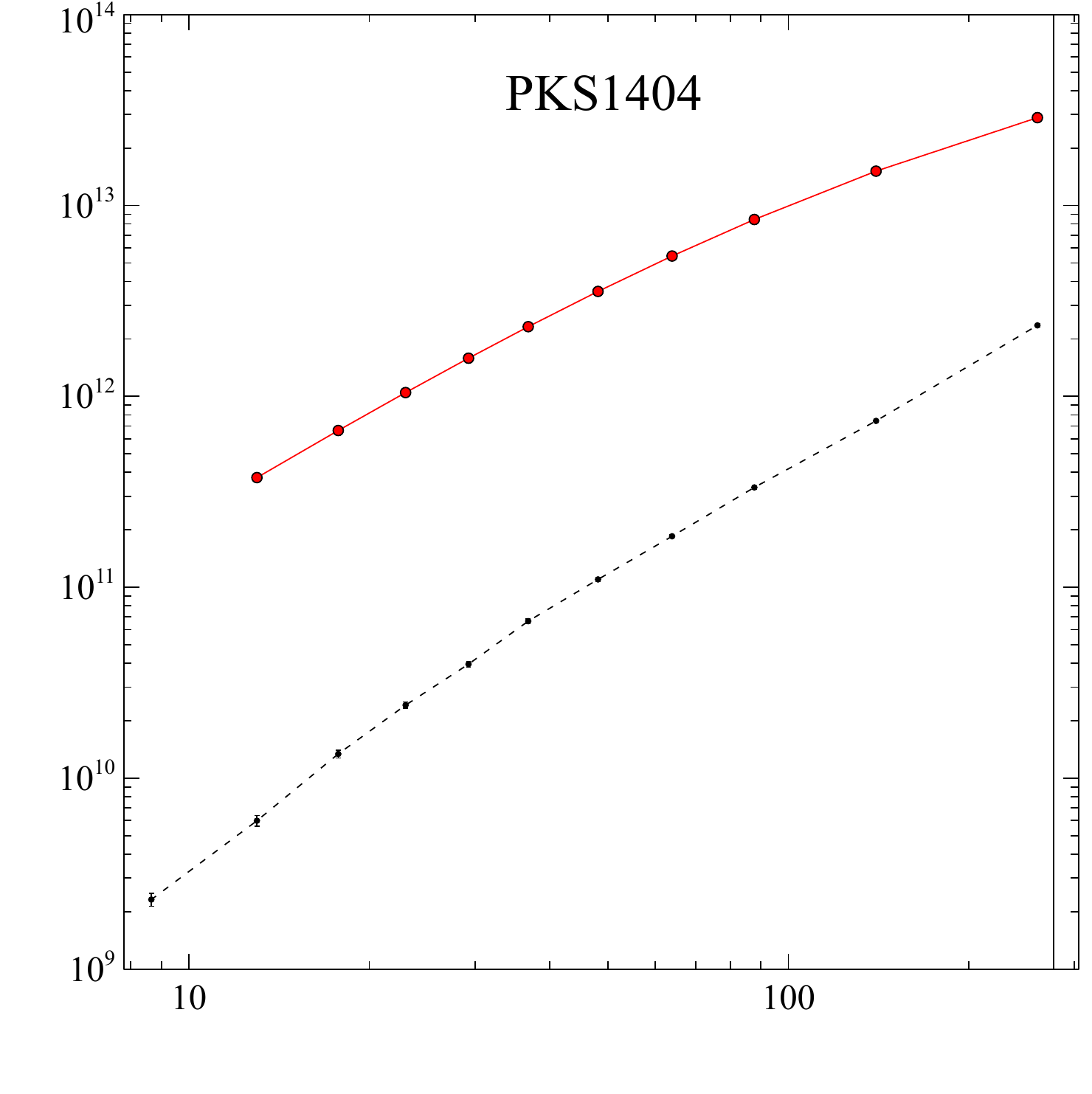}
\includegraphics[width=0.32\textwidth]{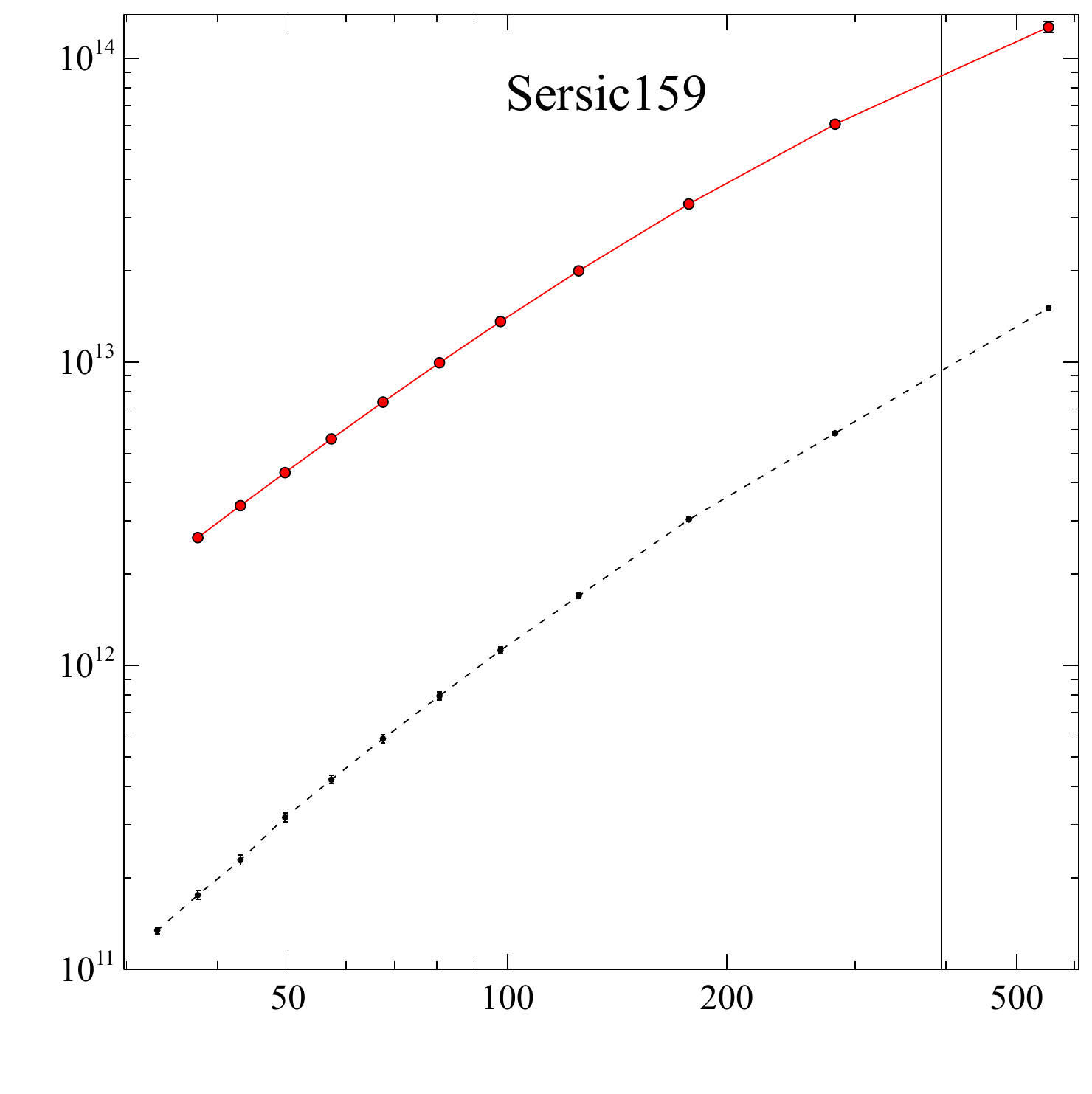}

\label{figure:massprofs2}

\caption{Continued from figure \ref{figure:massprofs1}}

\end{minipage}
\end{figure*}

\begin{figure*}
\begin{minipage}{1.0\textwidth}
\centering
 
\includegraphics[width=0.32\textwidth]{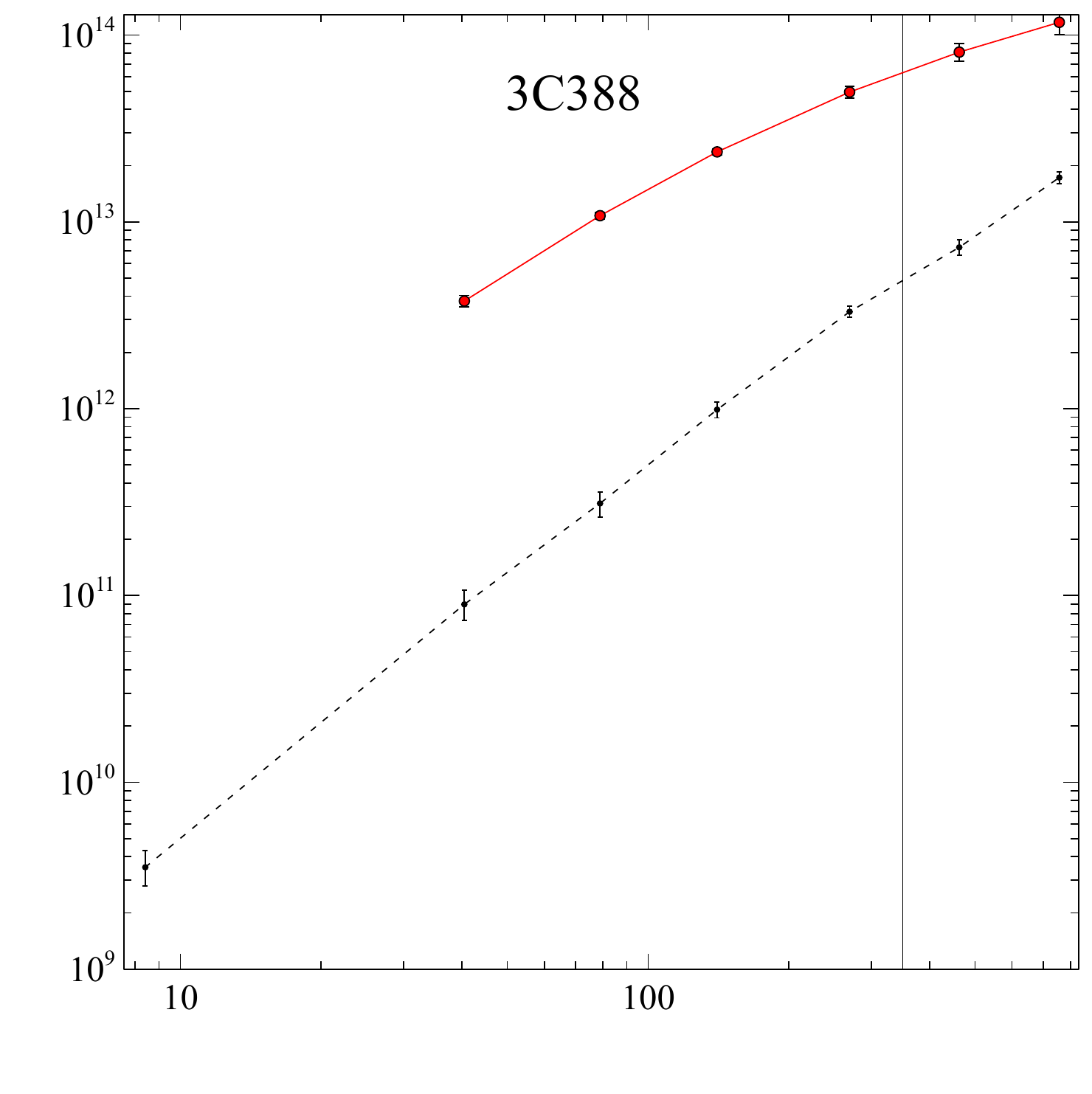}
\includegraphics[width=0.32\textwidth]{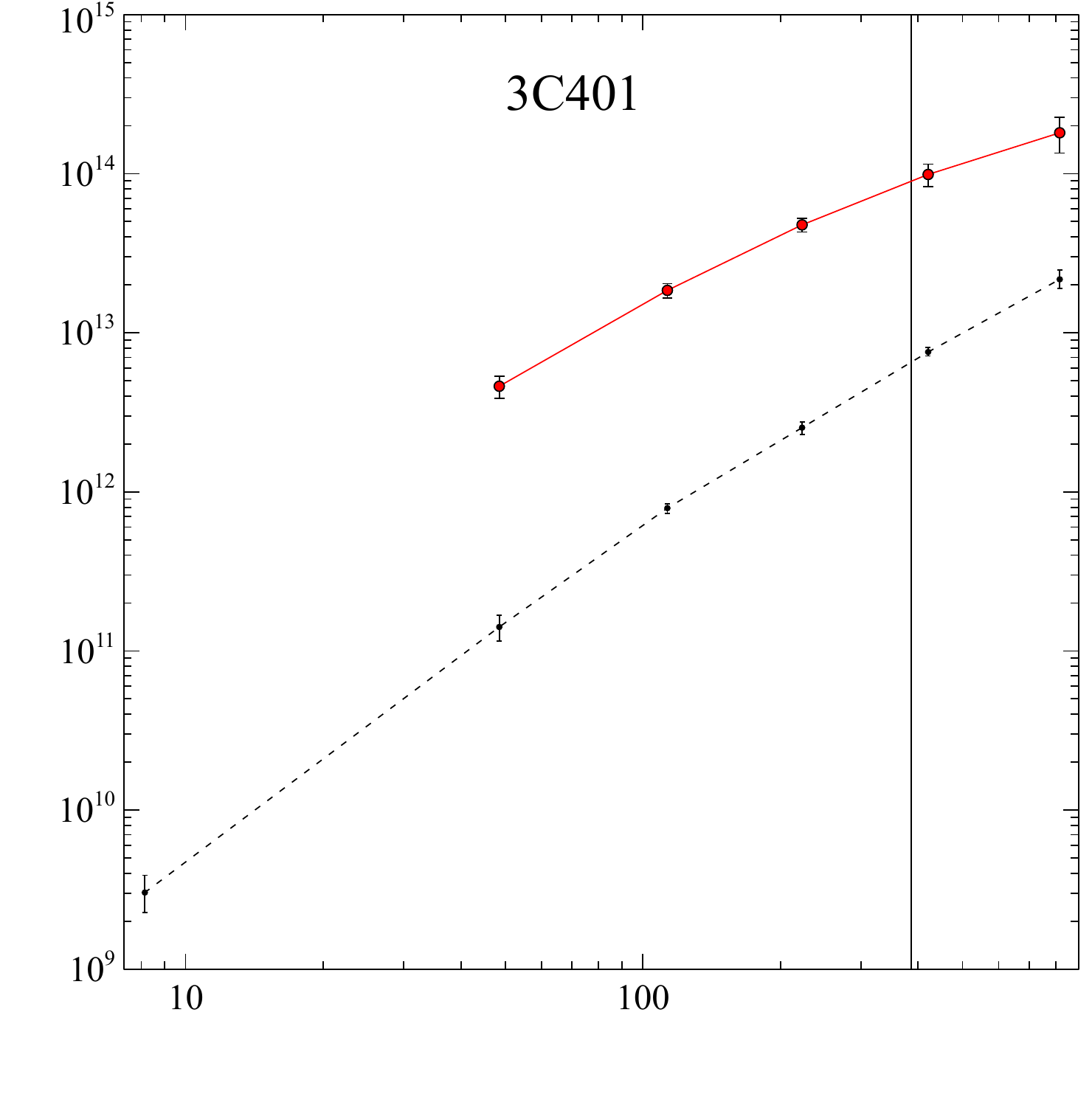}
\includegraphics[width=0.32\textwidth]{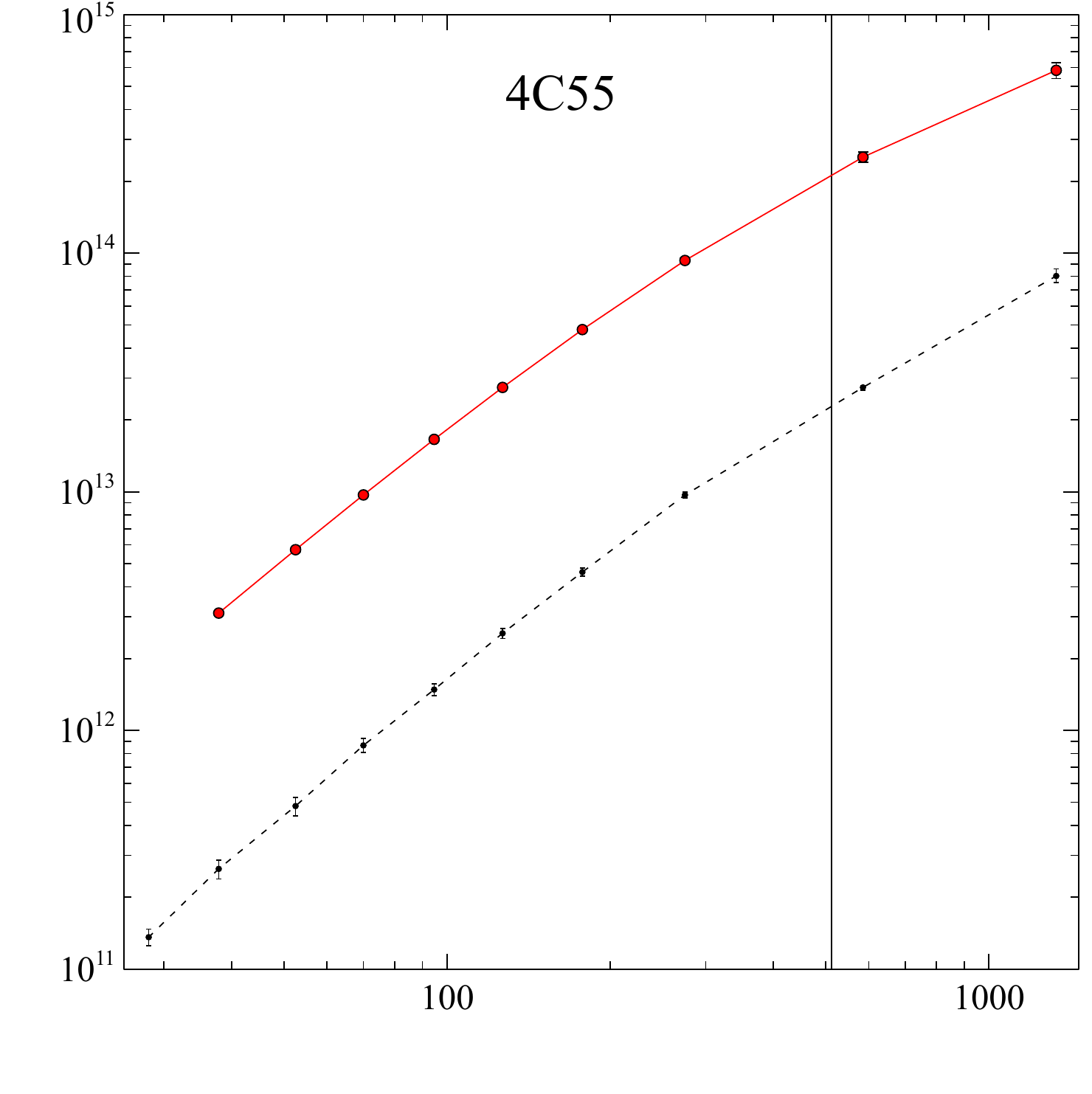} \\
\includegraphics[width=0.32\textwidth]{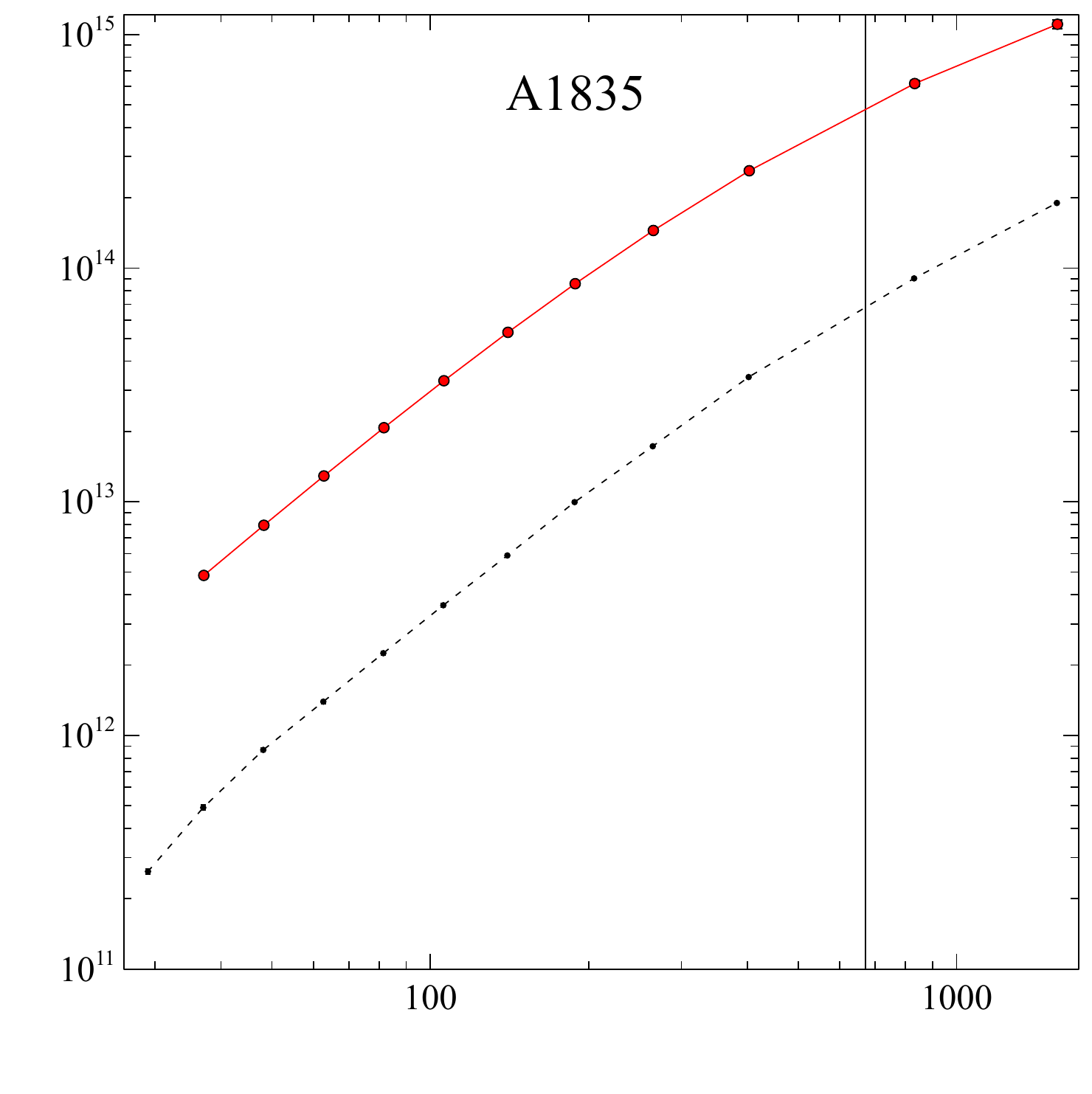}
\includegraphics[width=0.32\textwidth]{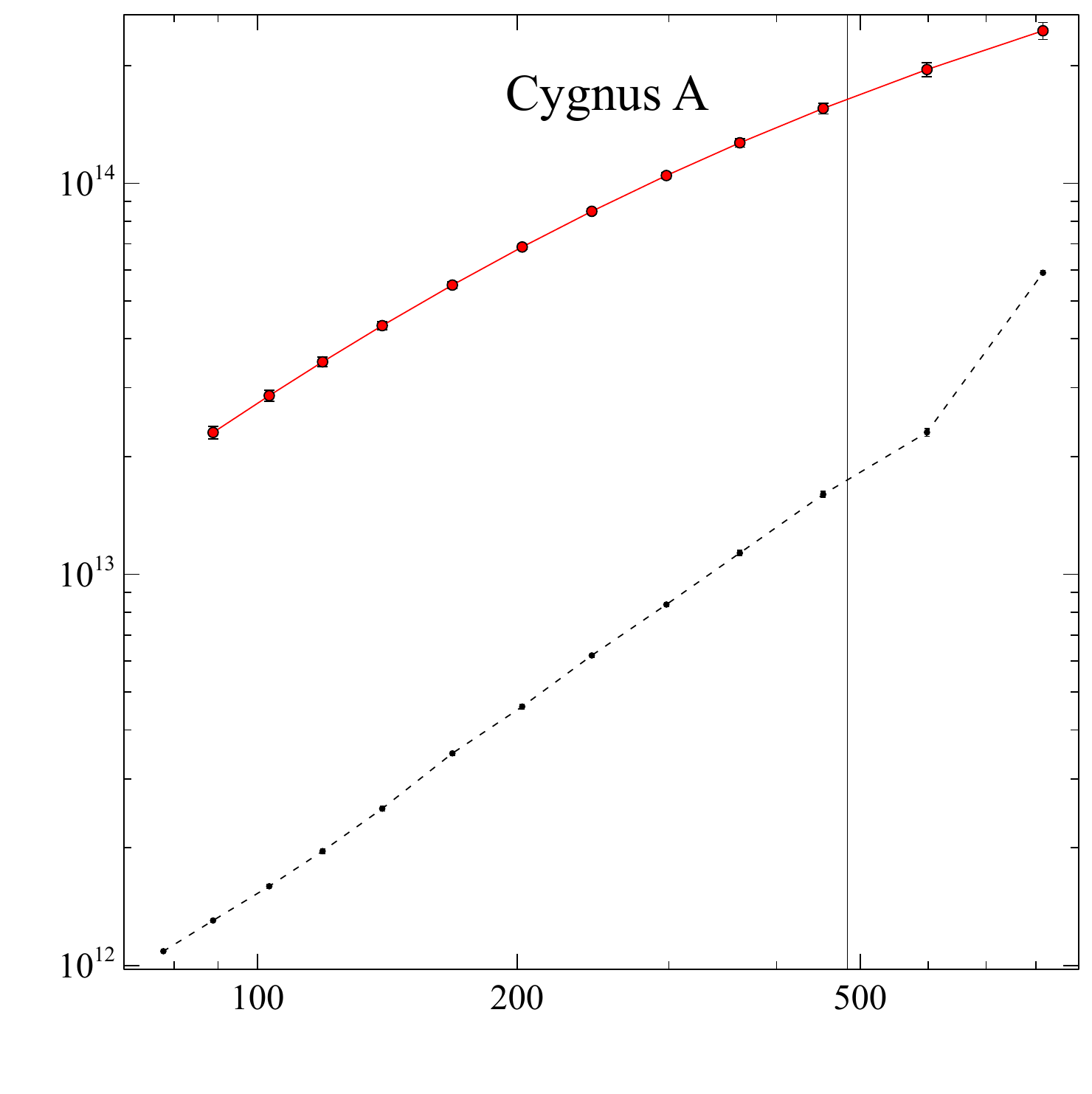}
\includegraphics[width=0.32\textwidth]{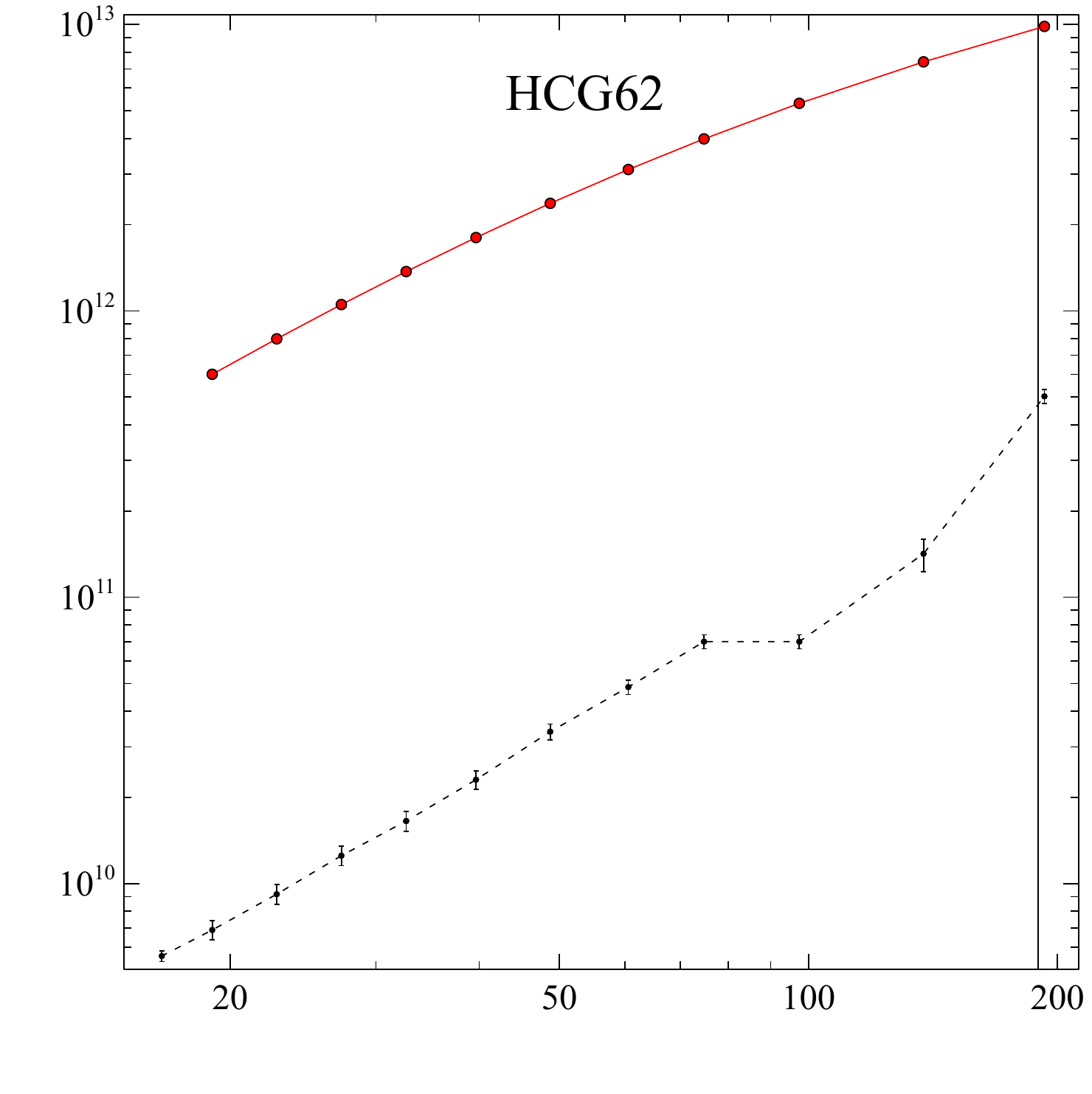} \\
\includegraphics[width=0.32\textwidth]{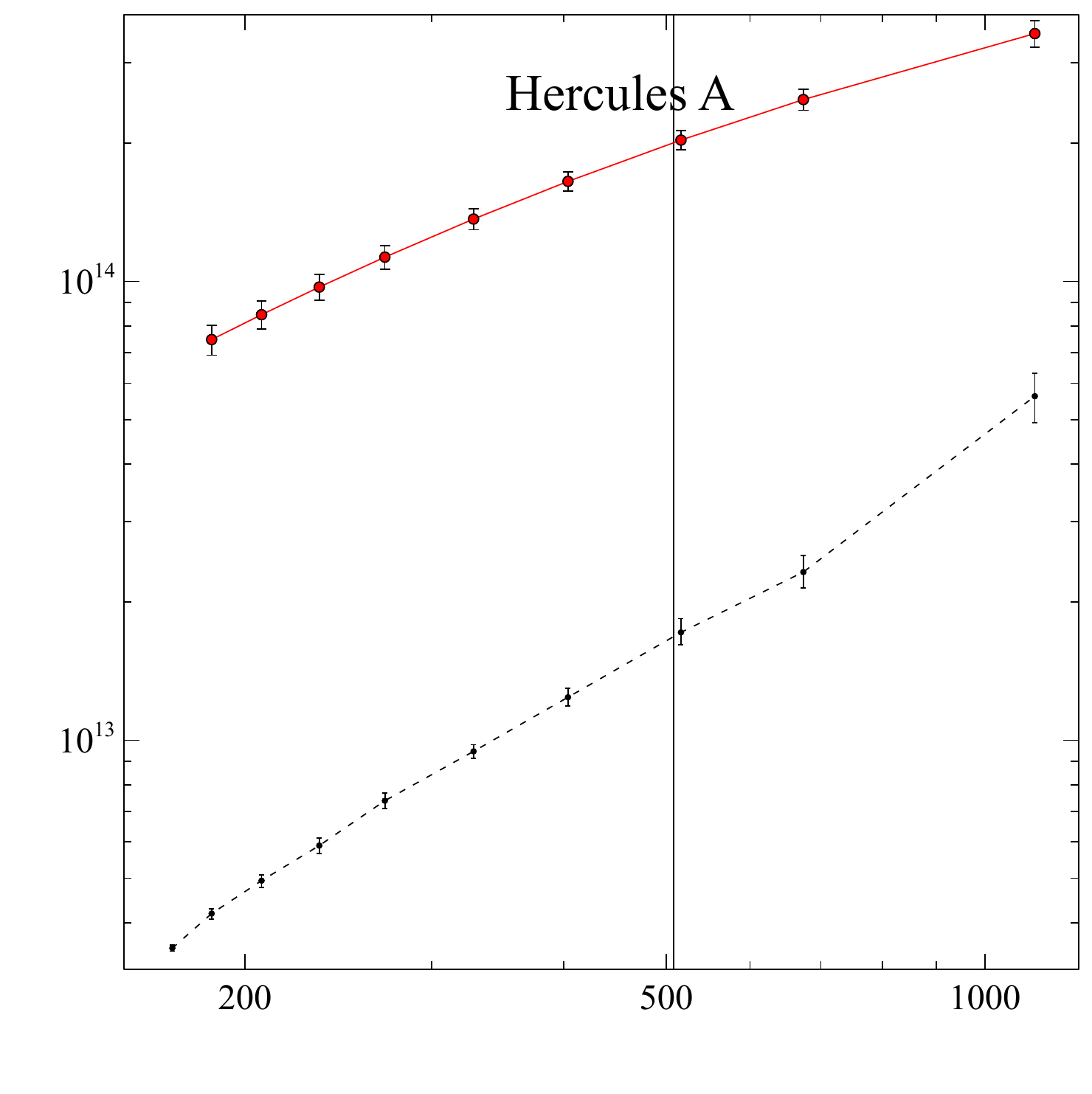}
\includegraphics[width=0.32\textwidth]{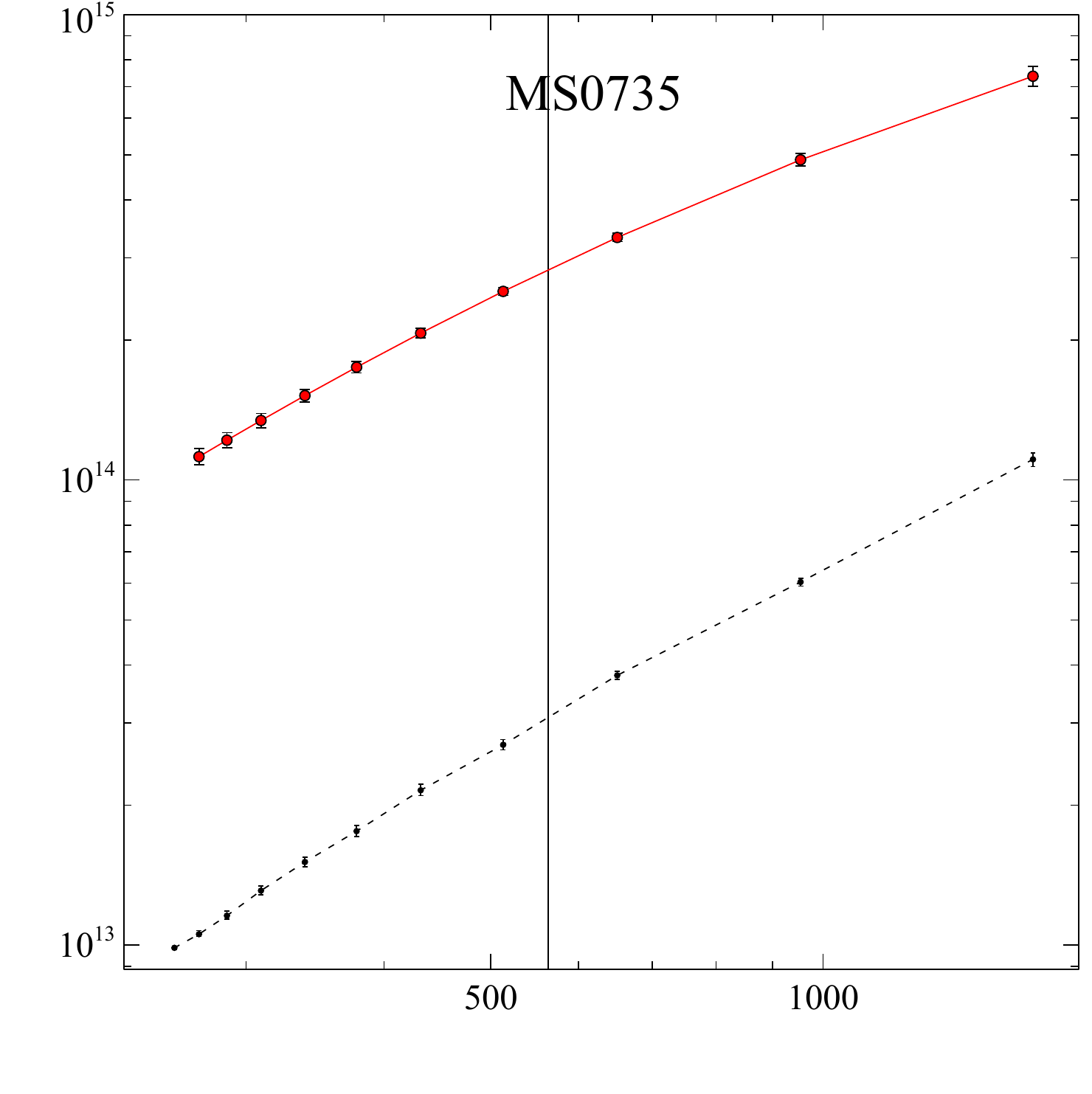}
\includegraphics[width=0.32\textwidth]{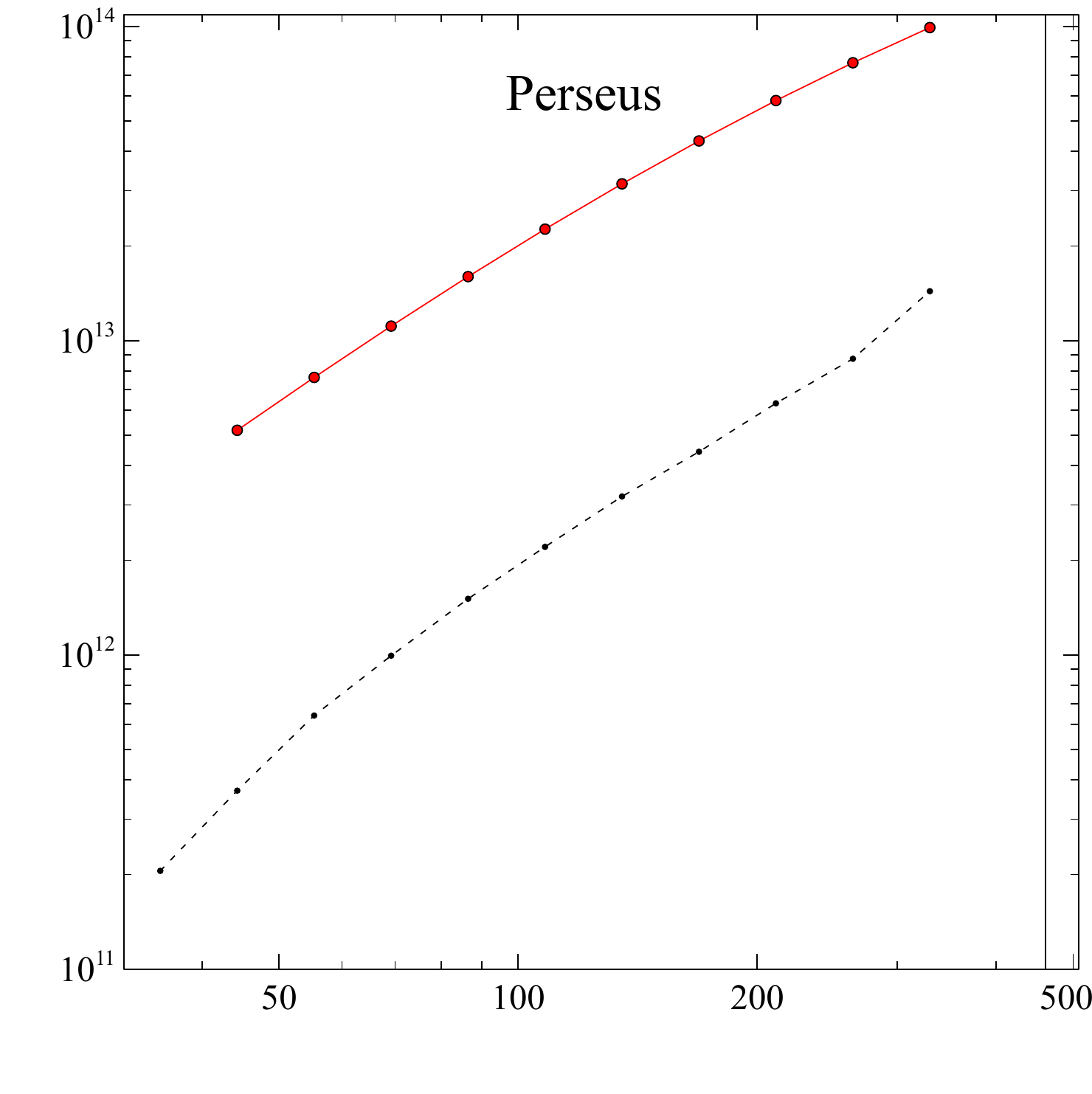} \\
\includegraphics[width=0.32\textwidth]{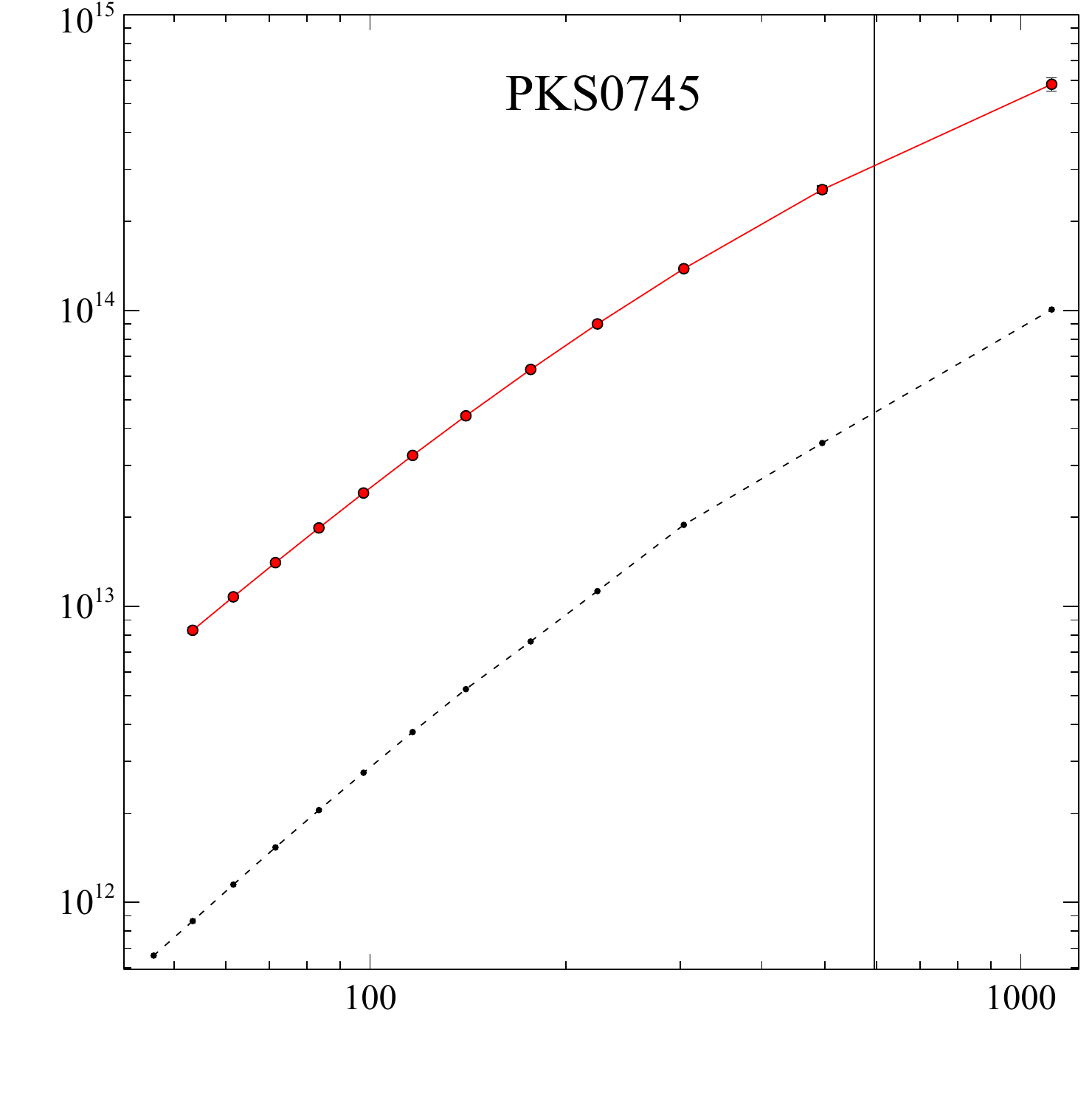}
\includegraphics[width=0.32\textwidth]{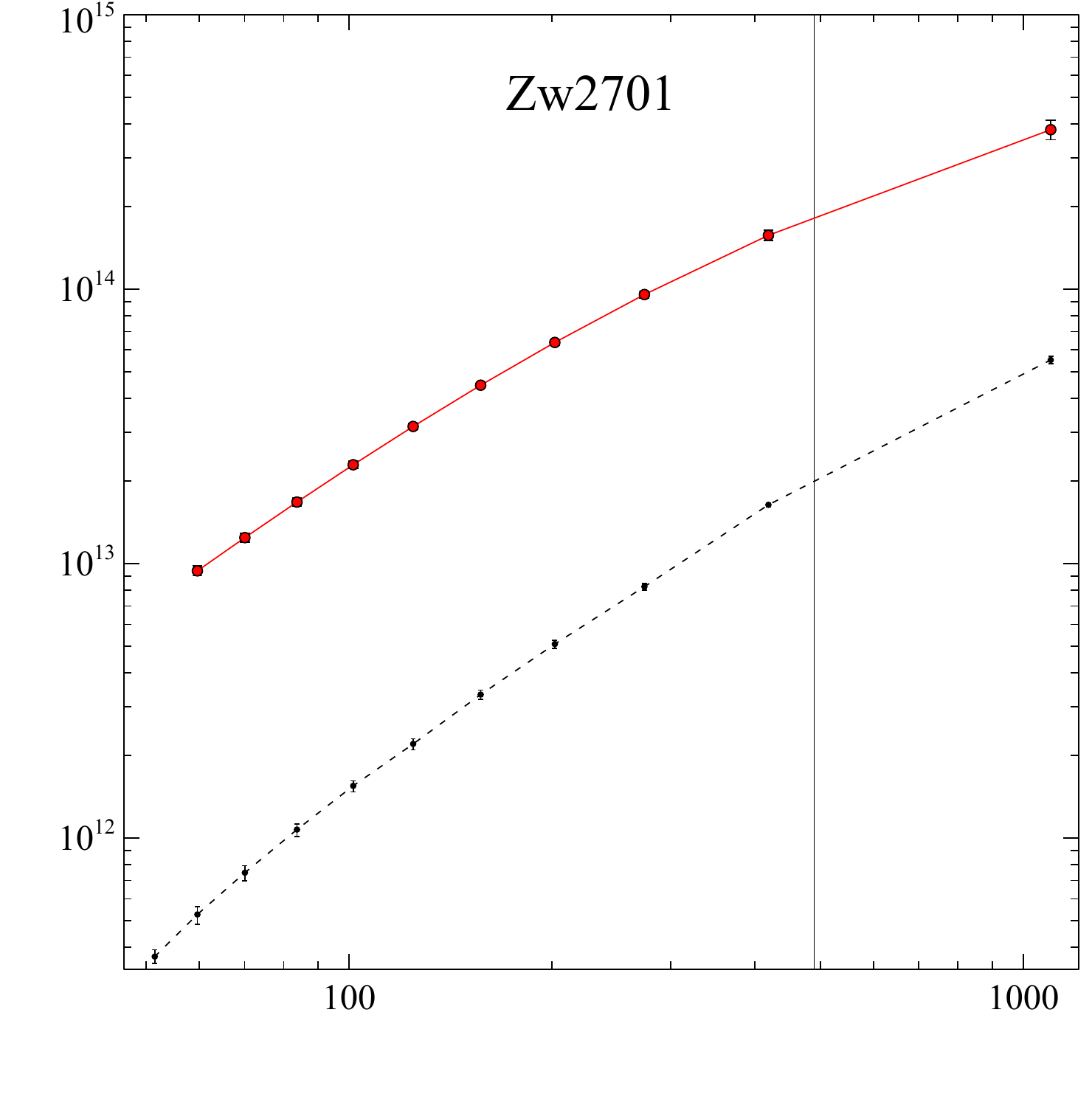}
\includegraphics[width=0.32\textwidth]{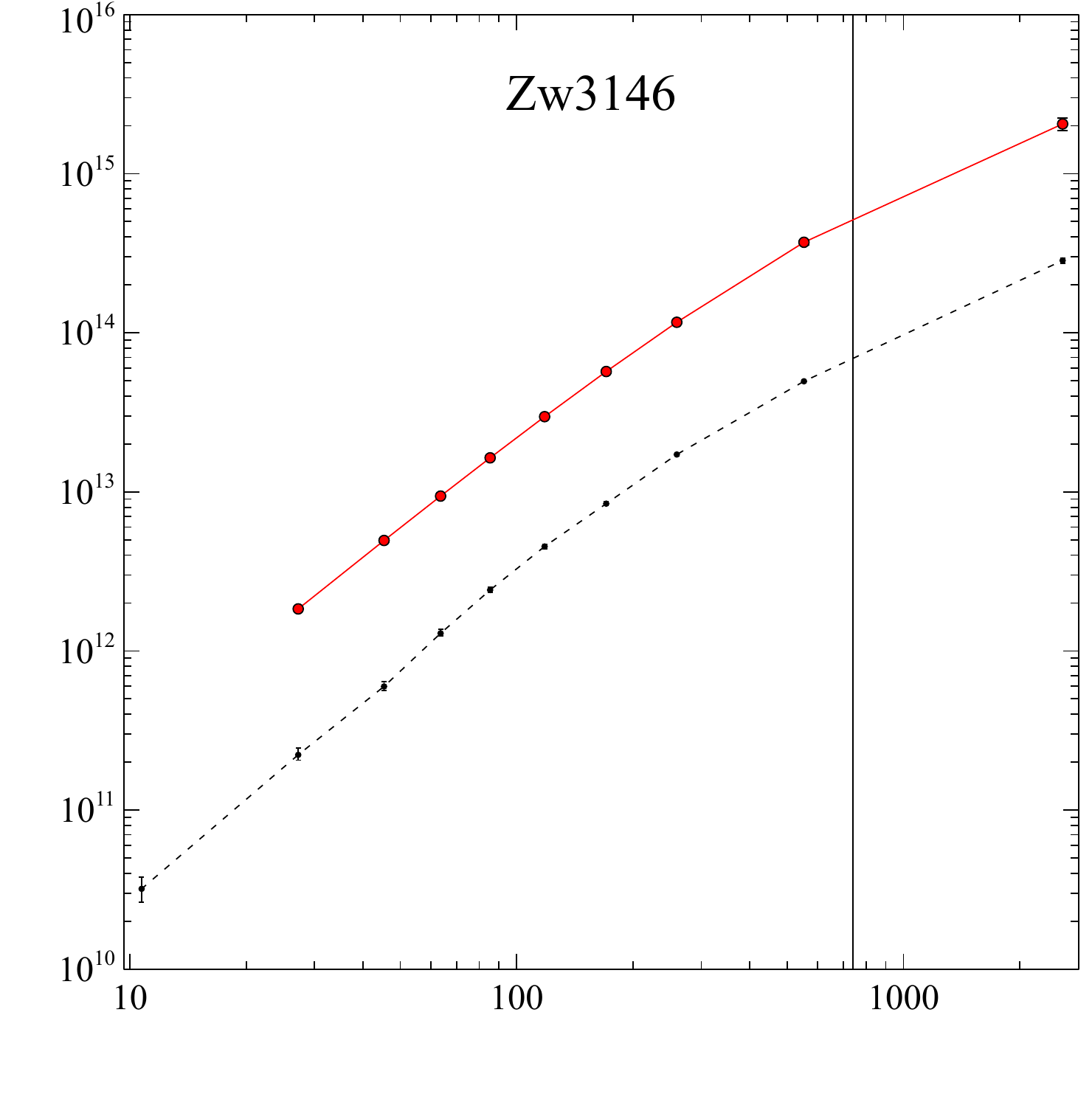}

\caption{Continued from figure \ref{figure:massprofs1}}

\label{figure:massprofs3}

\end{minipage}
\end{figure*}

\begin{figure*}
\begin{minipage}{1.0\textwidth}
\centering

\includegraphics[width=0.32\textwidth]{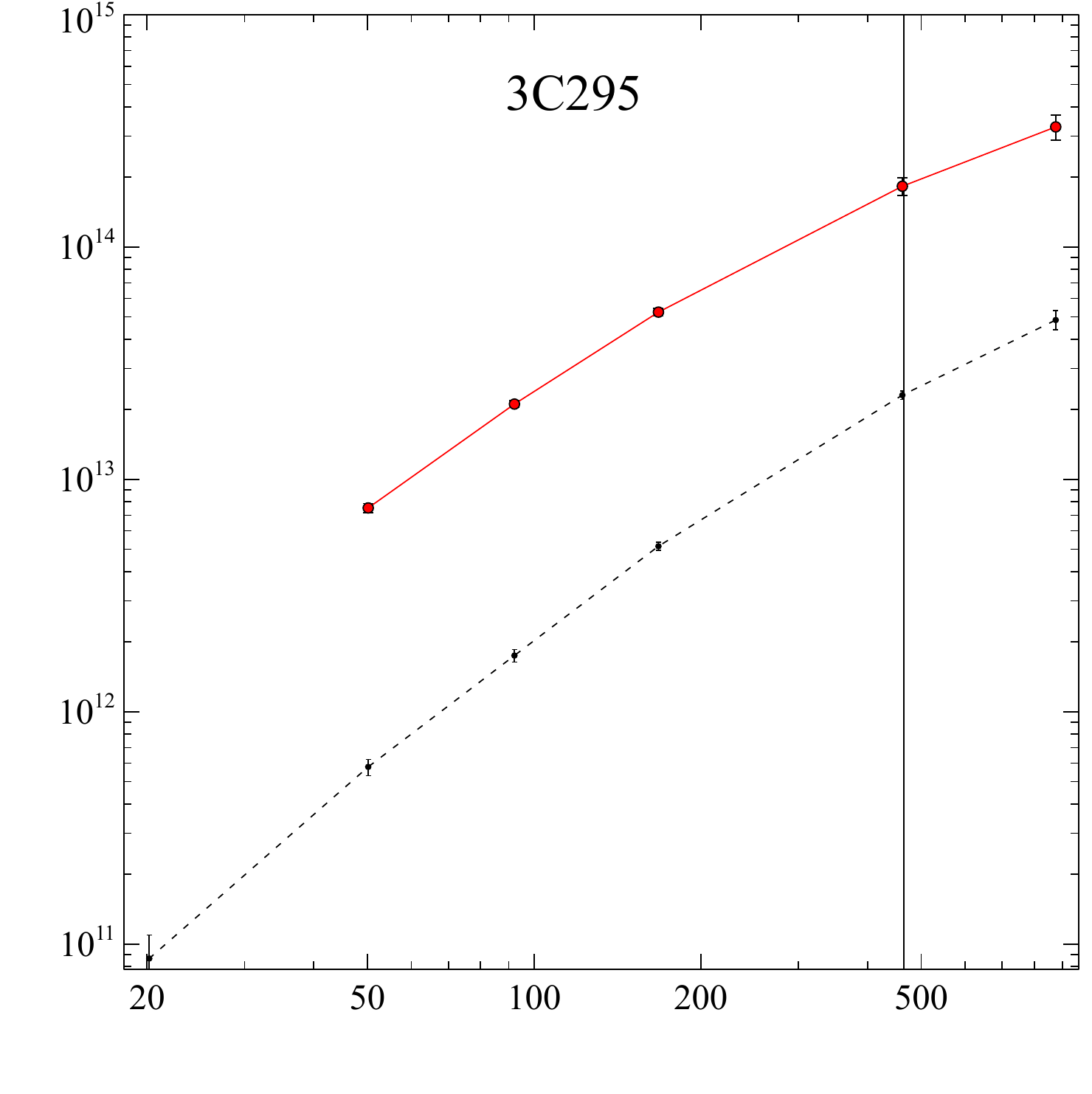} 
\includegraphics[width=0.32\textwidth]{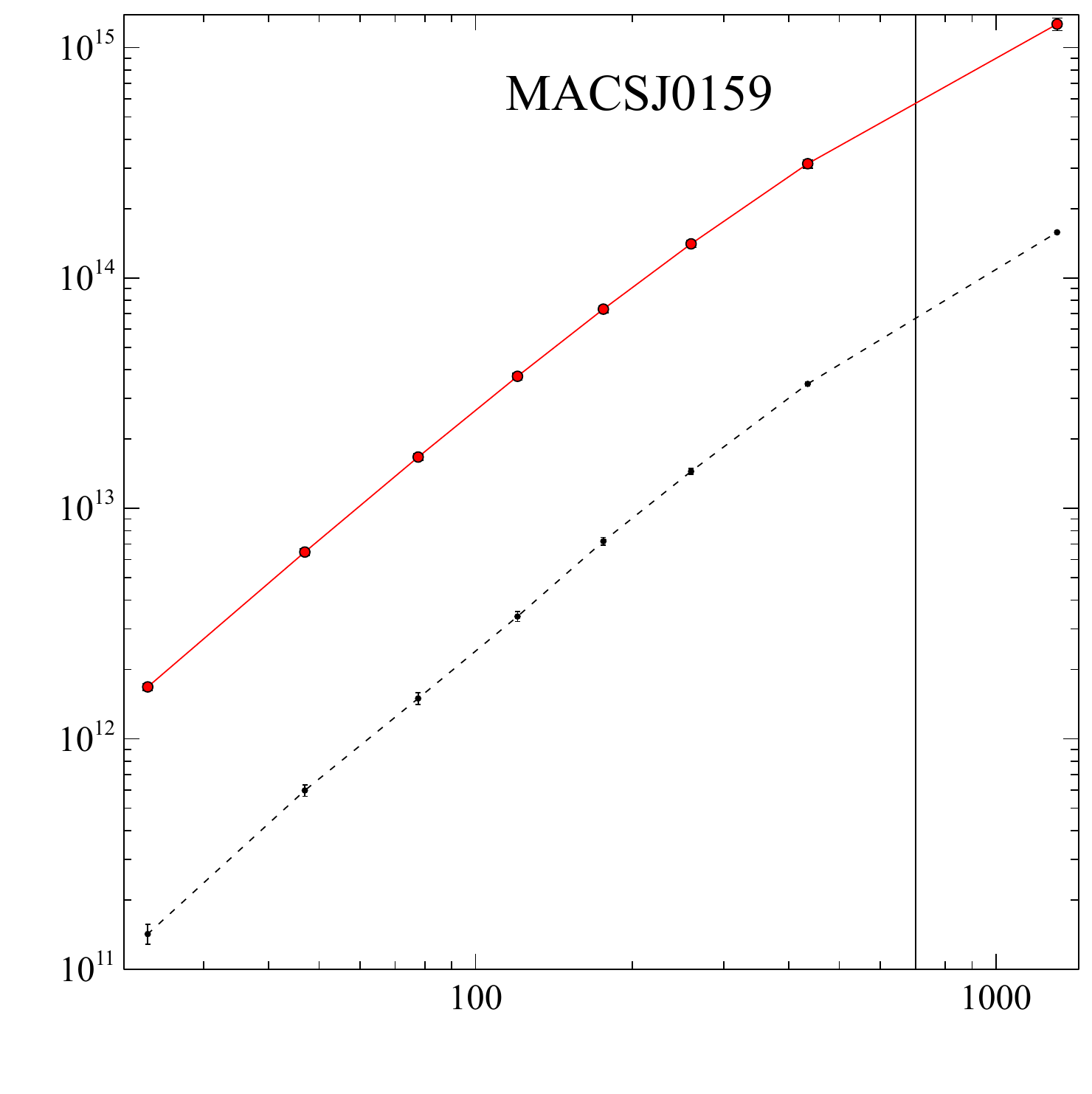} 
\includegraphics[width=0.32\textwidth]{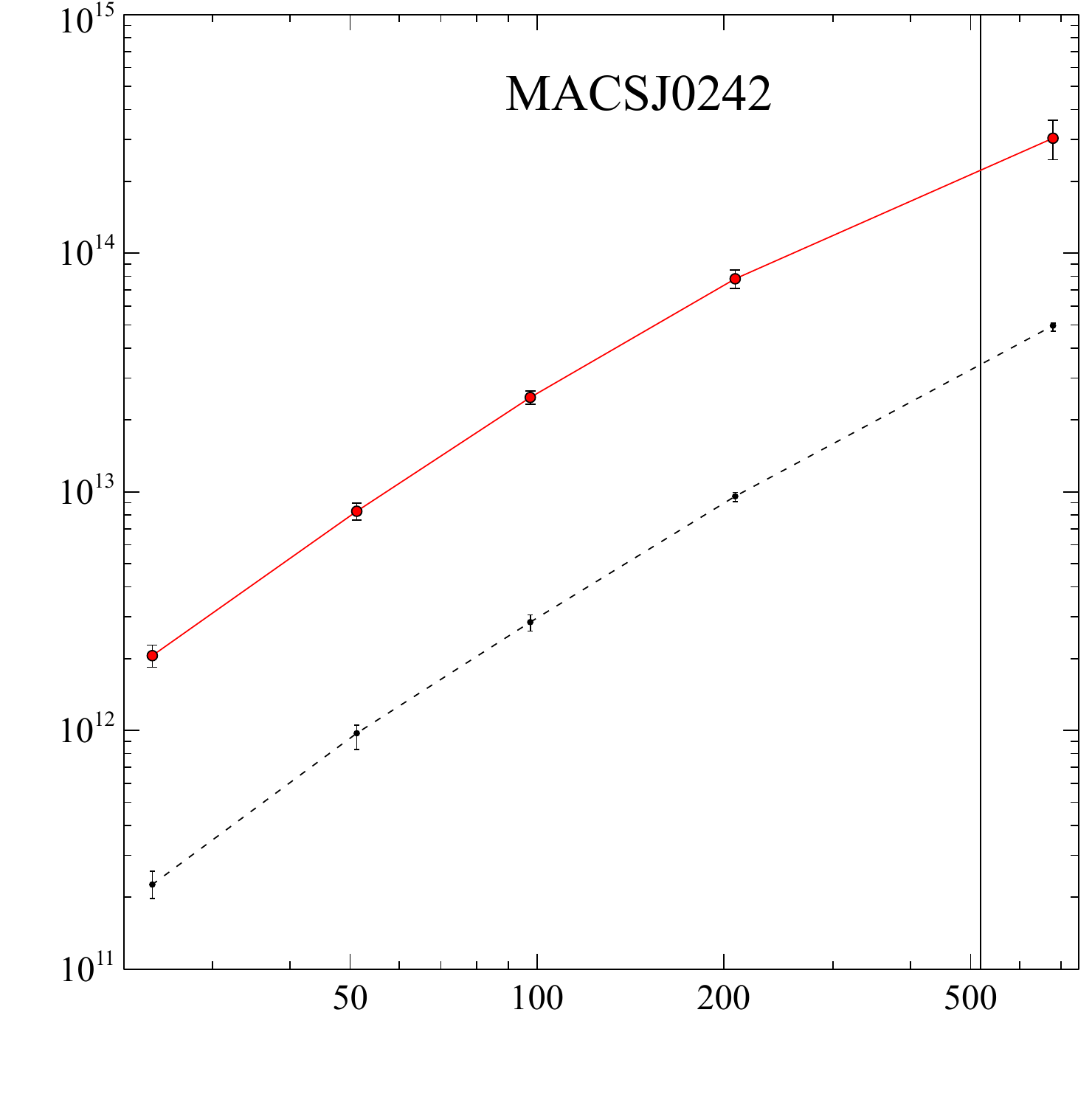} \\
\includegraphics[width=0.32\textwidth]{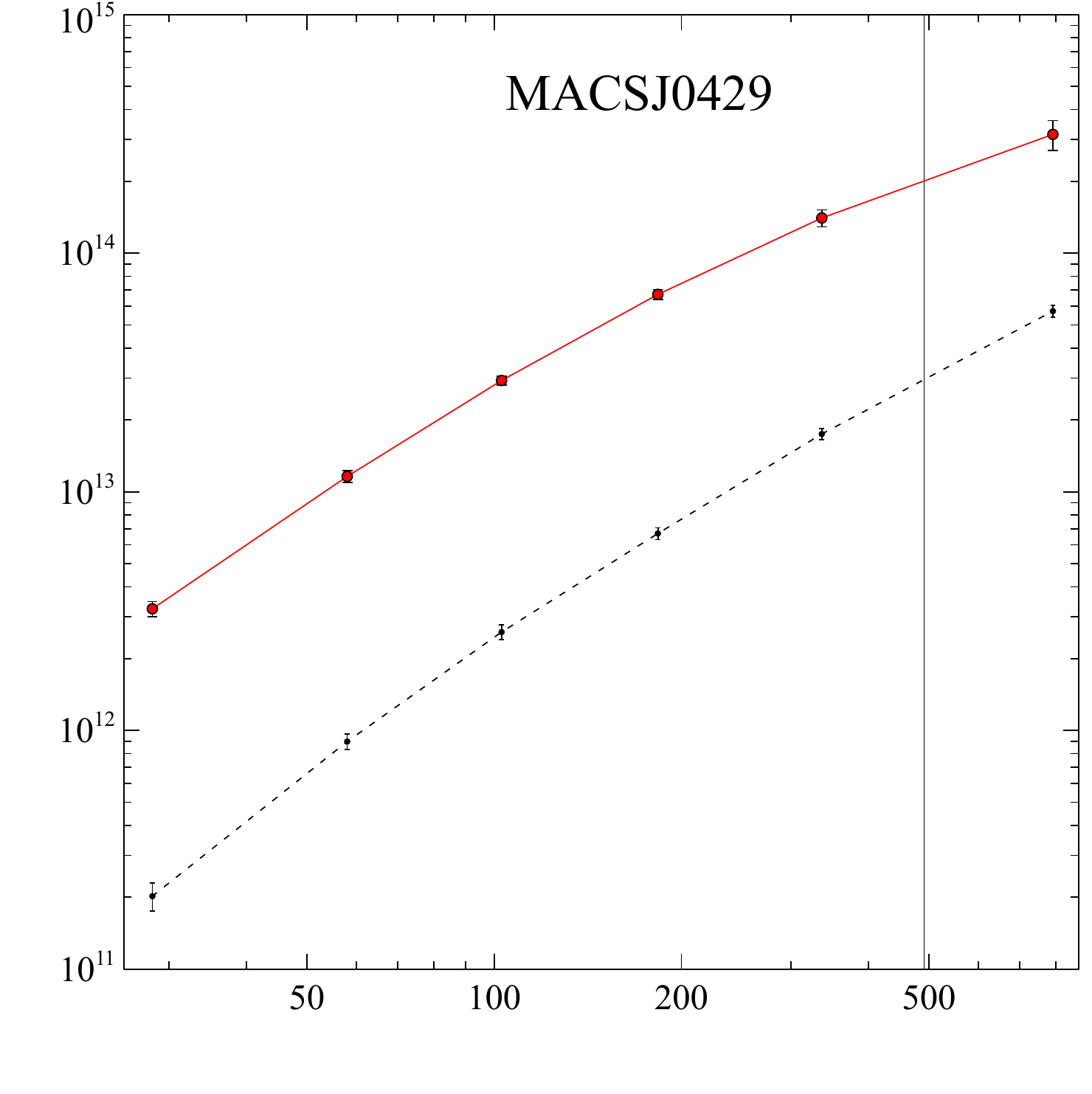}
\includegraphics[width=0.32\textwidth]{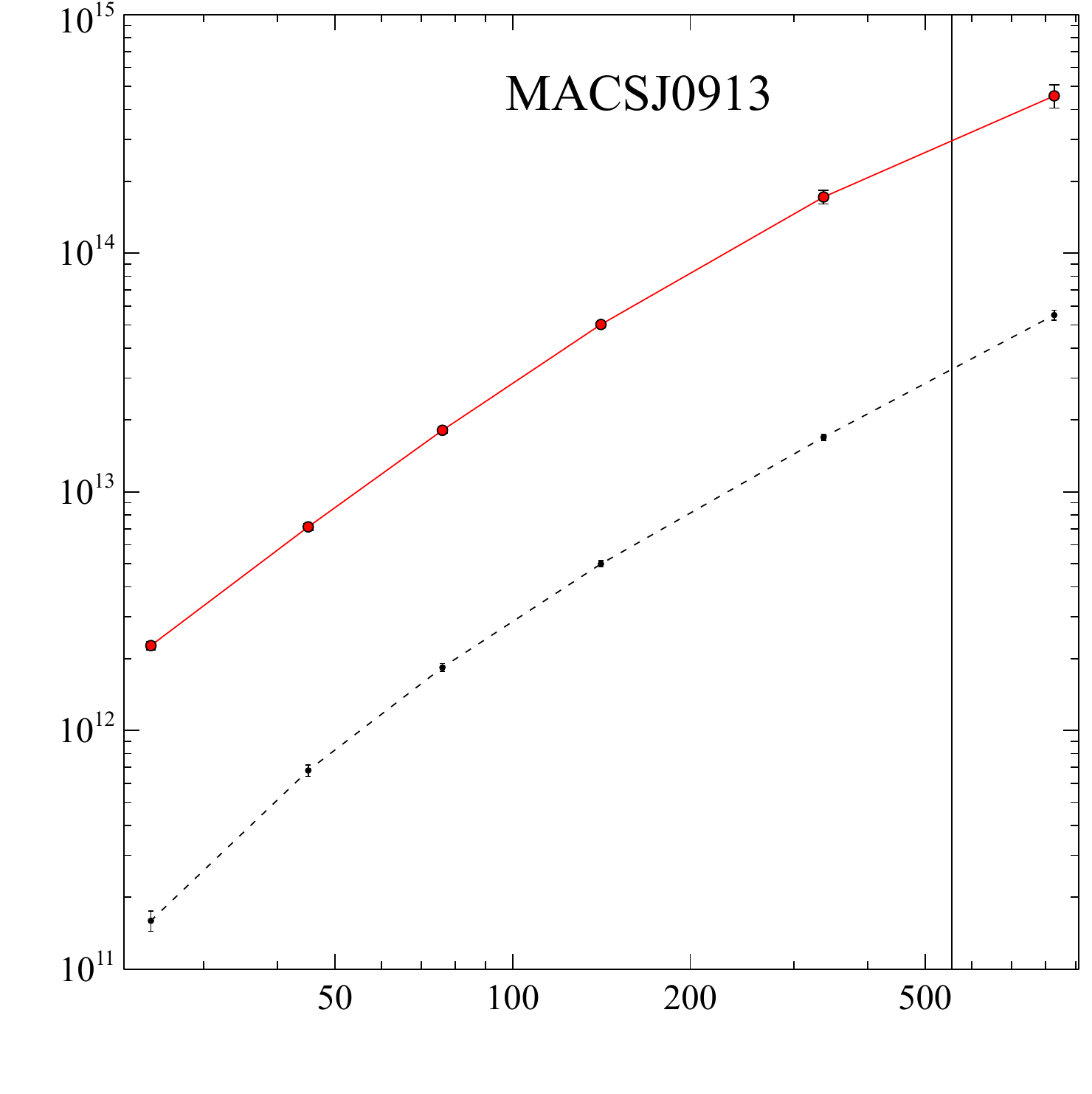}
\includegraphics[width=0.32\textwidth]{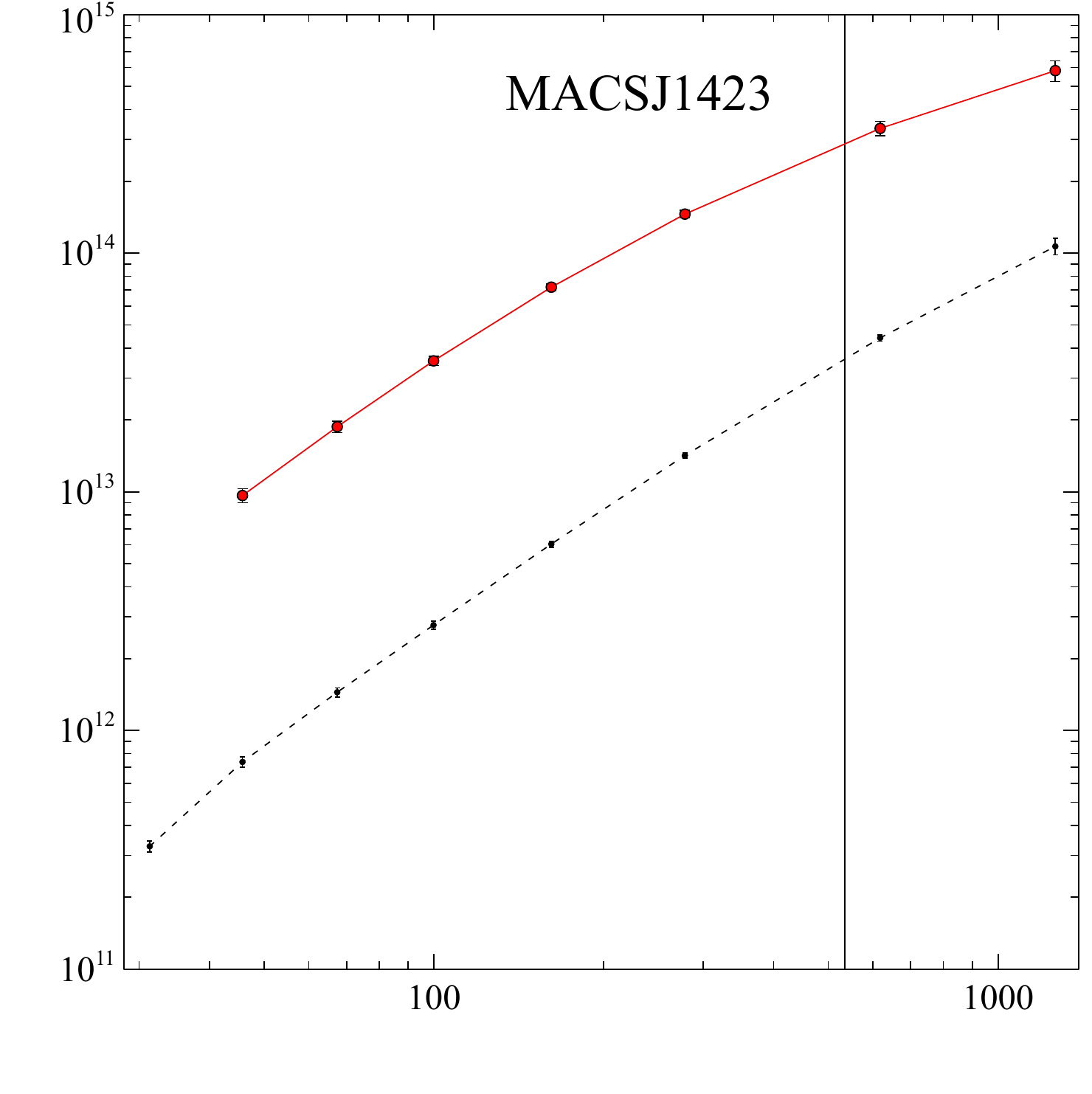} \\
\includegraphics[width=0.32\textwidth]{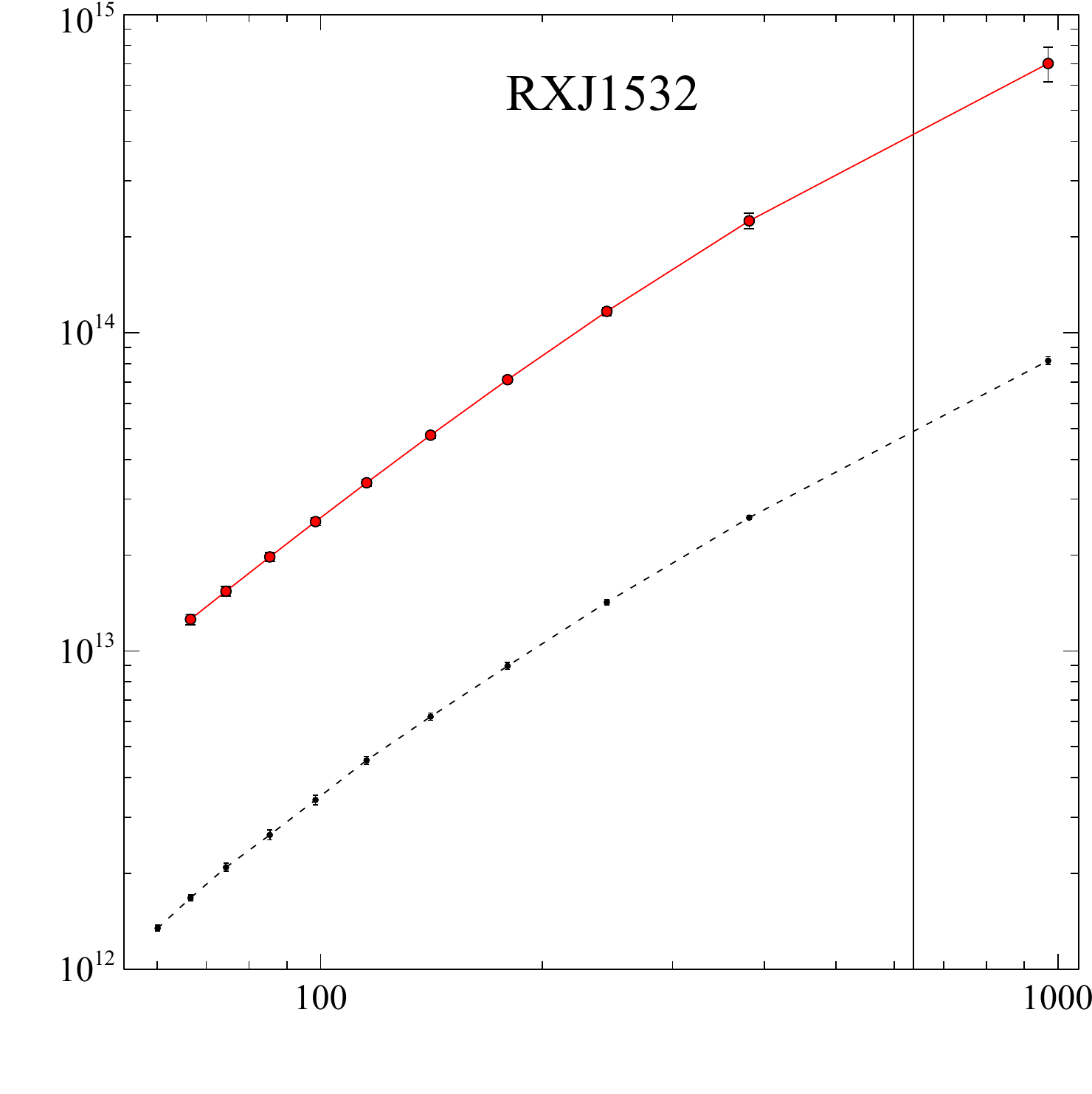}
\includegraphics[width=0.32\textwidth]{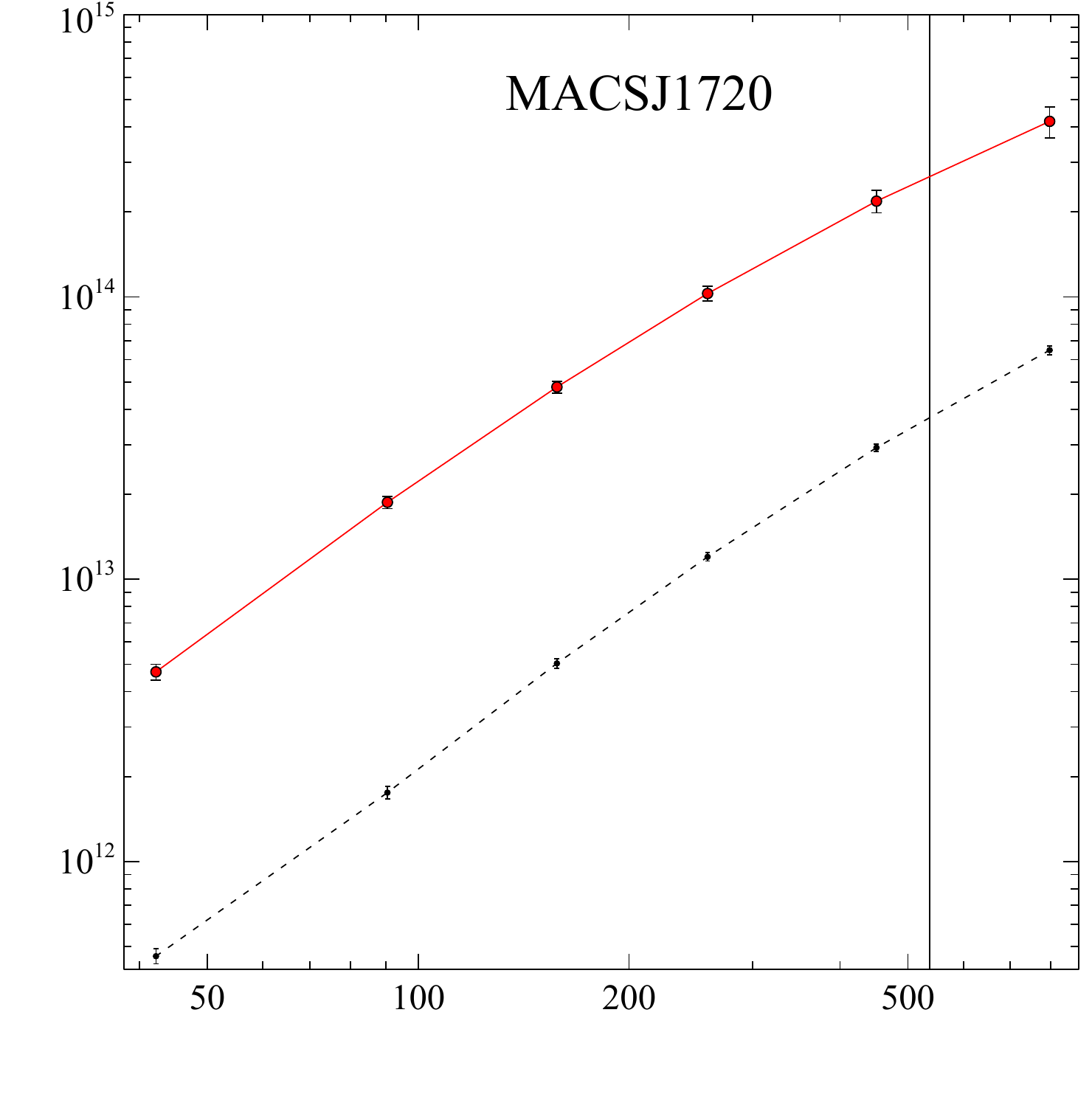} 
\includegraphics[width=0.32\textwidth]{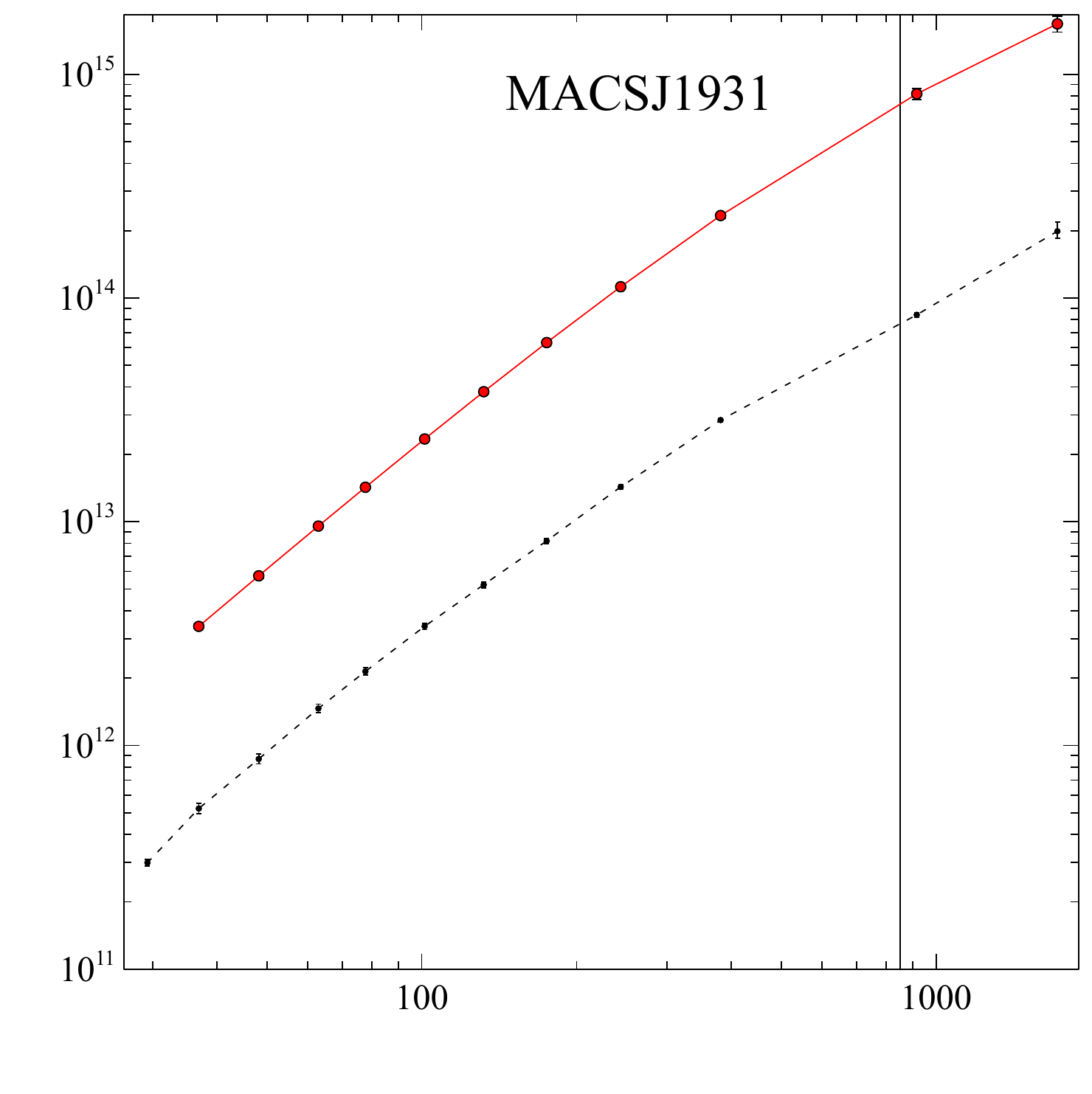} \\
\includegraphics[width=0.32\textwidth]{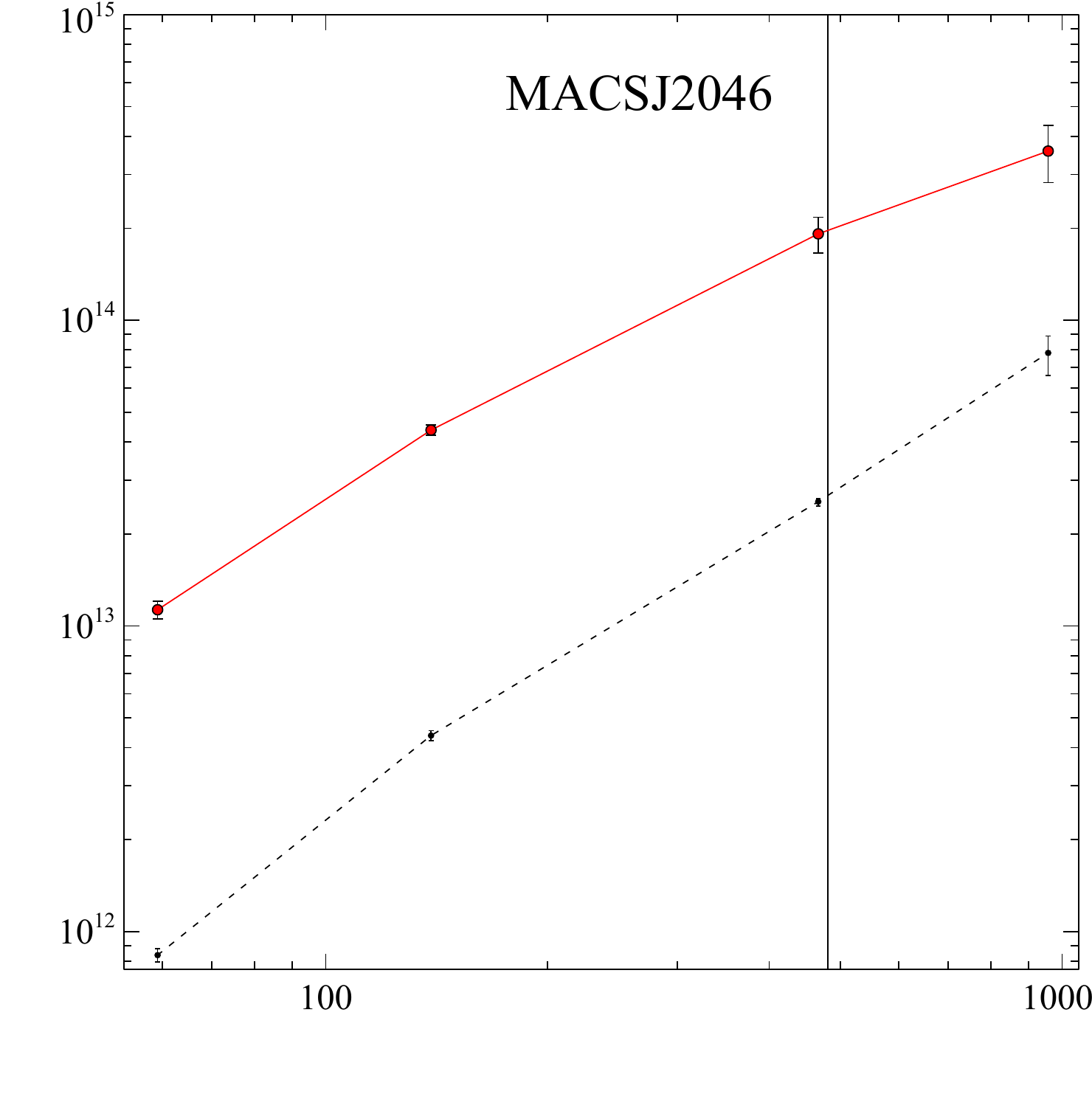}
\includegraphics[width=0.32\textwidth]{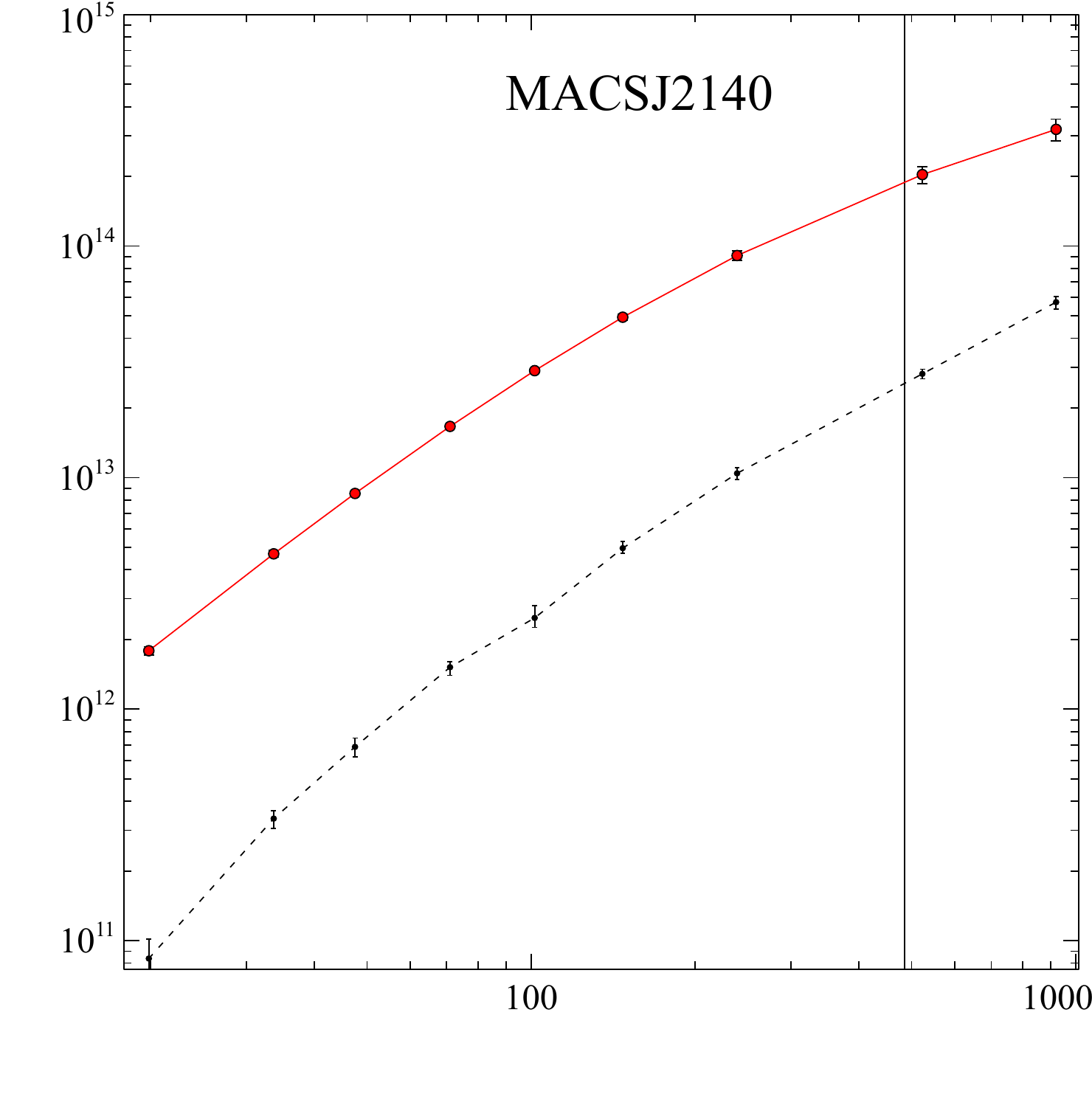}
\includegraphics[width=0.32\textwidth]{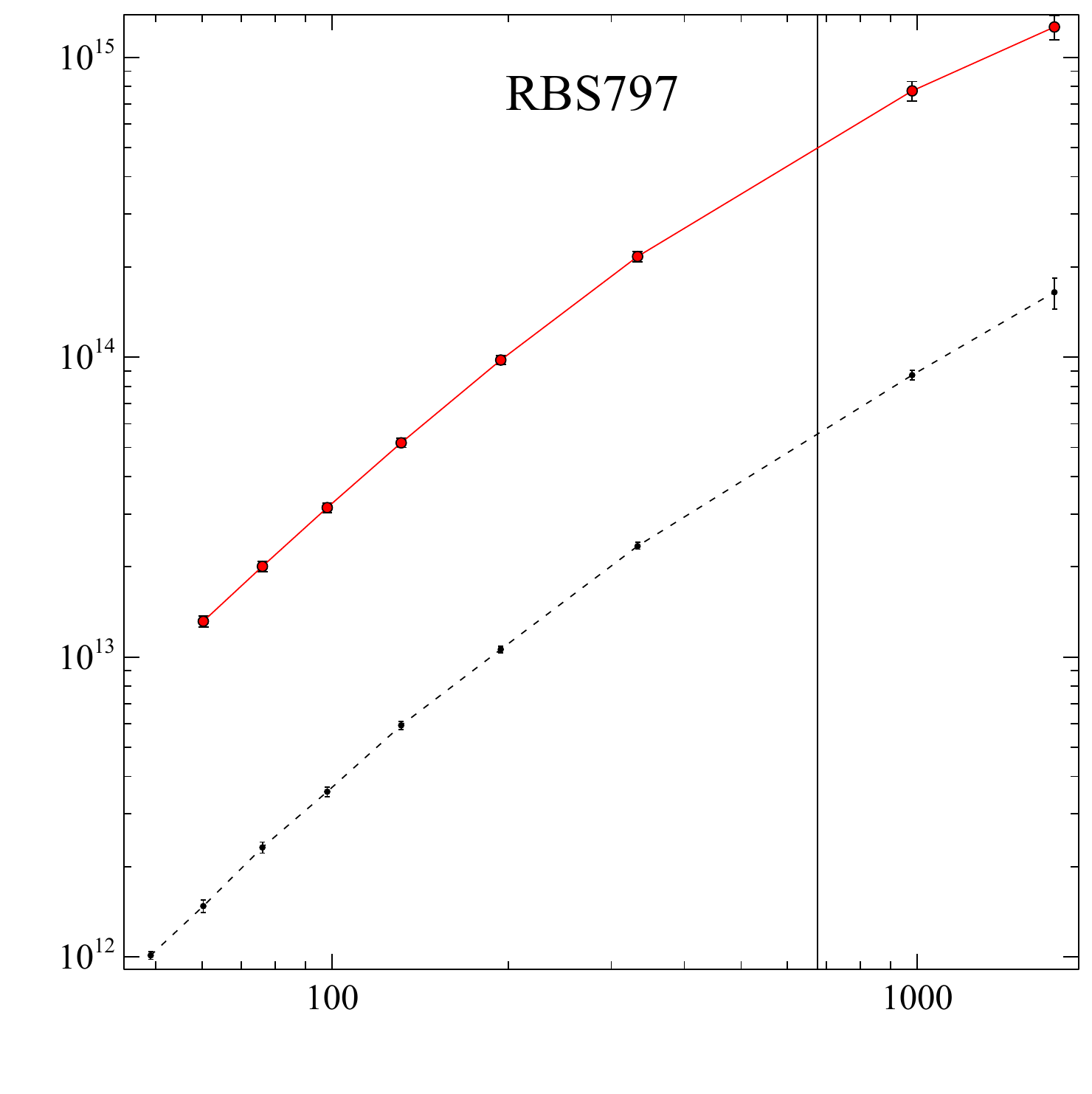}

\caption{Continued from figure \ref{figure:massprofs1}}

\label{figure:massprofs4}

\end{minipage}
\end{figure*}

\clearpage

\onecolumn

\landscape
\tiny
\setlength\LTleft{0pt minus 2pt}
\setlength\LTright{0pt minus 2pt}
\setlength{\tabcolsep}{1pt}
\begin{longtable}{llllllllllllllll}

\caption{Derived Mass Profiles }\\
\caption*{$A = 4\pi G \rho_{0} r_{s}^{2} \mu m_{H}$ is in units of keV, R$_{s}$ is given in units of kpc} \\
\hline
2A0335 & r (kpc)  & 27.2 & 30.4 & 34.3 & 39.0 & 45.1 & 53.1 & 63.4 & 76.8 & 94.2 & 116.8 & 150.8 & 252.8 & 517.8 \\ 
$\rho_{o}=31.0^{2.6}_{-2.1}$ & M$_{g}$ ($10^{11}$ M$_{\odot}$ )  & 1.21$^{0.014}_{-0.014}$  & 1.51$^{0.022}_{-0.022}$  & 1.95$^{0.031}_{-0.031}$  & 2.55$^{0.04}_{-0.04}$  & 3.36$^{0.05}_{-0.05}$  & 4.47$^{0.064}_{-0.063}$  & 6.03$^{0.08}_{-0.079}$  & 8.18$^{0.1}_{-0.1}$  & 11.6$^{0.13}_{-0.13}$  & 16.5$^{0.17}_{-0.17}$  & 25.5$^{0.23}_{-0.23}$  & 58.3$^{0.47}_{-0.47}$  & 159.0$^{2.1}_{-2.1}$  \\ 
$R_{s}=204.0^{28.0}_{-22.0}$ & M ($10^{13}$ M$_{\odot}$ ) & -  & 0.21$\pm0.0025$ & 0.262$\pm0.0028$ & 0.329$\pm0.0032$ & 0.426$\pm0.0035$ & 0.564$\pm0.0037$ & 0.763$\pm0.0037$ & 1.05$\pm0.0043$ & 1.45$\pm0.0081$ & 2.02$\pm0.018$ & 2.92$\pm0.04$ & 5.75$\pm0.14$ & 12.4$\pm0.51$ \\ 
\hline 
3C295 & r (kpc)  & 20.2 & 50.2 & 92.1 & 167.7 & 462.4 & 875.0 \\ 
$\rho_{o}=50.9^{9.9}_{-6.7}$ & M$_{g}$ ($10^{13}$ M$_{\odot}$ )  & 0.00866$^{0.0023}_{-0.0031}$  & 0.0579$^{0.0045}_{-0.0049}$  & 0.174$^{0.011}_{-0.011}$  & 0.515$^{0.02}_{-0.02}$  & 2.3$^{0.099}_{-0.099}$  & 4.85$^{0.47}_{-0.45}$  \\ 
$R_{s}=236.0^{96.0}_{-64.0}$ & M ($10^{13}$ M$_{\odot}$ ) & -  & 0.75$\pm0.035$ & 2.1$\pm0.073$ & 5.2$\pm0.2$ & 18.0$\pm1.6$ & 33.0$\pm4.0$ \\ 
\hline 
3C388 & r (kpc)  & 8.4 & 40.5 & 78.9 & 140.5 & 269.7 & 462.7 & 757.2 \\ 
$\rho_{o}=23.7^{2.5}_{-1.5}$ & M$_{g}$ ($10^{12}$ M$_{\odot}$ )  & 0.00351$^{0.00081}_{-0.00072}$  & 0.0896$^{0.017}_{-0.016}$  & 0.311$^{0.047}_{-0.048}$  & 0.989$^{0.094}_{-0.096}$  & 3.3$^{0.23}_{-0.21}$  & 7.31$^{0.7}_{-0.68}$  & 17.3$^{1.3}_{-1.3}$  \\ 
$R_{s}=127.0^{38.0}_{-27.0}$ & M ($10^{13}$ M$_{\odot}$ ) & -  & 0.38$\pm0.027$ & 1.1$\pm0.045$ & 2.4$\pm0.08$ & 5.0$\pm0.35$ & 8.1$\pm0.88$ & 12.0$\pm1.7$ \\ 
\hline 
3C401 & r (kpc)  & 8.1 & 48.5 & 113.2 & 223.0 & 420.7 & 816.2 \\ 
$\rho_{o}=30.3^{13.0}_{-5.8}$ & M$_{g}$ ($10^{12}$ M$_{\odot}$ )  & 0.00303$^{0.00086}_{-0.00076}$  & 0.141$^{0.026}_{-0.026}$  & 0.79$^{0.051}_{-0.058}$  & 2.53$^{0.23}_{-0.23}$  & 7.57$^{0.53}_{-0.43}$  & 21.7$^{3.2}_{-2.7}$  \\ 
$R_{s}=210.0^{180.0}_{-82.0}$ & M ($10^{13}$ M$_{\odot}$ ) & -  & 0.5$\pm0.073$ & 2.0$\pm0.19$ & 5.0$\pm0.47$ & 10.0$\pm1.6$ & 20.0$\pm4.6$ \\ 
\hline 
A85 & r (kpc)  & 26.3 & 38.6 & 53.4 & 71.2 & 92.4 & 118.0 & 146.1 & 182.6 & 230.1 & 306.1 & 458.0 & 640.4 \\ 
$\rho_{o}=51.1^{5.0}_{-4.0}$ & M$_{g}$ ($10^{12}$ M$_{\odot}$ )  & 0.0679$^{0.0029}_{-0.0029}$  & 0.168$^{0.0065}_{-0.0064}$  & 0.326$^{0.011}_{-0.011}$  & 0.6$^{0.018}_{-0.018}$  & 1.07$^{0.027}_{-0.026}$  & 1.74$^{0.039}_{-0.039}$  & 2.6$^{0.054}_{-0.054}$  & 3.97$^{0.074}_{-0.073}$  & 6.3$^{0.095}_{-0.095}$  & 10.5$^{0.13}_{-0.13}$  & 19.3$^{0.29}_{-0.29}$  & 37.6$^{0.46}_{-0.46}$  \\ 
$R_{s}=381.0^{66.0}_{-53.0}$ & M ($10^{13}$ M$_{\odot}$ ) & -  & 0.32$\pm0.0057$ & 0.58$\pm0.0089$ & 0.97$\pm0.012$ & 1.5$\pm0.015$ & 2.3$\pm0.018$ & 3.3$\pm0.023$ & 4.7$\pm0.04$ & 6.7$\pm0.082$ & 10.0$\pm0.18$ & 17.0$\pm0.47$ & 25.0$\pm0.9$ \\ 
\hline 
A133 & r (kpc)  & 54.1 & 63.5 & 75.8 & 90.5 & 109.6 & 132.5 & 150.8 & 183.8 & 230.0 & 299.3 & 410.2 & 654.1 \\ 
$\rho_{o}=36.4^{6.9}_{-3.6}$ & M$_{g}$ ($10^{12}$ M$_{\odot}$ )  & 0.247$^{0.0099}_{-0.0096}$  & 0.343$^{0.018}_{-0.026}$  & 0.442$^{0.056}_{-0.029}$  & 0.615$^{0.063}_{-0.044}$  & 0.894$^{0.071}_{-0.055}$  & 1.29$^{0.084}_{-0.074}$  & 1.65$^{0.1}_{-0.094}$  & 2.41$^{0.12}_{-0.11}$  & 3.58$^{0.15}_{-0.14}$  & 5.54$^{0.21}_{-0.2}$  & 9.64$^{0.3}_{-0.29}$  & 20.6$^{0.56}_{-0.56}$  \\ 
$R_{s}=314.0^{130.0}_{-75.0}$ & M ($10^{13}$ M$_{\odot}$ ) & -  & 0.66$\pm0.037$ & 0.89$\pm0.045$ & 1.2$\pm0.053$ & 1.7$\pm0.061$ & 2.3$\pm0.066$ & 2.8$\pm0.068$ & 3.8$\pm0.07$ & 5.2$\pm0.094$ & 7.4$\pm0.2$ & 11.0$\pm0.47$ & 18.0$\pm1.2$ \\ 
\hline 
A1795 & r (kpc)  & 53.1 & 59.0 & 66.1 & 73.9 & 84.0 & 97.2 & 115.7 & 141.5 & 177.6 & 228.2 & 314.2 & 494.9 & 892.2 \\ 
$\rho_{o}=55.0^{1.6}_{-1.6}$ & M$_{g}$ ($10^{12}$ M$_{\odot}$ )  & 0.453$^{0.003}_{-0.0035}$  & 0.571$^{0.0079}_{-0.0064}$  & 0.724$^{0.0097}_{-0.0088}$  & 0.935$^{0.012}_{-0.0094}$  & 1.2$^{0.013}_{-0.012}$  & 1.6$^{0.016}_{-0.014}$  & 2.21$^{0.019}_{-0.017}$  & 3.16$^{0.024}_{-0.023}$  & 4.73$^{0.032}_{-0.03}$  & 7.36$^{0.042}_{-0.041}$  & 12.7$^{0.06}_{-0.059}$  & 25.8$^{0.21}_{-0.22}$  & 57.5$^{1.7}_{-1.5}$  \\ 
$R_{s}=373.0^{21.0}_{-21.0}$ & M ($10^{13}$ M$_{\odot}$ ) & -  & 0.758$\pm0.0072$ & 0.93$\pm0.0084$ & 1.13$\pm0.0097$ & 1.42$\pm0.011$ & 1.83$\pm0.013$ & 2.46$\pm0.015$ & 3.43$\pm0.016$ & 4.93$\pm0.016$ & 7.2$\pm0.016$ & 11.3$\pm0.044$ & 20.2$\pm0.17$ & 38.1$\pm0.54$ \\ 
\hline 
A1835 & r (kpc)  & 29.1 & 37.0 & 48.2 & 62.6 & 81.4 & 105.9 & 140.1 & 188.0 & 264.7 & 402.7 & 830.4 & 1550.8 \\ 
$\rho_{o}=95.7^{3.4}_{-3.7}$ & M$_{g}$ ($10^{12}$ M$_{\odot}$ )  & 0.262$^{0.0074}_{-0.0073}$  & 0.491$^{0.013}_{-0.012}$  & 0.867$^{0.019}_{-0.018}$  & 1.4$^{0.028}_{-0.027}$  & 2.25$^{0.041}_{-0.041}$  & 3.61$^{0.059}_{-0.058}$  & 5.89$^{0.085}_{-0.085}$  & 9.97$^{0.12}_{-0.12}$  & 17.3$^{0.18}_{-0.18}$  & 34.2$^{0.32}_{-0.29}$  & 90.4$^{0.87}_{-0.86}$  & 190.0$^{3.2}_{-3.2}$  \\ 
$R_{s}=439.0^{29.0}_{-31.0}$ & M ($10^{13}$ M$_{\odot}$ ) & -  & 0.48$\pm0.0099$ & 0.79$\pm0.015$ & 1.3$\pm0.022$ & 2.1$\pm0.03$ & 3.3$\pm0.039$ & 5.3$\pm0.045$ & 8.6$\pm0.052$ & 14.0$\pm0.11$ & 26.0$\pm0.4$ & 62.0$\pm1.9$ & 110.0$\pm4.9$ \\ 
\hline 
A2029 & r (kpc)  & 7.2 & 20.3 & 32.0 & 43.8 & 57.9 & 73.4 & 93.6 & 119.9 & 154.0 & 201.8 & 273.4 & 431.1 & 969.7 \\ 
$\rho_{o}=68.8^{3.9}_{-3.4}$ & M$_{g}$ ($10^{12}$ M$_{\odot}$ )  & 0.00577$^{0.0003}_{-0.0003}$  & 0.0575$^{0.0015}_{-0.0015}$  & 0.164$^{0.0037}_{-0.0036}$  & 0.343$^{0.0062}_{-0.0062}$  & 0.629$^{0.01}_{-0.01}$  & 1.06$^{0.015}_{-0.015}$  & 1.77$^{0.021}_{-0.021}$  & 2.9$^{0.03}_{-0.03}$  & 4.63$^{0.044}_{-0.044}$  & 7.4$^{0.063}_{-0.063}$  & 13.0$^{0.091}_{-0.097}$  & 24.5$^{0.21}_{-0.21}$  & 89.5$^{1.1}_{-1.1}$  \\ 
$R_{s}=304.0^{30.0}_{-26.0}$ & M ($10^{13}$ M$_{\odot}$ ) & -  & 0.153$\pm0.0024$ & 0.365$\pm0.005$ & 0.651$\pm0.0079$ & 1.08$\pm0.011$ & 1.64$\pm0.014$ & 2.48$\pm0.017$ & 3.73$\pm0.02$ & 5.53$\pm0.024$ & 8.27$\pm0.044$ & 12.6$\pm0.1$ & 22.3$\pm0.3$ & 50.4$\pm1.1$ \\ 
\hline 
A2052 & r (kpc)  & 52.4 & 57.9 & 64.5 & 71.8 & 80.6 & 91.1 & 102.6 & 116.4 & 135.8 & 161.0 & 198.8 & 327.4 & 533.0 \\ 
$\rho_{o}=27.3^{0.48}_{-0.46}$ & M$_{g}$ ($10^{12}$ M$_{\odot}$ )  & 0.24$^{0.0028}_{-0.0028}$  & 0.287$^{0.0033}_{-0.0032}$  & 0.346$^{0.0037}_{-0.0038}$  & 0.421$^{0.0043}_{-0.0043}$  & 0.518$^{0.005}_{-0.005}$  & 0.648$^{0.0059}_{-0.0059}$  & 0.811$^{0.0069}_{-0.0069}$  & 1.03$^{0.0082}_{-0.0081}$  & 1.36$^{0.0099}_{-0.0098}$  & 1.82$^{0.013}_{-0.013}$  & 2.72$^{0.017}_{-0.018}$  & 6.4$^{0.14}_{-0.14}$  & 13.9$^{0.78}_{-0.76}$  \\ 
$R_{s}=198.0^{9.7}_{-9.6}$ & M ($10^{13}$ M$_{\odot}$ ) & -  & 0.588$\pm0.0057$ & 0.705$\pm0.0064$ & 0.841$\pm0.0071$ & 1.01$\pm0.0077$ & 1.23$\pm0.0084$ & 1.48$\pm0.0088$ & 1.79$\pm0.0091$ & 2.24$\pm0.0093$ & 2.85$\pm0.0095$ & 3.77$\pm0.012$ & 6.85$\pm0.045$ & 11.2$\pm0.12$ \\ 
\hline 
A2199 & r (kpc)  & 14.8 & 22.8 & 30.9 & 39.0 & 47.9 & 60.4 & 76.6 & 96.0 & 119.3 & 149.6 & 192.1 & 268.5 & 444.2 \\ 
$\rho_{o}=45.6^{2.8}_{-2.1}$ & M$_{g}$ ($10^{11}$ M$_{\odot}$ )  & 0.135$^{0.0048}_{-0.0047}$  & 0.351$^{0.013}_{-0.013}$  & 0.738$^{0.022}_{-0.021}$  & 1.27$^{0.032}_{-0.031}$  & 2.08$^{0.044}_{-0.044}$  & 3.47$^{0.061}_{-0.06}$  & 5.93$^{0.082}_{-0.081}$  & 9.29$^{0.11}_{-0.11}$  & 14.3$^{0.15}_{-0.14}$  & 22.1$^{0.19}_{-0.19}$  & 34.1$^{0.26}_{-0.26}$  & 66.9$^{0.34}_{-0.34}$  & 134.0$^{0.74}_{-0.74}$  \\ 
$R_{s}=363.0^{36.0}_{-26.0}$ & M ($10^{13}$ M$_{\odot}$ ) & -  & 0.109$\pm0.0013$ & 0.194$\pm0.0021$ & 0.3$\pm0.003$ & 0.439$\pm0.0039$ & 0.669$\pm0.005$ & 1.02$\pm0.0061$ & 1.52$\pm0.0067$ & 2.19$\pm0.0069$ & 3.17$\pm0.01$ & 4.68$\pm0.024$ & 7.64$\pm0.071$ & 14.8$\pm0.24$ \\ 
\hline 
A2204 & r (kpc)  & 129.9 & 140.8 & 153.9 & 169.6 & 190.0 & 216.6 & 248.4 & 293.1 & 364.4 & 478.6 & 672.7 & 1022.2 & 1581.2 \\ 
$\rho_{o}=97.9^{3.6}_{-2.7}$ & M$_{g}$ ($10^{13}$ M$_{\odot}$ )  & 0.55$^{0.0051}_{-0.0046}$  & 0.616$^{0.0094}_{-0.025}$  & 0.693$^{0.018}_{-0.029}$  & 0.811$^{0.019}_{-0.03}$  & 0.95$^{0.022}_{-0.032}$  & 1.14$^{0.024}_{-0.032}$  & 1.39$^{0.03}_{-0.036}$  & 1.77$^{0.035}_{-0.041}$  & 2.47$^{0.05}_{-0.052}$  & 3.68$^{0.064}_{-0.065}$  & 5.79$^{0.1}_{-0.1}$  & 9.98$^{0.2}_{-0.2}$  & 17.4$^{0.56}_{-0.56}$  \\ 
$R_{s}=305.0^{54.0}_{-45.0}$ & M ($10^{14}$ M$_{\odot}$ ) & -  & 0.684$\pm0.0077$ & 0.785$\pm0.0086$ & 0.91$\pm0.0097$ & 1.08$\pm0.011$ & 1.3$\pm0.013$ & 1.58$\pm0.015$ & 1.97$\pm0.018$ & 2.59$\pm0.023$ & 3.57$\pm0.031$ & 5.11$\pm0.046$ & 7.51$\pm0.076$ & 10.5$\pm0.12$ \\ 
\hline 
A2597 & r (kpc)  & 59.0 & 66.4 & 75.9 & 87.3 & 103.3 & 125.7 & 157.0 & 204.0 & 288.6 & 551.0 & 965.5 \\ 
$\rho_{o}=39.1^{1.3}_{-1.5}$ & M$_{g}$ ($10^{12}$ M$_{\odot}$ )  & 0.671$^{0.0051}_{-0.0051}$  & 0.819$^{0.01}_{-0.01}$  & 1.02$^{0.015}_{-0.015}$  & 1.29$^{0.02}_{-0.02}$  & 1.71$^{0.025}_{-0.025}$  & 2.33$^{0.033}_{-0.033}$  & 3.33$^{0.043}_{-0.043}$  & 4.89$^{0.058}_{-0.058}$  & 8.34$^{0.081}_{-0.08}$  & 18.6$^{0.28}_{-0.28}$  & 39.5$^{1.5}_{-1.5}$  \\ 
$R_{s}=234.0^{20.0}_{-25.0}$ & M ($10^{13}$ M$_{\odot}$ ) & -  & 0.943$\pm0.014$ & 1.18$\pm0.016$ & 1.49$\pm0.019$ & 1.95$\pm0.021$ & 2.64$\pm0.023$ & 3.68$\pm0.024$ & 5.3$\pm0.021$ & 8.26$\pm0.032$ & 16.7$\pm0.18$ & 27.3$\pm0.45$ \\ 
\hline 
A262 & r (kpc)  & 13.0 & 19.4 & 27.7 & 36.8 & 47.8 & 59.9 & 74.4 & 94.7 & 137.3 & 252.3 & 357.2 \\ 
$\rho_{o}=14.0^{0.52}_{-0.45}$ & M$_{g}$ ($10^{11}$ M$_{\odot}$ )  & 0.0624$^{0.0016}_{-0.0016}$  & 0.13$^{0.004}_{-0.004}$  & 0.252$^{0.0079}_{-0.0079}$  & 0.445$^{0.013}_{-0.013}$  & 0.799$^{0.019}_{-0.019}$  & 1.28$^{0.028}_{-0.028}$  & 2.07$^{0.04}_{-0.04}$  & 3.27$^{0.064}_{-0.064}$  & 7.62$^{0.11}_{-0.11}$  & 22.8$^{0.41}_{-0.4}$  & 49.6$^{0.8}_{-0.79}$  \\ 
$R_{s}=134.0^{11.0}_{-9.7}$ & M ($10^{12}$ M$_{\odot}$ ) & -  & 0.59$\pm0.012$ & 1.1$\pm0.02$ & 1.8$\pm0.027$ & 2.8$\pm0.033$ & 4.1$\pm0.034$ & 5.7$\pm0.032$ & 8.1$\pm0.034$ & 13.0$\pm0.12$ & 27.0$\pm0.57$ & 39.0$\pm1.1$ \\ 
\hline 
A4059 & r (kpc)  & 54.9 & 63.4 & 73.5 & 85.7 & 100.3 & 117.9 & 140.7 & 170.4 & 214.9 & 308.3 & 549.1 & 825.9 \\ 
$\rho_{o}=40.2^{1.7}_{-1.5}$ & M$_{g}$ ($10^{12}$ M$_{\odot}$ )  & 0.203$^{0.0058}_{-0.0084}$  & 0.288$^{0.0085}_{-0.012}$  & 0.415$^{0.011}_{-0.013}$  & 0.583$^{0.014}_{-0.016}$  & 0.803$^{0.017}_{-0.019}$  & 1.1$^{0.022}_{-0.023}$  & 1.54$^{0.028}_{-0.029}$  & 2.29$^{0.034}_{-0.035}$  & 3.34$^{0.047}_{-0.048}$  & 6.15$^{0.081}_{-0.081}$  & 13.7$^{0.3}_{-0.3}$  & 24.6$^{0.71}_{-0.73}$  \\ 
$R_{s}=237.0^{29.0}_{-25.0}$ & M ($10^{13}$ M$_{\odot}$ ) & -  & 0.89$\pm0.019$ & 1.1$\pm0.021$ & 1.5$\pm0.024$ & 1.9$\pm0.025$ & 2.4$\pm0.025$ & 3.2$\pm0.022$ & 4.2$\pm0.02$ & 5.8$\pm0.041$ & 9.2$\pm0.14$ & 17.0$\pm0.49$ & 25.0$\pm0.92$ \\ 
\hline 
A478 & r (kpc)  & 24.4 & 37.8 & 54.0 & 73.4 & 94.8 & 122.5 & 158.6 & 205.5 & 266.4 & 357.9 & 613.9 & 1228.5 \\ 
$\rho_{o}=86.7^{6.6}_{-5.5}$ & M$_{g}$ ($10^{12}$ M$_{\odot}$ )  & 0.0831$^{0.0039}_{-0.0038}$  & 0.277$^{0.0075}_{-0.0075}$  & 0.606$^{0.014}_{-0.014}$  & 1.18$^{0.023}_{-0.023}$  & 2.07$^{0.034}_{-0.034}$  & 3.43$^{0.049}_{-0.049}$  & 5.71$^{0.069}_{-0.069}$  & 9.14$^{0.099}_{-0.099}$  & 14.1$^{0.14}_{-0.14}$  & 22.8$^{0.21}_{-0.21}$  & 51.6$^{0.5}_{-0.5}$  & 118.0$^{2.8}_{-2.8}$  \\ 
$R_{s}=553.0^{68.0}_{-56.0}$ & M ($10^{13}$ M$_{\odot}$ ) & -  & 0.37$\pm0.0059$ & 0.73$\pm0.011$ & 1.3$\pm0.017$ & 2.0$\pm0.024$ & 3.2$\pm0.032$ & 5.0$\pm0.044$ & 7.8$\pm0.063$ & 12.0$\pm0.1$ & 18.0$\pm0.21$ & 38.0$\pm0.74$ & 83.0$\pm2.6$ \\ 
\hline 
A496 & r (kpc)  & 86.0 & 93.9 & 103.3 & 113.7 & 126.2 & 142.3 & 161.8 & 194.8 & 313.6 & 491.9 & 714.4 \\ 
$\rho_{o}=37.9^{0.66}_{-0.62}$ & M$_{g}$ ($10^{12}$ M$_{\odot}$ )  & 0.765$^{0.0065}_{-0.0064}$  & 0.884$^{0.011}_{-0.011}$  & 1.03$^{0.016}_{-0.016}$  & 1.22$^{0.02}_{-0.02}$  & 1.5$^{0.024}_{-0.024}$  & 1.87$^{0.029}_{-0.029}$  & 2.33$^{0.034}_{-0.035}$  & 2.95$^{0.053}_{-0.053}$  & 6.83$^{0.11}_{-0.1}$  & 14.0$^{0.46}_{-0.31}$  & 26.5$^{0.78}_{-0.91}$  \\ 
$R_{s}=131.0^{26.0}_{-21.0}$ & M ($10^{13}$ M$_{\odot}$ ) & -  & 2.2$\pm0.078$ & 2.5$\pm0.081$ & 2.9$\pm0.084$ & 3.3$\pm0.085$ & 3.8$\pm0.085$ & 4.5$\pm0.082$ & 5.6$\pm0.075$ & 9.2$\pm0.13$ & 14.0$\pm0.36$ & 18.0$\pm0.66$ \\ 
\hline 
Centaurus & r (kpc)  & 39.6 & 42.9 & 47.1 & 52.1 & 58.2 & 67.3 & 83.8 & 113.3 & 169.4 & 249.0 \\ 
$\rho_{o}=35.1^{0.27}_{-0.38}$ & M$_{g}$ ($10^{11}$ M$_{\odot}$ )  & 0.944$^{0.0047}_{-0.0047}$  & 1.06$^{0.0079}_{-0.0078}$  & 1.26$^{0.011}_{-0.011}$  & 1.49$^{0.014}_{-0.014}$  & 1.82$^{0.018}_{-0.018}$  & 2.32$^{0.026}_{-0.026}$  & 3.67$^{0.035}_{-0.035}$  & 6.63$^{0.053}_{-0.053}$  & 12.8$^{0.13}_{-0.13}$  & 31.7$^{0.21}_{-0.21}$  \\ 
$R_{s}=276.0$ & M ($10^{12}$ M$_{\odot}$ ) & -  & 3.67$\pm0.034$ & 4.35$\pm0.04$ & 5.22$\pm0.048$ & 6.35$\pm0.059$ & 8.19$\pm0.076$ & 11.9$\pm0.11$ & 19.5$\pm0.18$ & 36.3$\pm0.34$ & 62.3$\pm0.57$ \\ 
\hline 
Cygnus A & r (kpc)  & 77.8 & 88.8 & 103.2 & 119.0 & 139.5 & 168.3 & 202.8 & 244.2 & 298.0 & 362.6 & 453.0 & 597.7 & 814.7 \\ 
$\rho_{o}=45.2^{4.5}_{-2.1}$ & M$_{g}$ ($10^{12}$ M$_{\odot}$ )  & 1.09$^{0.0047}_{-0.0024}$  & 1.3$^{0.0083}_{-0.016}$  & 1.59$^{0.016}_{-0.021}$  & 1.96$^{0.025}_{-0.028}$  & 2.52$^{0.033}_{-0.036}$  & 3.49$^{0.042}_{-0.044}$  & 4.59$^{0.057}_{-0.059}$  & 6.2$^{0.066}_{-0.073}$  & 8.37$^{0.097}_{-0.1}$  & 11.3$^{0.17}_{-0.17}$  & 16.0$^{0.27}_{-0.26}$  & 23.1$^{0.55}_{-0.54}$  & 59.0$^{0.65}_{-0.64}$  \\ 
$R_{s}=145.0^{70.0}_{-43.0}$ & M ($10^{13}$ M$_{\odot}$ ) & -  & 2.3$\pm0.087$ & 2.9$\pm0.094$ & 3.5$\pm0.099$ & 4.3$\pm0.1$ & 5.5$\pm0.1$ & 6.9$\pm0.11$ & 8.5$\pm0.13$ & 10.0$\pm0.21$ & 13.0$\pm0.32$ & 16.0$\pm0.5$ & 20.0$\pm0.79$ & 25.0$\pm1.2$ \\ 
\hline 
HCG62 & r (kpc)  & 16.5 & 19.0 & 22.8 & 27.3 & 32.6 & 39.6 & 48.7 & 60.6 & 74.8 & 97.5 & 137.8 & 192.9 \\ 
$\rho_{o}=7.02^{1.9}_{-0.8}$ & M$_{g}$ ($10^{10}$ M$_{\odot}$ )  & 0.558$^{0.024}_{-0.024}$  & 0.688$^{0.053}_{-0.053}$  & 0.918$^{0.073}_{-0.074}$  & 1.25$^{0.097}_{-0.097}$  & 1.65$^{0.13}_{-0.13}$  & 2.3$^{0.17}_{-0.17}$  & 3.39$^{0.21}_{-0.21}$  & 4.86$^{0.28}_{-0.28}$  & 7.0$^{0.39}_{-0.39}$  & 7.0$^{0.39}_{-0.39}$  & 14.2$^{1.8}_{-1.9}$  & 50.3$^{2.9}_{-2.9}$  \\ 
$R_{s}=48.6^{37.0}_{-19.0}$ & M ($10^{12}$ M$_{\odot}$ ) & -  & 0.6$\pm0.011$ & 0.8$\pm0.014$ & 1.1$\pm0.017$ & 1.4$\pm0.021$ & 1.8$\pm0.025$ & 2.4$\pm0.029$ & 3.1$\pm0.034$ & 4.0$\pm0.04$ & 5.3$\pm0.056$ & 7.4$\pm0.11$ & 9.8$\pm0.21$ \\ 
\hline 
Hercules A & r (kpc)  & 170.8 & 186.0 & 207.3 & 235.1 & 271.1 & 328.8 & 403.8 & 516.3 & 673.8 & 1114.7 \\ 
$\rho_{o}=53.7^{14.0}_{-3.4}$ & M$_{g}$ ($10^{13}$ M$_{\odot}$ )  & 0.352$^{0.0054}_{-0.0054}$  & 0.419$^{0.01}_{-0.012}$  & 0.494$^{0.015}_{-0.017}$  & 0.589$^{0.023}_{-0.023}$  & 0.738$^{0.028}_{-0.029}$  & 0.945$^{0.032}_{-0.033}$  & 1.24$^{0.056}_{-0.052}$  & 1.72$^{0.12}_{-0.1}$  & 2.33$^{0.2}_{-0.18}$  & 5.62$^{0.69}_{-0.7}$  \\ 
$R_{s}=135.0^{190.0}_{-85.0}$ & M ($10^{14}$ M$_{\odot}$ ) & -  & 0.75$\pm0.056$ & 0.85$\pm0.059$ & 0.97$\pm0.063$ & 1.1$\pm0.066$ & 1.4$\pm0.072$ & 1.7$\pm0.08$ & 2.0$\pm0.098$ & 2.5$\pm0.13$ & 3.5$\pm0.23$ \\ 
\hline 
Hydra A & r (kpc)  & 55.3 & 63.6 & 74.5 & 88.7 & 105.6 & 126.0 & 150.5 & 179.8 & 223.8 & 315.9 & 473.8 & 789.7 \\ 
$\rho_{o}=35.3^{1.2}_{-0.65}$ & M$_{g}$ ($10^{12}$ M$_{\odot}$ )  & 0.472$^{0.0025}_{-0.0025}$  & 0.607$^{0.005}_{-0.005}$  & 0.787$^{0.0077}_{-0.0076}$  & 1.06$^{0.011}_{-0.011}$  & 1.39$^{0.015}_{-0.014}$  & 1.85$^{0.02}_{-0.02}$  & 2.53$^{0.026}_{-0.026}$  & 3.53$^{0.034}_{-0.034}$  & 5.44$^{0.042}_{-0.042}$  & 9.28$^{0.069}_{-0.069}$  & 17.1$^{0.15}_{-0.15}$  & 31.6$^{0.47}_{-0.47}$  \\ 
$R_{s}=341.0^{25.0}_{-24.0}$ & M ($10^{13}$ M$_{\odot}$ ) & -  & 0.598$\pm0.007$ & 0.792$\pm0.0083$ & 1.07$\pm0.0098$ & 1.44$\pm0.011$ & 1.93$\pm0.012$ & 2.57$\pm0.012$ & 3.38$\pm0.013$ & 4.68$\pm0.02$ & 7.56$\pm0.062$ & 12.5$\pm0.18$ & 21.6$\pm0.47$ \\ 
\hline 
MACSJ0159 & r (kpc)  & 23.4 & 47.0 & 77.5 & 120.4 & 176.0 & 259.5 & 434.9 & 1311.7 \\ 
$\rho_{o}=125.0^{100.0}_{-39.0}$ & M$_{g}$ ($10^{13}$ M$_{\odot}$ )  & 0.0142$^{0.0014}_{-0.0014}$  & 0.0596$^{0.0034}_{-0.0034}$  & 0.15$^{0.0086}_{-0.0086}$  & 0.339$^{0.017}_{-0.017}$  & 0.72$^{0.027}_{-0.027}$  & 1.45$^{0.042}_{-0.042}$  & 3.47$^{0.057}_{-0.058}$  & 15.8$^{0.27}_{-0.27}$  \\ 
$R_{s}=700.0^{730.0}_{-270.0}$ & M ($10^{13}$ M$_{\odot}$ )  & 0.17$\pm0.006$ & 0.65$\pm0.022$ & 1.7$\pm0.057$ & 3.7$\pm0.12$ & 7.3$\pm0.25$ & 14.0$\pm0.5$ & 31.0$\pm1.3$ & 130.0$\pm8.0$ \\ 
\hline 
MACSJ0242 & r (kpc)  & 24.0 & 51.2 & 97.6 & 208.7 & 679.0 \\ 
$\rho_{o}=57.9^{63.0}_{-12.0}$ & M$_{g}$ ($10^{12}$ M$_{\odot}$ )  & 0.226$^{0.031}_{-0.028}$  & 0.974$^{0.079}_{-0.14}$  & 2.84$^{0.21}_{-0.24}$  & 9.58$^{0.37}_{-0.47}$  & 49.7$^{1.1}_{-2.6}$  \\ 
$R_{s}=258.0^{230.0}_{-81.0}$ & M ($10^{13}$ M$_{\odot}$ )  & 0.21$\pm0.022$ & 0.83$\pm0.068$ & 2.5$\pm0.16$ & 7.8$\pm0.7$ & 30.0$\pm5.7$ \\ 
\hline 
MACSJ0429 & r (kpc)  & 28.2 & 58.0 & 102.8 & 183.3 & 336.3 & 791.1 \\ 
$\rho_{o}=54.5^{14.0}_{-6.8}$ & M$_{g}$ ($10^{13}$ M$_{\odot}$ )  & 0.0202$^{0.0028}_{-0.0027}$  & 0.0899$^{0.0068}_{-0.0066}$  & 0.258$^{0.019}_{-0.018}$  & 0.669$^{0.038}_{-0.038}$  & 1.75$^{0.096}_{-0.096}$  & 5.72$^{0.34}_{-0.34}$  \\ 
$R_{s}=202.0^{100.0}_{-51.0}$ & M ($10^{13}$ M$_{\odot}$ )  & 0.32$\pm0.024$ & 1.2$\pm0.064$ & 2.9$\pm0.12$ & 6.7$\pm0.32$ & 14.0$\pm1.1$ & 31.0$\pm4.5$ \\ 
\hline 
MACSJ0913 & r (kpc)  & 24.4 & 45.1 & 76.1 & 141.2 & 336.5 & 827.3 \\ 
$\rho_{o}=71.7^{18.0}_{-8.6}$ & M$_{g}$ ($10^{13}$ M$_{\odot}$ )  & 0.0159$^{0.0015}_{-0.0015}$  & 0.068$^{0.0039}_{-0.0038}$  & 0.184$^{0.0071}_{-0.007}$  & 0.499$^{0.015}_{-0.015}$  & 1.69$^{0.051}_{-0.051}$  & 5.51$^{0.27}_{-0.26}$  \\ 
$R_{s}=305.0^{120.0}_{-57.0}$ & M ($10^{13}$ M$_{\odot}$ )  & 0.23$\pm0.0093$ & 0.71$\pm0.023$ & 1.8$\pm0.046$ & 5.0$\pm0.15$ & 17.0$\pm1.1$ & 46.0$\pm5.0$ \\ 
\hline 
MACSJ1423 & r (kpc)  & 31.3 & 45.8 & 67.4 & 99.8 & 161.4 & 278.5 & 618.1 & 1263.4 \\ 
$\rho_{o}=70.6^{8.1}_{-6.0}$ & M$_{g}$ ($10^{13}$ M$_{\odot}$ )  & 0.0327$^{0.0017}_{-0.0017}$  & 0.0738$^{0.0036}_{-0.0035}$  & 0.145$^{0.0063}_{-0.0063}$  & 0.276$^{0.011}_{-0.011}$  & 0.604$^{0.018}_{-0.018}$  & 1.42$^{0.038}_{-0.038}$  & 4.42$^{0.14}_{-0.14}$  & 10.7$^{0.87}_{-0.84}$  \\ 
$R_{s}=211.0^{59.0}_{-46.0}$ & M ($10^{13}$ M$_{\odot}$ ) & -  & 0.97$\pm0.065$ & 1.9$\pm0.11$ & 3.5$\pm0.16$ & 7.2$\pm0.23$ & 15.0$\pm0.53$ & 33.0$\pm2.3$ & 58.0$\pm5.8$ \\ 
\hline 
MACSJ1720 & r (kpc)  & 42.2 & 90.4 & 157.8 & 258.9 & 450.9 & 796.6 \\ 
$\rho_{o}=67.8^{18.0}_{-16.0}$ & M$_{g}$ ($10^{13}$ M$_{\odot}$ )  & 0.0462$^{0.0029}_{-0.0028}$  & 0.176$^{0.0091}_{-0.0092}$  & 0.504$^{0.019}_{-0.019}$  & 1.2$^{0.041}_{-0.041}$  & 2.93$^{0.085}_{-0.087}$  & 6.48$^{0.24}_{-0.24}$  \\ 
$R_{s}=405.0^{170.0}_{-150.0}$ & M ($10^{14}$ M$_{\odot}$ )  & 0.047$\pm0.003$ & 0.19$\pm0.0094$ & 0.48$\pm0.023$ & 1.0$\pm0.063$ & 2.2$\pm0.2$ & 4.2$\pm0.52$ \\ 
\hline 
MACSJ1931 & r (kpc)  & 29.3 & 36.8 & 48.2 & 62.9 & 77.7 & 101.3 & 132.0 & 174.9 & 243.7 & 381.1 & 917.2 & 1721.4 \\ 
$\rho_{o}=126.0^{43.0}_{-22.0}$ & M$_{g}$ ($10^{12}$ M$_{\odot}$ )  & 0.299$^{0.0096}_{-0.0095}$  & 0.522$^{0.029}_{-0.027}$  & 0.87$^{0.048}_{-0.042}$  & 1.46$^{0.067}_{-0.064}$  & 2.14$^{0.084}_{-0.079}$  & 3.41$^{0.1}_{-0.11}$  & 5.23$^{0.15}_{-0.16}$  & 8.2$^{0.22}_{-0.23}$  & 14.3$^{0.31}_{-0.31}$  & 28.5$^{0.5}_{-0.5}$  & 84.1$^{2.1}_{-2.1}$  & 199.0$^{19.0}_{-14.0}$  \\ 
$R_{s}=851.0^{390.0}_{-200.0}$ & M ($10^{13}$ M$_{\odot}$ ) & -  & 0.34$\pm0.0087$ & 0.57$\pm0.014$ & 0.96$\pm0.021$ & 1.4$\pm0.029$ & 2.3$\pm0.042$ & 3.8$\pm0.06$ & 6.3$\pm0.092$ & 11.0$\pm0.19$ & 23.0$\pm0.61$ & 82.0$\pm4.7$ & 170.0$\pm14.0$ \\ 
\hline 
MACSJ2046 & r (kpc)  & 59.1 & 139.0 & 466.5 & 957.4 \\ 
$\rho_{o}=52.9^{16.0}_{-5.3}$ & M$_{g}$ ($10^{13}$ M$_{\odot}$ )  & 0.0838$^{0.0041}_{-0.0042}$  & 0.438$^{0.017}_{-0.016}$  & 2.55$^{0.055}_{-0.087}$  & 7.82$^{1.1}_{-1.2}$  \\ 
$R_{s}=210.0^{130.0}_{-45.0}$ & M ($10^{14}$ M$_{\odot}$ )  & 0.11$\pm0.0076$ & 0.44$\pm0.017$ & 1.9$\pm0.26$ & 3.6$\pm0.76$ \\ 
\hline 
MACSJ2140 & r (kpc)  & 19.9 & 33.7 & 47.5 & 71.0 & 101.6 & 147.4 & 239.0 & 523.1 & 920.7 \\ 
$\rho_{o}=51.1^{5.2}_{-2.7}$ & M$_{g}$ ($10^{12}$ M$_{\odot}$ )  & 0.0837$^{0.018}_{-0.018}$  & 0.337$^{0.027}_{-0.032}$  & 0.687$^{0.062}_{-0.065}$  & 1.52$^{0.086}_{-0.12}$  & 2.48$^{0.33}_{-0.22}$  & 4.96$^{0.34}_{-0.24}$  & 10.4$^{0.59}_{-0.63}$  & 28.1$^{1.3}_{-1.3}$  & 57.3$^{3.4}_{-3.9}$  \\ 
$R_{s}=176.0^{36.0}_{-30.0}$ & M ($10^{13}$ M$_{\odot}$ )  & 0.18$\pm0.0075$ & 0.47$\pm0.016$ & 0.85$\pm0.022$ & 1.7$\pm0.031$ & 2.9$\pm0.055$ & 4.9$\pm0.14$ & 9.1$\pm0.45$ & 20.0$\pm1.7$ & 32.0$\pm3.4$ \\ 
\hline 
MKW3S & r (kpc)  & 95.3 & 103.7 & 113.8 & 126.0 & 140.6 & 159.5 & 186.1 & 223.3 & 275.3 & 363.8 & 567.3 \\ 
$\rho_{o}=34.9^{12.0}_{-4.3}$ & M$_{g}$ ($10^{12}$ M$_{\odot}$ )  & 0.816$^{0.011}_{-0.0098}$  & 0.966$^{0.015}_{-0.018}$  & 1.13$^{0.025}_{-0.021}$  & 1.37$^{0.035}_{-0.032}$  & 1.68$^{0.044}_{-0.043}$  & 2.13$^{0.05}_{-0.051}$  & 2.76$^{0.056}_{-0.062}$  & 3.65$^{0.074}_{-0.078}$  & 5.12$^{0.095}_{-0.084}$  & 7.99$^{0.12}_{-0.12}$  & 14.1$^{0.19}_{-0.19}$  \\ 
$R_{s}=214.0^{210.0}_{-90.0}$ & M ($10^{13}$ M$_{\odot}$ ) & -  & 1.8$\pm0.092$ & 2.1$\pm0.093$ & 2.5$\pm0.091$ & 2.9$\pm0.086$ & 3.5$\pm0.078$ & 4.3$\pm0.077$ & 5.5$\pm0.13$ & 7.1$\pm0.27$ & 9.8$\pm0.58$ & 15.0$\pm1.5$ \\ 
\hline 
MS0735 & r (kpc)  & 258.3 & 271.9 & 288.2 & 309.5 & 339.2 & 377.8 & 431.9 & 513.0 & 650.9 & 954.2 & 1549.8 \\ 
$\rho_{o}=66.2^{13.0}_{-3.0}$ & M$_{g}$ ($10^{13}$ M$_{\odot}$ )  & 0.987$^{0.0083}_{-0.0085}$  & 1.06$^{0.019}_{-0.011}$  & 1.16$^{0.026}_{-0.02}$  & 1.31$^{0.029}_{-0.027}$  & 1.51$^{0.035}_{-0.035}$  & 1.76$^{0.049}_{-0.048}$  & 2.15$^{0.063}_{-0.056}$  & 2.7$^{0.071}_{-0.067}$  & 3.8$^{0.075}_{-0.072}$  & 6.03$^{0.11}_{-0.12}$  & 11.1$^{0.35}_{-0.39}$  \\ 
$R_{s}=368.0^{27.0}_{-140.0}$ & M ($10^{14}$ M$_{\odot}$ ) & -  & 1.1$\pm0.045$ & 1.2$\pm0.046$ & 1.3$\pm0.047$ & 1.5$\pm0.048$ & 1.7$\pm0.048$ & 2.1$\pm0.049$ & 2.5$\pm0.051$ & 3.3$\pm0.069$ & 4.9$\pm0.15$ & 7.4$\pm0.36$ \\ 
\hline 
NGC1550 & r (kpc)  & 16.6 & 20.8 & 25.9 & 32.4 & 40.5 & 50.7 & 63.3 & 79.2 & 99.0 & 123.7 & 154.6 & 193.3 & 241.6 \\ 
$\rho_{o}=9.62^{0.25}_{-0.16}$ & M$_{g}$ ($10^{11}$ M$_{\odot}$ )  & 0.0891$^{0.0023}_{-0.0022}$  & 0.13$^{0.0052}_{-0.0052}$  & 0.202$^{0.0086}_{-0.0083}$  & 0.308$^{0.013}_{-0.013}$  & 0.499$^{0.016}_{-0.018}$  & 0.762$^{0.021}_{-0.027}$  & 1.17$^{0.045}_{-0.04}$  & 1.83$^{0.082}_{-0.075}$  & 2.77$^{0.1}_{-0.098}$  & 4.15$^{0.13}_{-0.13}$  & 5.87$^{0.26}_{-0.19}$  & 7.82$^{0.48}_{-0.59}$  & 13.4$^{0.58}_{-0.73}$  \\ 
$R_{s}=58.7^{5.3}_{-3.7}$ & M ($10^{12}$ M$_{\odot}$ ) & -  & 0.84$\pm0.024$ & 1.2$\pm0.028$ & 1.7$\pm0.032$ & 2.4$\pm0.033$ & 3.2$\pm0.029$ & 4.3$\pm0.026$ & 5.7$\pm0.045$ & 7.3$\pm0.094$ & 9.2$\pm0.17$ & 11.0$\pm0.27$ & 14.0$\pm0.41$ & 17.0$\pm0.58$ \\ 
\hline 
NGC5044 & r (kpc)  & 18.7 & 23.4 & 29.2 & 36.5 & 45.6 & 57.1 & 71.3 & 89.1 & 111.4 & 139.3 & 174.1 & 217.7 \\ 
$\rho_{o}=6.97^{0.2}_{-0.16}$ & M$_{g}$ ($10^{11}$ M$_{\odot}$ )  & 0.12$^{0.0024}_{-0.0024}$  & 0.18$^{0.0051}_{-0.005}$  & 0.268$^{0.0067}_{-0.0078}$  & 0.396$^{0.01}_{-0.011}$  & 0.583$^{0.015}_{-0.016}$  & 0.806$^{0.029}_{-0.029}$  & 1.2$^{0.049}_{-0.052}$  & 1.82$^{0.085}_{-0.092}$  & 2.78$^{0.12}_{-0.12}$  & 3.99$^{0.2}_{-0.21}$  & 5.08$^{0.29}_{-0.36}$  & 8.99$^{0.45}_{-0.45}$  \\ 
$R_{s}=45.1^{5.4}_{-4.7}$ & M ($10^{12}$ M$_{\odot}$ ) & -  & 0.86$\pm0.021$ & 1.2$\pm0.023$ & 1.6$\pm0.022$ & 2.2$\pm0.017$ & 2.9$\pm0.014$ & 3.8$\pm0.034$ & 4.8$\pm0.073$ & 6.0$\pm0.13$ & 7.4$\pm0.2$ & 8.9$\pm0.29$ & 11.0$\pm0.4$ \\ 
\hline 
NGC507 & r (kpc)  & 20.1 & 24.8 & 30.9 & 39.5 & 49.0 & 62.2 & 79.4 & 100.0 & 126.8 & 167.0 \\ 
$\rho_{o}=10.4^{2.4}_{-1.5}$ & M$_{g}$ ($10^{11}$ M$_{\odot}$ )  & 0.0713$^{0.0095}_{-0.0094}$  & 0.109$^{0.022}_{-0.022}$  & 0.187$^{0.031}_{-0.031}$  & 0.329$^{0.044}_{-0.044}$  & 0.55$^{0.054}_{-0.054}$  & 0.861$^{0.072}_{-0.072}$  & 1.33$^{0.1}_{-0.1}$  & 2.1$^{0.14}_{-0.13}$  & 3.07$^{0.23}_{-0.22}$  & 6.63$^{0.26}_{-0.25}$  \\ 
$R_{s}=115.0^{77.0}_{-34.0}$ & M ($10^{12}$ M$_{\odot}$ ) & -  & 0.77$\pm0.027$ & 1.1$\pm0.033$ & 1.7$\pm0.038$ & 2.4$\pm0.038$ & 3.5$\pm0.042$ & 5.0$\pm0.089$ & 6.9$\pm0.2$ & 9.4$\pm0.39$ & 13.0$\pm0.76$ \\ 
\hline 
Perseus & r (kpc)  & 35.4 & 44.3 & 55.4 & 69.2 & 86.5 & 108.1 & 135.1 & 168.9 & 211.1 & 263.9 & 329.9 \\ 
$\rho_{o}=39.3^{0.5}_{-0.46}$ & M$_{g}$ ($10^{12}$ M$_{\odot}$ )  & 0.205$^{0.00057}_{-0.00056}$  & 0.37$^{0.0011}_{-0.0011}$  & 0.641$^{0.0016}_{-0.0016}$  & 0.994$^{0.0022}_{-0.0022}$  & 1.51$^{0.0029}_{-0.0029}$  & 2.21$^{0.0039}_{-0.0039}$  & 3.19$^{0.0054}_{-0.0054}$  & 4.43$^{0.0083}_{-0.0083}$  & 6.31$^{0.013}_{-0.013}$  & 8.75$^{0.023}_{-0.023}$  & 14.4$^{0.031}_{-0.031}$  \\ 
$R_{s}=205.0^{4.9}_{-4.6}$ & M ($10^{13}$ M$_{\odot}$ ) & -  & 0.518$\pm0.002$ & 0.763$\pm0.0025$ & 1.11$\pm0.003$ & 1.6$\pm0.0034$ & 2.26$\pm0.0039$ & 3.15$\pm0.0056$ & 4.32$\pm0.01$ & 5.8$\pm0.019$ & 7.66$\pm0.033$ & 9.91$\pm0.053$ \\ 
\hline 
PKS0745 & r (kpc)  & 46.5 & 53.4 & 61.6 & 71.5 & 83.4 & 97.7 & 116.3 & 140.4 & 176.5 & 223.6 & 303.5 & 495.3 & 1115.8 \\ 
$\rho_{o}=68.9^{5.3}_{-4.5}$ & M$_{g}$ ($10^{12}$ M$_{\odot}$ )  & 0.661$^{0.0061}_{-0.006}$  & 0.865$^{0.011}_{-0.011}$  & 1.15$^{0.016}_{-0.016}$  & 1.53$^{0.021}_{-0.021}$  & 2.05$^{0.026}_{-0.026}$  & 2.74$^{0.033}_{-0.033}$  & 3.77$^{0.041}_{-0.041}$  & 5.26$^{0.051}_{-0.05}$  & 7.62$^{0.067}_{-0.067}$  & 11.3$^{0.089}_{-0.089}$  & 18.9$^{0.12}_{-0.12}$  & 35.7$^{0.24}_{-0.24}$  & 101.0$^{1.5}_{-1.5}$  \\ 
$R_{s}=351.0^{50.0}_{-42.0}$ & M ($10^{13}$ M$_{\odot}$ ) & -  & 0.83$\pm0.016$ & 1.1$\pm0.02$ & 1.4$\pm0.023$ & 1.8$\pm0.027$ & 2.4$\pm0.03$ & 3.2$\pm0.033$ & 4.4$\pm0.036$ & 6.3$\pm0.051$ & 9.0$\pm0.1$ & 14.0$\pm0.24$ & 26.0$\pm0.78$ & 58.0$\pm3.0$ \\ 
\hline 
PKS1404 & r (kpc)  & 8.7 & 13.0 & 17.8 & 23.0 & 29.3 & 36.8 & 48.1 & 64.0 & 87.7 & 140.0 & 260.3 \\ 
$\rho_{o}=14.4^{0.41}_{-0.37}$ & M$_{g}$ ($10^{11}$ M$_{\odot}$ )  & 0.0231$^{0.0018}_{-0.0018}$  & 0.0599$^{0.0039}_{-0.0039}$  & 0.134$^{0.0065}_{-0.0064}$  & 0.241$^{0.0095}_{-0.0095}$  & 0.396$^{0.014}_{-0.014}$  & 0.665$^{0.019}_{-0.019}$  & 1.1$^{0.027}_{-0.027}$  & 1.85$^{0.038}_{-0.038}$  & 3.34$^{0.057}_{-0.056}$  & 7.44$^{0.13}_{-0.13}$  & 23.6$^{0.45}_{-0.43}$  \\ 
$R_{s}=98.2^{6.0}_{-5.4}$ & M ($10^{12}$ M$_{\odot}$ ) & -  & 0.375$\pm0.0062$ & 0.663$\pm0.0097$ & 1.05$\pm0.013$ & 1.59$\pm0.017$ & 2.32$\pm0.02$ & 3.55$\pm0.021$ & 5.44$\pm0.02$ & 8.45$\pm0.042$ & 15.1$\pm0.17$ & 28.8$\pm0.58$ \\ 
\hline 
RBS797 & r (kpc)  & 49.0 & 60.3 & 76.1 & 98.1 & 131.3 & 194.2 & 332.7 & 980.1 & 1715.1 \\ 
$\rho_{o}=102.0^{29.0}_{-15.0}$ & M$_{g}$ ($10^{13}$ M$_{\odot}$ )  & 0.101$^{0.003}_{-0.003}$  & 0.148$^{0.007}_{-0.0071}$  & 0.232$^{0.0098}_{-0.0098}$  & 0.356$^{0.013}_{-0.013}$  & 0.592$^{0.018}_{-0.018}$  & 1.06$^{0.027}_{-0.027}$  & 2.34$^{0.072}_{-0.057}$  & 8.72$^{0.34}_{-0.33}$  & 16.5$^{1.9}_{-2.0}$  \\ 
$R_{s}=421.0^{230.0}_{-110.0}$ & M ($10^{13}$ M$_{\odot}$ ) & -  & 1.3$\pm0.056$ & 2.0$\pm0.08$ & 3.2$\pm0.12$ & 5.2$\pm0.18$ & 9.8$\pm0.33$ & 22.0$\pm0.86$ & 77.0$\pm5.7$ & 130.0$\pm12.0$ \\ 
\hline 
RXJ1532 & r (kpc)  & 60.1 & 66.6 & 74.4 & 85.4 & 98.4 & 115.5 & 141.0 & 179.3 & 244.4 & 381.2 & 969.2 \\ 
$\rho_{o}=93.5^{44.0}_{-17.0}$ & M$_{g}$ ($10^{12}$ M$_{\odot}$ )  & 1.34$^{0.028}_{-0.028}$  & 1.67$^{0.037}_{-0.036}$  & 2.09$^{0.063}_{-0.061}$  & 2.64$^{0.094}_{-0.093}$  & 3.4$^{0.12}_{-0.12}$  & 4.53$^{0.13}_{-0.13}$  & 6.21$^{0.17}_{-0.16}$  & 8.96$^{0.22}_{-0.22}$  & 14.2$^{0.26}_{-0.29}$  & 26.2$^{0.45}_{-0.34}$  & 81.7$^{2.4}_{-2.3}$  \\ 
$R_{s}=502.0^{380.0}_{-150.0}$ & M ($10^{13}$ M$_{\odot}$ ) & -  & 1.3$\pm0.048$ & 1.5$\pm0.055$ & 2.0$\pm0.064$ & 2.5$\pm0.073$ & 3.4$\pm0.083$ & 4.8$\pm0.097$ & 7.1$\pm0.14$ & 12.0$\pm0.34$ & 22.0$\pm1.2$ & 70.0$\pm8.9$ \\ 
\hline 
Sersic159 & r (kpc)  & 33.1 & 37.6 & 43.0 & 49.5 & 57.3 & 67.5 & 80.7 & 97.8 & 125.3 & 177.4 & 281.7 & 552.9 \\ 
$\rho_{o}=29.6^{2.1}_{-1.8}$ & M$_{g}$ ($10^{12}$ M$_{\odot}$ )  & 0.134$^{0.0032}_{-0.0032}$  & 0.175$^{0.0061}_{-0.0056}$  & 0.229$^{0.0092}_{-0.0084}$  & 0.316$^{0.012}_{-0.01}$  & 0.421$^{0.015}_{-0.013}$  & 0.574$^{0.018}_{-0.018}$  & 0.794$^{0.025}_{-0.023}$  & 1.12$^{0.03}_{-0.028}$  & 1.7$^{0.037}_{-0.035}$  & 3.03$^{0.055}_{-0.053}$  & 5.83$^{0.089}_{-0.088}$  & 15.1$^{0.26}_{-0.25}$  \\ 
$R_{s}=233.0^{29.0}_{-26.0}$ & M ($10^{12}$ M$_{\odot}$ ) & -  & 2.6$\pm0.044$ & 3.4$\pm0.051$ & 4.3$\pm0.058$ & 5.6$\pm0.064$ & 7.4$\pm0.067$ & 9.9$\pm0.065$ & 14.0$\pm0.062$ & 20.0$\pm0.12$ & 33.0$\pm0.44$ & 61.0$\pm1.5$ & 130.0$\pm5.3$ \\ 
\hline 
Zw2701 & r (kpc)  & 51.5 & 59.6 & 70.1 & 83.7 & 101.4 & 124.5 & 156.8 & 201.9 & 274.2 & 418.8 & 1098.2 \\ 
$\rho_{o}=48.8^{6.1}_{-3.5}$ & M$_{g}$ ($10^{12}$ M$_{\odot}$ )  & 0.37$^{0.022}_{-0.02}$  & 0.526$^{0.035}_{-0.042}$  & 0.747$^{0.046}_{-0.05}$  & 1.07$^{0.05}_{-0.063}$  & 1.55$^{0.067}_{-0.077}$  & 2.2$^{0.097}_{-0.11}$  & 3.33$^{0.13}_{-0.13}$  & 5.09$^{0.17}_{-0.17}$  & 8.23$^{0.24}_{-0.24}$  & 16.4$^{0.3}_{-0.3}$  & 55.1$^{1.8}_{-1.8}$  \\ 
$R_{s}=248.0^{79.0}_{-49.0}$ & M ($10^{13}$ M$_{\odot}$ ) & -  & 0.94$\pm0.039$ & 1.2$\pm0.047$ & 1.7$\pm0.057$ & 2.3$\pm0.069$ & 3.2$\pm0.082$ & 4.5$\pm0.1$ & 6.4$\pm0.15$ & 9.5$\pm0.29$ & 16.0$\pm0.7$ & 38.0$\pm3.1$ \\ 
\hline 
Zw3146 & r (kpc)  & 10.7 & 27.2 & 45.4 & 63.6 & 85.4 & 118.1 & 170.4 & 259.4 & 552.9 & 2578.5 \\ 
$\rho_{o}=107.0^{33.0}_{-21.0}$ & M$_{g}$ ($10^{12}$ M$_{\odot}$ )  & 0.032$^{0.006}_{-0.0056}$  & 0.222$^{0.024}_{-0.017}$  & 0.598$^{0.041}_{-0.035}$  & 1.29$^{0.083}_{-0.051}$  & 2.42$^{0.1}_{-0.085}$  & 4.53$^{0.14}_{-0.14}$  & 8.43$^{0.23}_{-0.24}$  & 17.2$^{0.33}_{-0.35}$  & 49.5$^{1.1}_{-0.98}$  & 285.0$^{9.4}_{-12.0}$  \\ 
$R_{s}=738.0^{310.0}_{-190.0}$ & M ($10^{13}$ M$_{\odot}$ ) & -  & 0.18$\pm0.0048$ & 0.49$\pm0.012$ & 0.94$\pm0.022$ & 1.6$\pm0.036$ & 3.0$\pm0.062$ & 5.7$\pm0.12$ & 12.0$\pm0.28$ & 37.0$\pm1.5$ & 210.0$\pm19.0$ \\ 
\hline

\end{longtable}

\end{document}